\documentclass[12pt,thmsa,a4paper]{book}%
\usepackage{amsmath}

\usepackage{amsfonts}
\usepackage{amssymb}

\begin{document}

\title{The general structure and ergodic properties of quantum and classical
mechanics: \\A unified C*-algebraic approach}
\author{Rocco de Villiers Duvenhage
\and PhD thesis, Department of Mathematics and Applied Mathematics
\and University of Pretoria}
\date{25 May 2002}
\maketitle

\newpage

\begin{center}

{\Large Dankbetuigings}
\end{center}

\begin{quotation}
Dankie aan almal wat my direk of indirek in hierdie projek
ondersteun het, in besonder my promotor Prof. Anton Str\"{o}h.
Gedurende my studies het gesprekke met die volgende mense ook
gehelp om my idees in verband met kwantum meganika te vorm: Conrad
Beyers, James Chalmers, Richard de Beer, Prof. Johan Malherbe,
Prof. Elem\'{e}r Rosinger, Prof. Johan Swart en Gusti van Zyl.

Thanks also to Prof. L\'{a}szl\'{o} Zsid\'{o} for mathematical support
relating to ergodic theory, and to Chris Fuchs for encouraging remarks
concerning the use of information theoretic ideas in quantum mechanics.

Lastly I thank the National Research Foundation and the Mellon Foundation
Mentoring Programme for financial support, among other things to visit
Universit\`{a} degli Studi di Roma ``Tor Vergata'' in November/December 2000
and again in November/December 2001.
\end{quotation}

\tableofcontents

\section{List of symbols and terms}

\noindent\textbf{General symbols}

$\varnothing$ is the empty set.

$A\subset B$ means that the set $A$ is contained in the set $B$, with $A=B$ allowed.

$\overline{S}$ is the closure of a set $S$ in a topological space.

$\mathbb{C}$ is the set of complex numbers.

$\mathbb{N}=\{1,2,3,...\}$.

$\mathbb{R}$ is the set of real numbers.

\bigskip$f^{\ast}=\overline{f}$ is the complex-conjugate of a complex-valued
function $f$.

$\lambda$ is the Lebesgue measure on $\mathbb{R}^{2n}$.

$\mathfrak{L}(V)$ is the space of bounded linear operators $V\rightarrow V$ on
a normed space $V$.

Tr denotes the trace of a bounded linear operator on a Hilbert space (see
\textbf{[Mu]}).

\bigskip\noindent\textbf{Symbols defined in the text}

$A\leq B$ in a $\ast$-algebra, 2.7

$B_{\infty}(F)$, 1.7.2

$B_{\infty}(\mathbb{R}^{2n})$, 1.3

$B_{\infty}(\mathbb{\Sigma})$, 1.7, 2.1

$L(V)$, 2.2

tr, 1.6

$\chi_{A}$, 1.3

$x\otimes y$, 2.3

$\left\|  \cdot\right\|  _{\varphi}$ where $\varphi$ is a state on a unital
$\ast$-algebra, 2.2

$[\cdot]$, 2.3

\bigskip\noindent\textbf{Terms}

accurate, precise, 1.4

bounded quantum system, 1.7.3

Cauchy-Schwarz inequality for states, 2.2.2, 2.5.6 (in the proofs)

conditional probability, 1.3, 1.5, 1.6.1

constant energy surface, 3.2.8

density operator, 1.2

ergodic, 2.3.2

factor, finite factor, 1.8.2

finite von Neumann algebra, 1.7

faithful, 1.7, 1.7.4, 2.7.1

flow, 1.3

Hamiltonian flow, 1.3

ideal measurement, 1.5.1

information, 1.4(b), 1.6.1

measurement, 1.1.2, 1.6

noncommutative information, 1.6, 1.6.1

normal state, 1.7.4

phase space, phase point, 1.3

pure state, 1.3, 1.4(b), 1.7.4, 1.9.2

$\ast$-dynamical system, 2.3.1

spectral projection, 1.4.2, 1.2, 1.3

state, 1.2, 1.4, 2.2

unital $\ast$-algebra, 2.1

\bigskip All Hilbert spaces are assumed complex.

We will use units in which $\hbar=h/2\pi=1$, where $h$ here denotes Planck's constant.\newpage

\section{Introduction}

The main motivation for this thesis is to gain a deeper understanding of the
structure and nature of quantum mechanics. This will be achieved by a careful
analysis of the relationship between quantum mechanics and classical
mechanics. Quantum mechanics is inherently a statistical theory, while
classical mechanics is not. The essential idea is therefore to study the
general structure of statistical mechanics in a mathematical framework that
applies to both quantum mechanics and classical mechanics. The language of
abstract C*-algebras is ideally suited for this, since it provides a unified
formulation of quantum mechanics and classical mechanics, with classical
mechanics then viewed as a special case of quantum mechanics where we have
commutativity. The concrete realizations of the C*-algebras in quantum
mechanics consist of linear operators on Hilbert spaces, which are
mathematical objects that differ very much from the measurable functions that
make up the concrete realizations of the C*-algebras in classical mechanics.
For this reason the abstract approach clarifies the general structure of
mechanics (quantum and classical), enabling the above mentioned unified
formulation of mechanics. This is discussed in detail in Sections 1.1 to 1.5
of Chapter 1.

From a mathematical point of view the general structure of classical mechanics
to be presented is nothing more than probability theory (or, a probabilistic
description of information) with dynamics, while the general structure of
quantum mechanics is noncommutative probability theory (or, a probabilistic
description of ``noncommutative information'') with dynamics. From a physical
point of view the information referred to here is the information an observer
has regarding the state of the physical system in question, while the dynamics
describes the time-evolution of the system. The mathematics then suggests an
interpretation of quantum mechanics in terms of the idea of noncommutative
information, which clarifies several conceptual problems surrounding the
measuring process. This interpretation is discussed in Section 1.6.

As is implied above, our view of statistical mechanics is as a description of
situations where we have incomplete information about the state of a physical
system (quantum or classical). In practice this is generally the case, since
exact measurements are impossible, except for some simple quantum systems
whose observables have discrete values which are separated enough to be
distinguished by our measurements. If an observable has a continuous spectrum
of values, then the best we can hope for when measuring the observable, is to
obtain an interval of values containing the ``actual'' value of the observable
( if we do not measure an observable of a quantum system exactly, then it does
not really make sense to say that the observable has an actual precise value,
unlike in classical mechanics where it is possible to think of an observable
as having an exact value, even if we did not measure it exactly). For
classical mechanics the most important observables (like energy, momentum and
position) are not discrete but continuous, the major exception being the
``number of particles'' which is important in the statistical mechanics of
large systems, but usually not exactly determinable, simply because in this
case there is typically a huge number of very small particles involved. For
this reason we view the statistical nature of physics as fundamental, even for
classical mechanics. Mathematically, the case where we do have complete
information is simply a special case of statistical mechanics, and hence is
covered by our work. We will therefore usually refer simply to ``mechanics''
(quantum or classical), rather than ``statistical mechanics''. When we do use
the term ``statistical mechanics'', it will be in the traditional sense,
namely to refer to large systems where there are too many parts (usually small
particles) for each to be measured individually (so we do not know the
position, momentum and so on of each particle). In this case only a small
number of parameters referring to the system as a whole (or to pieces of the
system much larger than its individual parts) can in practice be measured, for
example the temperature, volume, mass and pressure of a gas confined to some container.

Having set up a unified framework for quantum and classical mechanics, we
proceed to consider recurrence and ergodicity. These concepts originated
respectively in Poincar\'{e}'s work on celestial mechanics and in Boltzmann's
work on classical statistical mechanics, and now form part of what is known as
ergodic theory. We want to study recurrence and ergodicity in our unified
framework for mechanics to gain some insight into the properties of quantum
mechanics as opposed to classical mechanics.

The notion of Poincar\'{e} recurrence in classical mechanics is quite
well-known. Roughly it means that within experimental error a classical system
confined to a finite volume in phase space will eventually return to its
initial state. This happens because of Liouville's Theorem, which states that
Lebesgue measure is invariant under the Hamiltonian flow in the phase space
$\mathbb{R}^{2n}$. Ergodicity in classical mechanics refers to the situation
where for every observable and (almost) every pure state of a system, the time
mean of the observable (for that pure state) is equal to its average value on
the constant energy surface containing the pure state, in which case the
system is called ergodic. Again Liouville's Theorem is an implicit ingredient,
since it induces a time-invariant measure on the constant energy surface (see
Remark 3.2.8 for a brief discussion). It should be noted that ergodicity is of
some importance in physics, since it forms the starting point of many
developments of statistical mechanics (see for example\textbf{\ [Rue}, Section
1.1\textbf{]}). To study recurrence and ergodicity in quantum mechanics, we
can expect from these remarks that we will need a quantum mechanical analogue
of Liouville's Theorem. We propose such an analogue in Section 1.7 of Chapter
1, and in the process we are naturally led to consider finite von Neumann algebras.

Recurrence does in fact occur in quantum mechanics. One approach to recurrence
in quantum mechanics has been through the theory of almost periodic functions
(see for example \textbf{[BL]}, \textbf{[HH]} and \textbf{[Perc]}). Another
line of research, involving coherent states, along with possible applications
of quantum recurrence, can be traced in \textbf{[SLB]} and references therein.
However, these methods differ considerably from the measure theoretic
techniques employed to study recurrence in classical mechanics. In Section 3.1
of Chapter 3 we will see how recurrence (in a probabilistic sense) in quantum
mechanics can be cast in a mathematical form that looks the same as the
classical case, using our unified formulation of mechanics. More precisely,
the quantum case is a noncommutative extension of the classical case.

The mathematical aspects of recurrence and ergodicity is the subject of
Chapter 2, where one clearly sees that these concepts are not really measure
theoretic in nature, as it might seem from numerous books (for example
\textbf{[Pet]} and \textbf{[Wa]}), but rather $\ast$-algebraic, with the basic
tools being some Hilbert space techniques. The idea is to study recurrence and
ergodicity in the most general mathematical setting possible. This then
includes our unified framework for mechanics as a special case. In Chapter 3
we look at a few physical aspects of recurrence and ergodicity, including some
speculation on the relevance of these ideas in quantum mechanics.

The original inspiration for this thesis came from \textbf{[NSZ]}, where
recurrence is studied in a C*-algebraic framework from a purely mathematical
point of view. The work presented here is for the most part based on
\textbf{[D2]},\textbf{\ [D3]} and \textbf{[DS]}.

\chapter{A C*-algebraic formulation of mechanics}

In this chapter we formulate quantum mechanics and classical mechanics in the
language of C*-algebras. The exposition is based on \textbf{[D2]} and
\textbf{[D3]}, but contains some additional material. As we shall see, the
general structure of quantum mechanics and classical mechanics are identical,
except for commutativity, when both are viewed purely in C*-algebraic terms.
We therefore obtain a unified framework for mechanics which will be seen to be
very natural for studying some ergodic properties of quantum and classical
mechanics in Chapter 3.

Sections 1.1 to 1.5 present general aspects of mechanics in a statistical
framework, and in Section 1.6 an interpretation of quantum mechanics inspired
by the mathematical setting is discussed. The physical concepts are introduced
gradually in the sense that certain ideas are initially only used intuitively,
since their formal presentation can only be given once the framework has at
least been partially erected. Sections 1.7 to 1.9 treat specialized topics to
be used in Chapter 3; these topics do not apply to mechanical systems in general.

\section{Yes/no experiments}

We start with two simple definitions that apply to both quantum mechanics and
classical mechanics:

\bigskip\noindent\textbf{1.1.1 Definition. }\textit{An }\textbf{observable}%
\textit{\ of a physical system is any attribute of the system which results in
a real number when measured, where this measurement must be verifiable, in
other words, if the measurement is repeated immediately (so no disturbance or
time-evolution of the system occurs between the measurements) then it results
in the same real number. We call this real number the }\textbf{value}%
\textit{\ of the observable during the measurement.}

\bigskip If a measurement is not verifiable in the sense of Definition 1.1.1,
then there is no well-defined value of whatever it is that we measured, and
hence we do not consider it to be a measurement of an observable.

\bigskip\noindent\textbf{1.1.2 Remark.} By a \emph{measurement} (or
\emph{observation}) we mean that an observer receives information regarding
the physical system. The verifiability of a measurement essentially says that
the information obtained in the measurement is correct, since it means that if
we could repeat the measurement then we would with probability one get the
same result. This is the type of measurement we will deal with in this thesis.
We can therefore also refer to a measurement as a \emph{preparation}. This is
an idealization of reality (also see \textbf{[Om}, p. 82\textbf{]} on defining
an ideal measurement in terms of verifiability). In the worst cases a system
might in practice even be destroyed by a measurement (for example a particle
absorbed by a detector), and then a repetition of the measurement would not be
possible. Without idealization however, it would be impossible to do physics.
After an ideal framework has been set up, non-ideal situations can be
understood in terms of the idealization. See Sections 1.5 and 1.6 for more on
ideal measurements. It is very important to mention that we will view all
(ideal) measurements as yes/no experiments, defined below, which means that an
ideal measurement does not necessarily supply complete information, but only
correct information (also see Section 1.4). The idea of a single value in
Definition 1.1.1 should therefore be viewed only as preliminary, to help us to
build up the statistical ideas used later on.$\blacksquare$

\bigskip\noindent\textbf{1.1.3 Definition. }\textit{Consider any observable of
a physical system, and any Borel set }$S\subset\mathbb{R}$. \textit{We now
perform an experiment on the system which results in a ``yes'' if the value of
the observable lies in }$S$ \textit{during the experiment, and a ``no''
otherwise; the experiment gives no further information. We call this a
}\textbf{yes/no experiment}\textit{.}

\bigskip Definition 1.1.3 seems justified, since in practice there are always
experimental errors during measurements, in other words we always get a range
of values (namely $S$ in Definition 1.1.3) rather than a single value.

\bigskip\noindent\textbf{1.1.4 Remark.} In quantum mechanics one should be
careful in interpreting Definition 1.1.3. While in classical mechanics the
mathematical framework allows us to assume (if we want to) that there is some
objective single value of an observable at the time of a measurement (even
though we only get a set of values), this view can not be held in quantum
mechanics. In quantum mechanics the different values in the set correspond to
orthogonal state vectors (for simplicity we assume for the moment that the
observable's spectrum is discrete), but the system need not be in any of these
states, it can also be in a superposition of them, meaning that none of the
values in the set is the ``actual'' objective value of the observable.

So, if in quantum mechanics a measurement returns a set of values, then we
cannot view any one of these values as being the actual objective value of the
observable. However, a series of measurements of the same observable (assuming
there's no time-evolution, measurements of other observables, or outside
influences on the system) should at least be consistent with each other, in
the sense that the intersection of the sets obtained in the measurements
should be non-empty.$\blacksquare$

\bigskip Typically a measurement gives an interval which contains the value of
the observable being measured. For example, a measuring instrument with a
``digital'' read-out possessing only four digits might read $1.520$, which
means\ that the value lies in the interval $[1.5195,1.5205]$. We now assume
that this is then the only information\ we have concerning the value (for
example, we do not have a non-constant probability distribution for where in
the interval the value lies). The interval $[1.5195,1.5205]$ here plays the
role of $S$ in Definition 1.1.3.

To clarify the interpretation of Definition 1.1.3, we give another example.
Let's say we measure the $x$-coordinate of a given particle in some physical
system (quantum or classical) and we obtain the interval $[a,b]$. Then we view
this as the yes/no experiment ``Does the $x$-coordinate of the given particle
lie in $[a,b]$?'' performed on the system, and that it resulted in a ``yes''.
Similarly for any other observable of a system, and any Borel set $S$ instead
of $[a,b]$. Hence we can view a measurement of any observable as a yes/no experiment.

Since Definition 1.1.3 is stated for arbitrary Borel sets $S$, rather than
just the special case of intervals, it covers a much wider class of situations
than the examples above. For example, instead of an interval $[a,b]$, an
experiment might give us some union of possibly unbounded intervals. We can
mention that since we will use measure theory in any case (especially when
dealing with classical mechanics), the introduction of Borel sets at this
stage does not cause any extra effort later on.

We now want to show how the yes/no experiments can themselves be viewed as
observables. Consider any property that the system may or may not have that
can be verified or negated by a verifiable measurement (in the sense of
Definition 1.1.1) which results in the value $1$ if the system has this
property, and the value $0$ otherwise. Then we can view this property as an
observable which can have the value $1$ or $0$. Now consider the yes/no
experiment resulting in a ``yes'' if the value of the observable is $1$ (i.e.
the value lies in some Borel set containing $1$ but not $0$, for example
$\{1\}$), and a ``no'' otherwise. Then the observable and the yes/no
experiment are really one and the same thing, with the yes/no experiment
merely relabelling the values $1$ and $0$ as ``yes'' and ``no'' respectively.
An example of a property as discussed here is ``The $x$-coordinate of the
particle lies in $[x_{1},x_{2}]$, the $y$-coordinate in $[y_{1},y_{2}]$ and
the $z$-coordinate in $[z_{1},z_{2}]$'' for a given particle in a physical
system (where in this example we use Cartesian coordinates).

It should therefore now be clear that the observable in Definitions 1.1.1 and
1.1.3 may be some property (as above) comprising a combination of other
observables together with sets in which their values might lie, such as the
three position coordinates (each an observable) and the three intervals in the
last example. The point we are trying to make is that Definition 1.1.3 is very
general, applying to any property as discussed above. It must be stressed
though, that the verifiability of the measurement of such a property is
essential here. Say for example we consider the property ``The particle's
position lies in $[q_{1},q_{2}]$, and its momentum in $[p_{1},p_{2}]$'' of a
system consisting of a single particle confined to a straight line. In the
case of quantum mechanics this property is not an observable in the sense
described above, since it turns out that if we measure the position and
momentum, and then immediately measure them again, their values need not be
the same as during the first measurement, that is to say the measurement of
the property is not verifiable. This ``odd'' behaviour is the classic example
of how quantum mechanics differs from classical mechanics, since in the latter
this property is in fact an observable. It is usually expressed as saying that
the position and the momentum can not be measured simultaneously (or that the
two observables are not compatible). In principle we can study this type of
behaviour for a property constructed from an arbitrary set of observables of a
physical system. In Sections 1.2 to 1.6 we will see that the only real
difference between quantum and classical mechanics is that the latter is
commutative (or abelian) while the former is not (the meaning of this will
become clear in Sections 1.2 to 1.6). Therefore the noncommutativity of
quantum mechanics must be responsible for its ``odd'' behaviour as compared to
classical mechanics.

\bigskip\noindent\textbf{1.1.5 Remark.} The idea of yes/no experiments (and
their projections; refer to Section 1.2) can be traced back to \textbf{[vN1]},
where yes/no experiments are viewed as ``propositions'' stating various
possible properties of the system, a property being verified if we obtain a
``yes'' in the corresponding yes/no experiment. In classical mechanics the
first hint at yes/no experiments seems to be \textbf{[vN2]} where von Neumann
asks the question ``Does $P$ belong to $\theta$ or not?'', $P$ being the pure
state of the system as a point in the phase space, and $\theta$ a measurable
set in the phase space. (We will return to this very question in Section 1.3,
but in terms of Definition 1.1.3 and its interpretation explained above.) The
idea was further developed in \textbf{[BvN]} for both quantum and classical
mechanics.$\blacksquare$

\section{Quantum mechanics}

Let's look at the C*-algebraic formulation of quantum mechanics (also see
\textbf{[Ha]}). Consider any quantum mechanical system. We represent the
observables of the system by a unital C*-algebra $\mathfrak{A}$, called the
\emph{observable algebra} of the system, and the state of the system by a
\emph{state} $\omega$ on $\mathfrak{A}$, that is to say $\omega$ is a
normalized positive linear functional on $\mathfrak{A}$. (By \emph{normalized}
we mean that $\omega(1)=1$, and by \emph{positive} that $\omega(A^{\ast}A) $
for all $A\in\mathfrak{A}$.) At this stage we attach the intuitive meaning to
the term ``the state of the system''; we will return to this in Section 1.4.
$\mathfrak{A}$ contains the spectral projections of the system's observables
rather than the observables themselves. By this we mean the following: To any
yes/no experiment that we can perform on the system, there corresponds a
projection $P$ in $\mathfrak{A}$ such that $\omega(P)$ is the probability of
getting a ``yes'' during the experiment for any state $\omega$ of the system.
We will refer to $P$ as \emph{the} projection of the yes/no experiment.

We will only consider yes/no experiments for which the experimental setup is
such that at least in the case of a ``yes'' the system survives the experiment
(for example, it is not absorbed by a detector), so further experiments can be
performed on it. What does the system's state look like after such an
experiment? Consider for the moment the Hilbert space setting for quantum
mechanics. Here the (pure) states of a system are represented by non-zero
vectors, called \emph{state vectors}, in a Hilbert space $\mathfrak{H}$,
called the \textit{state space }of the system. Suppose the state is given by
the unit vector $x$ in $\mathfrak{H}$. After a yes/no experiment the state is
given by the projection of $x$ on some Hilbert subspace of $\mathfrak{H}$.
Denoting the projection operator onto the subspace in case of a ``yes'' by $Q
$, we see that the system's state after the experiment would then be given by
the unit vector $Qx/\left\|  Qx\right\|  $, according to the \emph{projection
postulate} (``collapse of the wave function''). It is clear that $Q$ is the
projection of the experiment, since $\left\|  Qx\right\|  ^{2}=\langle
x,Qx\rangle$ is exactly the probability of getting a ``yes''. (Here the state
$\theta$ on the C*-algebra $\mathfrak{L(H)}$ of all bounded linear operators
on $\mathfrak{H}$, given by $\theta(A)=\langle x,Ax\rangle$, is the
C*-algebraic representation of the state $x$, in the sense of $\omega$ above,
with $\mathfrak{L(H)}$ serving as the observable algebra.)

Returning to our system with observable algebra $\mathfrak{A}$, we know by the
GNS-construction (see Section 2.2, or for example \textbf{[BR}, Section
2.3.3\textbf{]}) that there exists a (cyclic) representation of $(\mathfrak{A}%
,\omega)$, namely a Hilbert space $\mathfrak{H}$, a $\ast$-homomorphism
$\pi:\mathfrak{A\rightarrow L(H)}$, and a unit vector $\Omega$ in
$\mathfrak{H}$, such that
\begin{equation}
\omega(A)=\langle\Omega,\pi(A)\Omega\rangle\tag{2.1}%
\end{equation}
for all $A$ in $\mathfrak{A}$. This looks like the usual expression for the
expectation value of an observable (here represented by $\pi(A)$) for a system
in the state $\Omega$ in the Hilbert space setting (compare $\theta$ above).
On a heuristic level we therefore regard $\mathfrak{H}$ as the state space of
the system, and $\Omega$ as its state. Say the result of the yes/no experiment
with projection $P$ is ``yes''. On the basis of the Hilbert space setting
described above, it would now be natural to expect that after the experiment
the state is represented by the unit vector $\Omega^{\prime}=\pi
(P)\Omega/\left\|  \pi(P)\Omega\right\|  $, since $\pi(P)$ is the projection
of the experiment in the Hilbert space setting in the same way as $Q$ above
(and hence $\pi(P)$ here plays the role of $Q$). Note that $\left\|
\pi(P)\Omega\right\|  ^{2}=\omega(P)>0$ since this is exactly the probability
of getting the result ``yes''. We now replace $\Omega$ in (2.1) by
$\Omega^{\prime}$ to get a new expectation functional $\omega^{\prime}$
defined by
\[
\omega^{\prime}(A)=\langle\Omega^{\prime},\pi(A)\Omega^{\prime}\rangle
\]
for all $A$ in $\mathfrak{A}$. Clearly $\omega^{\prime}(A)=\omega
(PAP)/\omega(P)$, so $\omega^{\prime}(1)=1$, which implies that $\omega
^{\prime}$ is a state on $\mathfrak{A}$. Based on these arguments we give the
following postulate:

\bigskip\noindent\textbf{1.2.1 Postulate. }\textit{Consider a quantum
mechanical system in the state }$\omega$\textit{\ on its observable algebra
}$\mathfrak{A}$\textit{. Suppose we get a ``yes'' during a yes/no experiment
performed on the system. After the experiment the state of the system is then
given by the state }$\omega^{\prime}$\textit{\ on }$\mathfrak{A}%
$\textit{\ defined by}
\begin{equation}
\omega^{\prime}(A)=\omega(PAP)/\omega(P) \tag{2.2}%
\end{equation}
\textit{for all }$A$\textit{\ in }$\mathfrak{A}$\textit{, where }%
$P$\textit{\ is the projection of the yes/no experiment.}

\bigskip Suppose the state is expressed in terms of a density operator $\rho$
on a Hilbert space $\mathfrak{H}$, namely $\omega(A)=$ Tr$(\rho A)$ for any
bounded linear operator $A$ on the Hilbert space. (Here \emph{density
operator} refers to a positive operator $\rho\in\mathfrak{L}(\mathfrak{H})$
with Tr$(\rho)=1$.) From Postulate 1.2.1 it then follows that after the
experiment the density operator is given by
\begin{equation}
\rho^{\prime}=\frac{P\rho P}{\text{Tr}(\rho P)} \tag{2.3}%
\end{equation}
in the case of a ``yes''. This is sometimes referred to as the L\"{u}ders rule
(see \textbf{[Hu}, p. 274\textbf{]} or \textbf{[Lu]}), and by the arguments
above we see that this rule can be viewed as the projection postulate applied
to a vector in a ``bigger'' Hilbert space, in which $\rho$ is represented by
this vector. The equivalence of (2.2) and (2.3), assuming we only consider
states given by density operators, follows from the fact that if Tr$(\rho
_{1}A)=$ Tr$(\rho_{2}A)$ for all $A\in\mathfrak{L(H)}$ for two density
operators $\rho_{1}$ and $\rho_{2}$ on $\mathfrak{H}$, then setting
$A=\rho_{1}-\rho_{2}$ gives
\[
\left\|  (\rho_{1}-\rho_{2})^{2}\right\|  _{1}=\text{Tr}((\rho_{1}-\rho
_{2})^{2})=0
\]
where $\left\|  \cdot\right\|  _{1}$ denotes the trace-class norm; see
\textbf{[Mu}, p. 63 and 65\textbf{]}. Hence $(\rho_{1}-\rho_{2})^{2}=0$ and
therefore $\left\|  \rho_{1}-\rho_{2}\right\|  ^{2}=\left\|  (\rho_{1}%
-\rho_{2})^{2}\right\|  =0$, where $\left\|  \cdot\right\|  $ denotes the
usual operator norm. So $\rho_{1}=\rho_{2}$, proving the equivalence, namely
that $\rho^{\prime}$ is the \emph{unique} density operator insuring that
$\omega^{\prime}(A)=$ Tr$(\rho^{\prime}A)$ satisfies (2.2).

Lastly we mention that the time-evolution of the system is described by a
one-parameter $*$-automorphism group $\tau$ of $\mathfrak{A}$, so if the
projection of a yes/no experiment is $P$ at time $0$, then at time $t$ the
projection of the same yes/no experiment will be $\tau_{t}(P)$.

\section{Classical mechanics}

Now we turn to the C*-algebraic formulation of classical mechanics. We can
represent the pure state of a classical system by a point in its \emph{phase
space} $\mathbb{R}^{2n}$, where $n$ of the entries are the generalized
position coordinates, and the other $n$ their conjugate momenta. This point is
called the \emph{phase point} of the system. This is somewhat restrictive
since such a point represents exact knowledge of the state of the system,
which is impossible in practice. Therefore we rather represent the state of
the system by a Borel measure $\mu$ on $\mathbb{R}^{2n}$ such that $\mu(S)$ is
the probability that the system's phase point is somewhere in the Borel set
$S\subset\mathbb{R}^{2n}$. In particular we have $\mu(\mathbb{R}^{2n})=1$.

We view each observable of the system as a Borel function $f:\mathbb{R}%
^{2n}\rightarrow\mathbb{R}$. This simply means that if the system's phase
point is $x\in\mathbb{R}^{2n}$, then the value of the observable is $f(x)$. If
we perform a yes/no experiment to determine if $f$'s value lies in the Borel
set $V\subset\mathbb{R}$, then the probability of getting ``yes'' is clearly
\[
\mu\left(  f^{-1}(V)\right)  =\int\chi_{f^{-1}(V)}d\mu
\]
where $\chi$ denotes characteristic functions (i.e. for any set $A$, the
function $\chi_{A}$ assumes the value $1$ on $A$, and zero everywhere else).
We can view $\chi_{f^{-1}(V)}$ as a spectral projection of the observable $f$,
and we will refer to it as \textit{the} projection of the yes/no experiment,
just as in the quantum mechanical case. Note that $\chi_{f^{-1}(V)}$ is a
projection in the C*-algebra $B_{\infty}(\mathbb{R}^{2n}) $ of all bounded
complex-valued Borel functions on $\mathbb{R}^{2n}$, where the norm of
$B_{\infty}(\mathbb{R}^{2})$ is the sup-norm, its operations are defined
pointwise, and its involution is given by complex conjugation (we will use the
$\ast$-algebraic notation $g^{\ast}=\overline{g}$ for the complex conjugate of
a complex-valued function $g$). We can define a state $\omega$ on the
C*-algebra $B_{\infty}(\mathbb{R}^{2n})$ by
\begin{equation}
\omega(g)=\int gd\mu\tag{3.1}%
\end{equation}
for all $g$ in $B_{\infty}(\mathbb{R}^{2n})$. Then we see that the probability
of getting a ``yes'' in the above mentioned yes/no experiment is $\omega
(\chi_{f^{-1}(V)})$. So we can view $\omega$ as representing the state of the
system in exactly the same way as in quantum mechanics, where now $B_{\infty
}(\mathbb{R}^{2n})$ is the unital C*-algebra representing the observables of
the system. For this reason we call $B_{\infty}(\mathbb{R}^{2n})$ the
\emph{observable algebra} of the system.

Postulate 1.2.1 then holds for the classical case as well, as we now explain.
Let $S\subset\mathbb{R}^{2n}$ be a Borel set. The probability for the system's
phase point to be in both $S$ and $f^{-1}(V)$ is merely the probability for it
to be in $S\cap f^{-1}(V)$, which is $\mu(S\cap f^{-1}(V))$. A ``yes'' in the
above mentioned yes/no experiment would mean that the system's phase point is
in $f^{-1}(V)$, and the probability of this is $\mu(f^{-1}(V))$. Denote by
$\mu^{\prime}(S)$ the so-called \emph{conditional} probability that the
system's phase point is in $S$, given that the phase point is in $f^{-1}(V)$.
Hence we should have
\begin{equation}
\mu(f^{-1}(V))\mu^{\prime}(S)=\mu(S\cap f^{-1}(V))\text{.} \tag{3.2}%
\end{equation}
It follows that if a ``yes'' is obtained in the experiment, then we can
describe the system's state after the experiment by the measure $\mu^{\prime}$
given by
\[
\mu^{\prime}(S)=\mu(S\cap f^{-1}(V))/\mu(f^{-1}(V))
\]
for all Borel sets $S\subset\mathbb{R}^{2n}$. It is easily verified that
$\mu^{\prime}$ is indeed a Borel measure on $\mathbb{R}^{2n}$. As for the case
of $\mu$ and $\omega$ in (3.1), $\mu^{\prime}$ corresponds to the state
$\omega^{\prime}$ on $B_{\infty}(\mathbb{R}^{2n})$ given by
\[
\omega^{\prime}(g)=\int gd\mu^{\prime}=\omega(\chi_{f^{-1}(V)}g\chi
_{f^{-1}(V)})/\omega(\chi_{f^{-1}(V)})
\]
(the second equality follows using standard measure theoretic arguments, i.e.
first prove it for $g$ a characteristic function and then use Lebesgue
convergence; refer to \textbf{[Rud]}). This is exactly what Postulate 1.2.1
says if we replace the word ``quantum'' by ``classical''.

For the time-evolution of a classical system we need the concept of a flow.
Consider a measure space $(X,\Sigma,\mu)$, where $\mu$ is a measure defined on
a $\sigma$-algebra $\Sigma$ of subsets of the set $X$. A \textit{flow} on
$(X,\Sigma,\mu)$ is a mapping $t\mapsto T_{t}$ on $\mathbb{R} $ with the
following properties: $T_{t}$ is a function defined on $X$ to itself, $T_{0}$
is the identity on $X$ (i.e. $T_{0}(x)=x$), $T_{s}\circ T_{t}=T_{s+t}$, and
$T_{t}(S)\in\Sigma$ and $\mu(T_{t}(S))=\mu(S)$ for all $S$ in $\Sigma$. We
denote this flow simply by $T_{t}$.

The time-evolution of our classical system is given by a flow $T_{t}$ on
$(\mathbb{R}^{2n},\mathcal{B},\lambda)$, where $\mathcal{B}$ is the $\sigma
$-algebra of Borel sets of $\mathbb{R}^{2n}$, and $\lambda$ is the Lebesgue
measure on $\mathbb{R}^{2n}$. Note that this statement contains Liouville's
theorem, namely $\lambda(T_{t}(S))=\lambda(S)$ for all $S$ in $\mathcal{B}$.
We call $T_{t}$ the \textit{Hamiltonian flow}. It simply means that if at time
$0$ the system's phase point is $x\in\mathbb{R}^{2n}$, then at time $t$ its
phase point is $T_{t}(x)$.

As in the C*-algebraic approach to quantum mechanics, we want the
time-evolution to act on the observable algebra rather than on the states.
Suppose the system's phase point is $x$ at time $0$. Consider an observable
given by the function $f$ at time $0$. Then the value of the observable at
time $0$ is $f(x)$, and hence at time $t$ its value must be $f(T_{t}%
(x))=(f\circ$ $T_{t})(x)$, where on the left hand side of the equation the
time-evolution is applied to the phase point, and on the right hand side it is
applied to the observable. So it is clear that an observable given by $f$ at
time $0$, will be given by $f\circ T_{t}$ at time $t$ if the time-evolution
acts on the observables rather than on the states (this is the well-known
Koopman construction, \textbf{[Ko]}). This is equivalent to the action of
$T_{t}$ on the spectral projections of $f$, since $\chi_{(f\circ T_{t}%
)^{-1}(V)}=\chi_{f^{-1}(V)}\circ T_{t}$ for all Borel sets $V\subset
\mathbb{R}$. We explain the meaning of this in more detail: Suppose the state
of the system is described as in (3.1). At time $t$ we perform the yes/no
experiment ``Does the value of the given observable lie in $V$?''. Let's say
that $x\in\mathbb{R}^{2n}$ is the phase point of the system at time $0$. The
value of the observable is in $V$ at time $t$ if and only if $f(T_{t}(x))\in
V$, in other words if and only if $x\in(f\circ T_{t})^{-1}(V) $. The
probability for this being the case (in other words, the probability of
getting a ``yes'' in the experiment) is
\[
\mu\left(  (f\circ T)^{-1}(V)\right)  =\omega(\chi_{(f\circ T_{t})^{-1}%
(V)})=\omega(\chi_{f^{-1}(V)}\circ T_{t})
\]
as explained at the beginning of this section. This means that at time $t$ the
projection of the yes/no experiment is given by $\chi_{f^{-1}(V)}\circ T_{t}$.
It is easily seen that if we define $\tau$ by
\begin{equation}
\tau_{t}(g)=g\circ T_{t} \tag{3.3}%
\end{equation}
for all $g$ in $B_{\infty}(\mathbb{R}^{2n})$, then $\tau$ is a one-parameter
$\ast$-automorphism group of the C*-algebra $B_{\infty}(\mathbb{R}^{2n})$. So
the time-evolution is described in exactly the same way as in quantum
mechanics when viewed in C*-algebraic terms.

We have now obtained a C*-algebraic formulation of classical mechanics. Note
that $B_{\infty}(\mathbb{R}^{2n})$ is an abelian C*-algebra. Replacing
$B_{\infty}(\mathbb{R}^{2n})$ by an arbitrary abelian unital C*-algebra would
give us an abstract C*-algebraic formulation of classical mechanics. From our
discussion above it is clear that if in the C*-algebraic formulation of
quantum mechanics described in Section 1.2 we assume that $\mathfrak{A}$ is
abelian, then we get exactly this abstract C*-algebraic formulation of
classical mechanics. Setting $\mathfrak{A}=B_{\infty}(\mathbb{R}^{2n})$ would
make it concrete. In this sense the C*-algebraic formulation of quantum
mechanics actually contains classical mechanics as a special case.

\bigskip\noindent\textbf{1.3.1 Remark.} Here we used $B_{\infty}%
(\mathbb{R}^{2n})$ as the classical observable algebra. Other choices are
possible in certain approaches to statistical mechanics. For example some
C*-algebra of continuous functions on the phase space (see for example
\textbf{[Rue}, Section 7.1\textbf{]}), but in general this precludes
projections and will therefore not do for our purposes.$\blacksquare$

\section{The general structure of mechanics}

We now summarize our work thus far to gain some perspective.

In a mathematical description of a physical system (quantum or classical), we
need to describe four things:

\bigskip(a) The observables of the system (as defined in 1.1.1).

(b) The state of the system, by which we mean the observer's information
regarding the system. (We assume that the observer knows what the system is,
i.e. he knows what the observables are.) The case of maximal information is
called a \emph{pure state}. We can say that by definition the \emph{state} of
the system is a mathematical object which for each possible outcome of each
measurement that can be performed on the system, provides the observer with
the probability for obtaining that outcome when performing that measurement.
We can then also say that the observer's \emph{information} about the system
is by definition this state. (Note that the state of the system is not an
objective property of the system, but depends on the observer.) The state of
the system must be constructed from data gained during measurements previously
performed on the system. Of course, we have to assume that the measurements
are \emph{accurate} (i.e. the data is correct, also see Remark 1.1.2), even
though they may \emph{not be precise} (i.e. the data is incomplete), for
example when we measure the position of a classical particle we get a set of
possible values rather than a single value, but the value of the position
during the measurement is contained in this set.

(c) The measuring process. This is clearly closely connected to (a) and (b),
since the observables are exactly that which is measured, while the result of
a measurement gives the observer new information regarding the system, that is
to say a measurement changes the state. We can view all measurements of the
observables as yes/no experiments, as explained in Section 1.1.

(d) The time-evolution of the system (dynamics). In other words, how the
probabilities mentioned in (b) change as we move forward (or backward) in time.

\bigskip The results of Sections 1.2 and 1.3 (for a quantum or classical
mechanical system) are:

\bigskip(i) We describe the observables by an observable algebra
$\mathfrak{A}$ which for each point in time contains a projection
corresponding to each yes/no experiment that can be performed on the system
(at that point in time). (These projections are referred to as \emph{spectral
projections}.) $\mathfrak{A}$ is taken as a unital C*-algebra.

(ii) The state of the system is described by a state $\omega$ on
$\mathfrak{A}$ (in the C*-algebraic sense defined in Section 1.2), such that
for every yes/no experiment, $\omega(P)$ is the probability of getting
``yes'', where $P$ is the projection of the yes/no experiment at the time at
which it is performed. (Obviously this implies that the probability of getting
``no'' is $1-\omega(P)=\omega(1-P)$.)

(iii) Regarding the measurement process we just have to describe how the state
is changed by a yes/no experiment. This is given by Postulate 1.2.1, which
also holds for a classical mechanical system as explained in Section 1.3. That
is to say, if a ``yes'' is obtained in the yes/no experiment, then after the
experiment the state of the system is given by the state $\omega^{\prime}$\ on
$\mathfrak{A}$\ defined by
\[
\omega^{\prime}(A)=\omega(PAP)/\omega(P)
\]
for all $A$\ in $\mathfrak{A}$, where $P$\ is the projection of the yes/no
experiment. (We will have more to say about the measuring process in the next
two sections.)

(iv) The time-evolution is given by a one-parameter $\ast$-automorphism group
$\tau$ of $\mathfrak{A}$, such that if at time $0$ the projection of a given
yes/no experiment is $P$, then at time $t$ the projection of the same yes/no
experiment will be $\tau_{t}(P)$. (The choice of when time $0$ is, is
arbitrary, since $\tau$ is a group.)

\bigskip This is \emph{the general structure of mechanics}. As will be
discussed in more detail in the next two sections, this general structure is
nothing more than probability theory (actually, it is a noncommutative
generalization of classical probability theory). It is a mathematical
framework for dealing with information. When applied to a physical system,
this information is the observer's information regarding the system, in other
words, the system's state.

\bigskip\noindent\textbf{1.4.1 Remark on hidden variables.} We have now seen
that quantum and classical mechanics have the same general structure, from a
probabilistic point of view, with classical mechanics being the special case
where the observable algebra is abelian. Suppose that there is some classical
theory underlying quantum mechanics (a hidden variable theory) and that
quantum behaviour is the result of our ignorance of these ``hidden
variables''. A good guess would then be that this underlying theory has the
general structure given above, the observable algebra being \emph{abelian},
where we lack precise information about the physical system being studied
(also see \textbf{[Ma}, pp. 180-184\textbf{]} and references therein). But
this fails to explain the noncommutative behaviour of quantum mechanics in a
simple way. It would therefore seem that a hidden variable theory would be a
complicated way of ``explaining'' the fact that quantum mechanics is simply a
noncommutative generalization of classical probability theory. Hidden
variables are then excised by Occam's razor.$\blacksquare$

\bigskip\noindent\textbf{1.4.2 Remark on spectral projections.} For a quantum
mechanical observable represented by a (possibly unbounded) self-adjoint
linear operator $A$ in the state space $\mathfrak{H}$, the projection of the
yes/no experiment ``Is the value of $A$ in $V$?'' can be taken as the spectral
projection $\chi_{V}(A)$ in terms of the Borel functional calculus on
self-adjoint operators; refer to \textbf{[SZ}, 9.9 to 9.13, and 9.32\textbf{]}
for the construction and properties of this calculus. Loosely speaking, this
projection represents the part of $A$ whose spectrum is contained in the Borel
subset $V$ of $\mathbb{R}$. It is interesting to note that this is very
similar to the classical case in Section 1.3, where we used $\chi_{f^{-1}%
(V)}=\chi_{V}\circ f$ instead of $\chi_{V}(A)$. We can write $\chi
_{V}(f):=\chi_{V}\circ f$ to complete the analogy, where more generally
$g(f):=g\circ f$ defines a Borel functional calculus on the measurable
functions $f:F\rightarrow\mathbb{R}$ for Borel measurable $g:\mathbb{R}%
\rightarrow\mathbb{C}$. Here the classical observable $f$ is also
self-adjoint, namely $f^{\ast}:=\overline{f}=f$ since it is
real-valued.$\blacksquare$

\section{Measurements and conditional probabilities}

The quantum mechanical projection postulate for the state vectors, which we
used in Section 1.2, often seems somewhat mysterious. However, within the
general structure of mechanics it is quite natural, as we now explain.

In Section 1.2 we extended this projection postulate to arbitrary states on an
abstract observable algebra to obtain Postulate 1.2.1. This was done using a
very natural heuristic argument based on the projection postulate for state
vectors. In Section 1.3 we motivated Postulate 1.2.1 for a classical
mechanical system by using the idea of a conditional probability.

By a \emph{conditional probability} we mean the probability for some event $A
$ to occur, given the information that some event $B$ has occurred. Denote
this probability by $p(A|B)$. Denote by $p(A)$ the probability for an event
$A$ to occur if no information regarding occurrences of other events are
available. Denote by $A\cap B$ the event where the events $A$ and $B$ both
occur. Then it is intuitively clear for any two events $A$ and $B$ that
\begin{equation}
p(B)p(A|B)=p(A\cap B)\text{.} \tag{5.1}%
\end{equation}
This is exactly what we used in equation (3.2).

To understand the intuition behind this, consider for example the case of a
finite number (of equally probable) sample points, say the six faces of a fair
die. Let $S$ be the set of sample points (we call it the sample space), then
events are represented by subsets of $S$. (Hence the notation $A\cap B$ above;
it is just the usual intersection of sets.) Suppose $S$ contains $n$ points,
and let $A$ and $B$ be events containing $a$ and $b$ sample points
respectively, while $A\cap B$ contains $c$ sample points. Then
\begin{equation}
p(A)=\frac{a}{n}\text{,\qquad}p(B)=\frac{b}{n}\text{\qquad and\qquad}p(A\cap
B)=\frac{c}{n}\text{.} \tag{5.2}%
\end{equation}
If we have the information that $B$ occurred, then our sample space collapses
to the set $B$. Event $A$ now consists of its sample points in $B$, in other
words it is given by $A\cap B$. Hence the probability of $A$ is now
\begin{equation}
p(A|B)=\frac{c}{b}\text{.} \tag{5.3}%
\end{equation}
From (5.2) and (5.3) we obtain (5.1).

The same argument can be applied to the case where the sample space $S$ is a
flat bounded surface with one of its point marked in some way, but we don't
know which point. Then the probability for a subset $A\subset S$ to contain
the marked point is given by $($area of $A)/($area of $S)$, and hence $A$ and
$B$ should be Lebesgue measurable. So, the probability of an event is the
``size'' of the set representing the event. It is essentially this measure
theoretic idea that is used in Section 1.3, where the phase point is the
marked point.

Refer to \textbf{[Fe]} for more on probability, including sample spaces and
conditional probabilities.

In Section 1.3 we saw that in the case of classical mechanics, Postulate 1.2.1
is simply another way of expressing (5.1) in the measure theoretic setting for
probability theory. Hence, in quantum mechanics, Postulate 1.2.1 can be viewed
as a ``noncommutative conditional probability''. (Also see \textbf{[Bu]}.) So
the ``mysterious'' projection postulate of quantum mechanics is mathematically
merely a noncommutative extension of the conditional probability encountered
in classical mechanics. (Also see \textbf{[Petz]} for a short survey of the
closely related idea of noncommutative conditional expectations, or refer to
\textbf{[OP]}.) It should of course be kept in mind that the physical
consequences of the quantum projection postulate differs surprisingly from
that of classical mechanics, with the Uncertainty Principle as the
archetypical example (it essentially states that the position and momentum of
a particle in one dimension can not be measured simultaneously, as was also
mentioned in Section 1.1).

We can now formalize the idea of an ideal measurement (see Remark 1.1.2):

\bigskip\noindent\textbf{1.5.1 Ideal measurements.} Postulate 1.2.1 can be
viewed as the definition of an \emph{ideal} measurement in quantum mechanics.
Replacing the word ``quantum'' by ``classical'', Postulate 1.2.1 defines an
\emph{ideal} measurement in classical mechanics. In short one can say that an
ideal measurement in mechanics is defined by (iii) in Section 1.4. So an ideal
measurement is a change in the observer's information regarding the system,
via a (possibly noncommutative) conditional probability. (Note that by
``ideal'' we do not mean ``precise''. In classical mechanics ``ideal'' means
that the system is not disturbed by the measurement. The same interpretation
can be used in quantum mechanics, as will be seen in Section
1.6.)$\blacksquare$

\section{An interpretation of quantum mechanics}

There are several problems surrounding the interpretation of quantum
mechanics, mainly involving the measuring process. What does the collapse of
the wave function mean? What causes it? And so on. In this section we argue
that these problems are essentially present in classical mechanics as well. In
classical mechanics a measurement is nothing strange. It is merely an event
where the observer obtains information about the system (we consider the case
of an ideal measurement as in 1.5.1). A measurement therefore changes the
observer's information. One can then ask: What does the change in the
observer's information mean? What causes it? And so on. These questions
correspond to the questions above, but now they seem tautological rather than
mysterious, since our intuitive idea of information tells us that the change
in the observer's information simply means that he has received new
information, and the change is caused by the reception of the new information.
We will see that the quantum case is no different, except that the nature of
information in quantum mechanics differs from that in classical mechanics. We
now first describe the basic idea, and afterwards we show how it is actually
an outgrowth of the mathematical framework we've been developing.

Let's say an observer has information regarding the phase point of a classical
system, but not necessarily complete information (this is the typical case, as
discussed in Sections 1.1 and 1.3). This information was of course obtained by
measurements the observer performed on the system (remember, by definition a
measurement is the reception of information by the observer). Now the observer
performs a measurement on the system to obtain new information (for example he
might have information regarding a particle's position, now he measures the
particle's momentum). The observer's information after this measurement then
differs from his information before the measurement. In other words, a
measurement ``disturbs'' the observer's information.

In classical mechanics we know that an observer's information isn't merely
disturbed, but is actually increased by a measurement (assuming the
measurement provides new information). We will view this as an assumption
regarding the nature of information which does not hold in quantum mechanics.
On an operational level, this can be seen as the essential difference between
quantum mechanics and classical mechanics: In both quantum and classical
mechanics the observer's information is disturbed (changed) by a measurement
if the measurement provides new information, but in classical mechanics the
observer's information before the measurement is still valid after the
measurement, while in quantum mechanics this is not necessarily the case.

In Section 1.5 we saw that the projection postulate of quantum mechanics is
essentially a noncommutative conditional probability which contains the
classical conditional probability as a special case. In fact, the general
structure of classical mechanics described in Sections 1.3 and 1.4 is nothing
more than probability theory (together with a time-evolution). One can shift
the perspective somewhat by saying that this general structure is a
probabilistic description of information. Since quantum mechanics has exactly
the same the general structure, except that it is noncommutative, the
mathematics seem to tell us that the general structure of quantum mechanics is
a probabilistic description of \emph{noncommutative information}. This
noncommutative nature of information in quantum mechanics is what causes the
essential difference between quantum mechanics and classical mechanics
mentioned above. (Also see \textbf{[D3]}, on which this section is based.)

\bigskip\noindent\textbf{1.6.1 Information. }We can view (i)-(iv) of Section
1.4 as the abstract axioms for a probabilistic description of information,
where the information can be noncommutative. Axiom (iii) is then a
(noncommutative) conditional probability describing how information changes
when new data (the result of a measurement in the case of physics) is
received. Here we define \emph{information} as being a state on an observable
algebra (or as the probabilities given by the state), with the information
called \emph{noncommutative} if it changes via the noncommutative conditional
probability. If we were to add the assumption that the observable algebra is
commutative, then we get an abstract formulation of classical probability
theory with the usual conditional probability. The algebras $B_{\infty
}(\mathbb{R}^{2n})$ (or more generally $B_{\infty}(F)$ for a phase space $F$;
see Remark 1.7.2) and $\mathfrak{L(H)}$ are nothing more than convenient
representations (of the commutative and noncommutative cases respectively),
suitable for doing physics (in the way explained in Sections 1.2 and
1.3).$\blacksquare$

\bigskip Interpreting quantum mechanics in this way implies that an (ideal)
measurement disturbs the information regarding a system's state, rather than
disturbing the system itself as is often argued (see for example \textbf{[Sc},
Section 1.6\textbf{]}). (In \textbf{[I]} a similar remark is made: ``a
measurement produces an uncontrollable disturbance in the \emph{potentiality}
for different results to be obtained in later measurements'' (p. 165), but
this remark becomes much clearer in the present setting in terms of
information.) This then renders many problems surrounding measurements in
quantum mechanics no more difficult than in classical mechanics. The answer to
both question at the beginning of this section is simply that the observer
received new information (i.e., the observer made a measurement), exactly as
for the corresponding classical questions. (In particular this means that
consciousness has no role to play in the measuring process. The observer could
be a computer connected to a measuring instrument, or the measuring instrument
itself, as long as it can receive information from the system.) We give a few
more examples:

\bigskip\noindent\textbf{1.6.2} The Heisenberg cut. This refers to an
imaginary dividing line between the observer and the system being observed
(see for example \textbf{[vN1]} and \textbf{[Ha]}). It can seen as the place
where information crosses from the system to the observer, but it leads to the
question of where exactly it should be; where does the observer begin? In
practice it's not really a problem: It doesn't matter where the cut is. It is
merely a philosophical question which is already present in classical
mechanics, since in the classical case information also passes from the system
to the observer and one could again ask where the observer begins. The
Heisenberg cut is therefore no more problematic in quantum mechanics than in
classical mechanics.$\blacksquare$

\bigskip\noindent\textbf{1.6.3} When does the collapse of the wave function
take place and how long does it take? (See for example \textbf{[Su}, p.
212\textbf{]}.) This is essentially the Heisenberg cut with space replaced by
time. One can pose the question as follows: When does an observer ``absorb''
the information received from a measurement (i.e., when does the measurement
take place), and how long does it take? Again the quantum case is no different
from the classical case, and moreover, in practice it is no more of a problem
than in the classical case.$\blacksquare$

\bigskip\noindent\textbf{1.6.4} Continuous observation (see \textbf{[Su]} and
\textbf{[Ho]}). The ideal measurement discussed in Remarks 1.1.2 and 1.5.1
refers to a single measurement made at some point in time. It can therefore
not be applied directly to continuous observation, i.e. when the observer's
information is continually changing. However, in classical mechanics this is
not considered a conceptual problem, since one could in principle describe
such a situation as a continual change in the probability distribution
(probability measure) describing the information, even though it might be a
difficult technical problem in practice. The same is true in quantum
mechanics, with the probability distribution replaced by a state representing
noncommutative information. (In quantum mechanics however, the idea of
continuous observation is probably an idealization, for example watching
something without blinking your eyes is not a continuous measurement, since
the photons registered by your retina are discrete.)

The ``paradox of the watched pot that never boils'' (called \emph{Zeno's
paradox} by \textbf{[MS]}) is resolved by noting that if an observer
continuously measures a certain observable, then the system can still evolve
in time to produce other values for the observable if the measurement is not
precise (as is typically the case). Say the observer measures an observable
$A$ which has a discrete spectrum, and he can only determine its value up to
some interval containing (at a point in time) a number of eigenvalues of the
observable, say $a_{1},...,a_{n}$. Then the state vector is projected onto the
subspace spanned by the eigenstates (at that point in time) corresponding to
$a_{1},...,a_{n}$, in other words, onto the subspace which at that point in
time corresponds to the interval (keep in mind that time-evolution acts on the
observable algebra, and hence on the eigenstates of the observable). This
happens according to postulate (iii); see for example \textbf{[CDL}, Section
III.E.2.b\textbf{]}. To clarify our argument, we assume here that before the
continuous measurement starts, the observer has maximal information, i.e. his
information is a state vector [the general case does not differ significantly,
since it is still handled with the same projection postulate (iii)]. Note that
the state is now still a state vector, and not a mixture of the eigenstates
corresponding to $a_{1},...,a_{n}$. The interval which is measured (and hence
the eigenvalues of $A$ contained in it) can change in the course of time (for
example it can drift up and down the real line), simply because of the lack of
precision in the continuous measurement. Therefore the value of $A$ can change
within this drifting interval, in turn allowing the drifting interval's
average location to change accordingly, which is what the observer sees. In
the mathematics this looks as follows: The continuous measurement confines the
state vector via the projection postulate to the ``drifting'' subspace
corresponding to the drifting interval. The observable's eigenstates are
evolving in time, but since this drifting subspace contains many eigenstates
of the observable at any point in time, the projection postulate does not
cause the state vector to be ``dragged along'' by one of the time-evolving
eigenstates. Also, since the interval is drifting, eigenstates are moving in
and out of the subspace. Therefore the state vector can be projected onto
subspaces containing new eigenstates (corresponding to new eigenvalues), with
eigenstates brought closer to the state vector by time-evolution having higher
probability. (This argument becomes somewhat clearer in the Schr\"{o}dinger
picture, where the eigenstates are fixed, but the subspace is still drifting.)

If the continuous measurement is precise enough, then quantum mechanics indeed
predict that ``a watched pot never boils'' if the observable's eigenvalues are
discrete (precise measurement of a continuous observable is impossible in
practice). This happens because a quantum measurement can invalidate previous
information (i.e. the state vector can change by projection) which then
``cancels out'' the changes due to time-evolution acting on the observable
algebra (and thus on the observable's eigenvectors onto which projection of
the state vector occurs). In effect the state vector is dragged along by the
time-evolving eigenstate corresponding to the measured value. In classical
mechanics on the other hand, previous information is not invalidated by
measurement, hence the values of observables can change as time-evolution acts
on the observable algebra while the pure state of the system stays put. Note
that this is true even if the classical observable being observed is discrete
(for example ``number of particles in the left half of the container''). So no
matter how closely we watch a classical pot, it can still boil.$\blacksquare$

\bigskip\noindent\textbf{1.6.5} The EPR ``paradox.'' Einstein, Podolsky and
Rosen \textbf{[EPR]} described a now famous experiment in which two particles
are created together (or interact) and then move away from each other (which
ends any interaction between them) before a measurement is performed on one of
the particles. This measurement then gives corresponding information about the
other particle as well. [This is the result of an entanglement of the two
particles' states (for example due to a conservation law), which can occur
since the state space is the tensor product of the two particles' state
spaces.] EPR\ argued that this means that the second particle simultaneously
has values for two noncommuting observables like position and momentum, since
only the first particle is measured (either its position or its momentum is
measured, but not both), and hence quantum mechanics must be incomplete, since
it says that a particle does not simultaneously have values for position and
momentum. They based this on the idea that a measurement on the first particle
does not disturb the second. However, we have viewed a measurement as the
reception of information by the observer; it has nothing to do with the
observer ``directly'' observing (and disturbing) the system. Measuring the
first particle gives the observer information regarding the second particle as
well (and hence \emph{is} a measurement of the second particle), which is
mathematically described by the second particle's state vector (representing
the observer's \emph{noncommutative} information about this particle) now
being in an eigenspace of the observable which was measured. This is no
different from the analogous situation in classical mechanics where for
example conservation of momentum can give the second particle's momentum when
the first particle's momentum is measured, except that in this case
information is commutative.

We can even have two observers A and B measuring the same observable of the
two particles respectively (as in \textbf{[I]} for example). A's measurement
is then also a measurement of the value B will get (A receives \ information
about what B's result will be) and so there's nothing strange in them getting
correlated results (say opposite values for momentum; or opposite values for
spin $z$, where the particles have spin half as in Bohm's version of the
EPR\ experiment, \textbf{[Bo]}). No signal need travel faster than the speed
of light to B's particle to ``tell'' it to have the opposite value to A's
result, in the same way that no such signal is needed in the classical case.
From A's point of view, B is part of the system along with the two particles,
and so this experiment is really no different from the original one observer
EPR\ experiment above. The particles \emph{along with} B are in a
superposition of states from A's point of view until A measures his particle,
which reduces (by projection) the state vector of the combined system of
particles and B, with B then in the eigenspace ``B gets the opposite
value''.$\blacksquare$

\bigskip\noindent\textbf{1.6.6 }System and observer as a combined system (see
\textbf{[I]} for a clear exposition). Here the time-evolution of the combined
system is supposed to account for the projection postulate of quantum
mechanics. This is not possible in a natural way, since time-evolution is the
result of a one-parameter $\ast$-automorphism group. In classical mechanics
the combined system evolves according to classical dynamics (the observer
being thought of as a classical system in this case), and this then similarly
would have to account for the change in the observer's information via a
conditional probability due to a measurement he performs on the system. Again
this is not possible in a natural way, since here too we have the same
projection postulate, namely the conditional probability (iii) in Section 1.4
acting on the state (of the system without observer), while the time-evolution
acts as a one-parameter $\ast$-automorphism group on the observable algebra.
The solution is that the state of the combined system has to contain from the
start the fact that the observer will perform a measurement on the system at a
given point in time and will subsequently experience a change of information
(this change is a physical process in the observer, described by the combined
system's time-evolution, for example some neural activity in a human
observer's brain), otherwise such a measurement and the change of information
would not take place. This is clear, since time-evolution does not act on the
state, but on the observable algebra, hence the state of the combined system
is the state ``for all time'' and does not change when the observer performs a
measurement. Exactly the same is true for quantum mechanics (where the
observer is then also viewed as a quantum system). The (noncommutative)
conditional probability, that is to say the projection postulate, is only
relevant when the observer is not considered to be part of the system, in
which case the conditional probability says what the change in the observer's
information will be, it does not describe the physical process taking place in
the observer to accommodate (or store) the new information.$\blacksquare$

\bigskip In connection with the two-slit experiment we mention the following:

\bigskip\noindent\textbf{1.6.7 The two-slit experiment. }Assume that the
probability distribution for the position of detection of a particle on the
screen in the two-slit experiment is given by an interference pattern when no
measurement is performed at the two open slits (this is due to the wave nature
of quantum particles, which is not accounted for by the abstract concept of
noncommutative information (in 1.6.1) by itself, but rather follows from the
specific form of dynamics of quantum mechanics). This distribution represents
the observer's information about where on the screen the particle will be
detected. In the light of our discussion thus far, it should then not be too
surprising that this distribution (i.e. the observer's information) can be
invalidated via the noncommutative conditional probability (iii) in Section
1.4, if the observer does measure through which slit the particle goes (i.e.
if the observer receives new information), giving a completely different
probability distribution at the screen. This is unlike the classical case
where a measurement at the slits gives the observer more information, rather
than invalidating previous information. (Also see \textbf{[Bu]}.)$\blacksquare$

\bigskip The point we attempt to make with examples 1.6.2 to 1.6.6 is that,
even though there might be certain problems surrounding the measuring process,
quantum mechanics does not introduce any new conceptual problems not already
present in classical mechanics when one considers a single observer performing
measurements on a physical system, as long as we assume that information is
noncommutative in quantum mechanics.

We can also consider the case of more than one observer touched upon in 1.6.5:

\bigskip\noindent\textbf{1.6.8 Thought experiment. }Say three observers A, B
and C are observing the same system, but B and C are not aware of each other
or of A. B and C measure two noncommuting observables $P$ and $Q$
respectively, in the order $P$, $Q$, $P$, and A in turn measures B and C's
results in this order ( he ``sees'' each of their results at the time they
obtain them). We ignore the time-evolution of the system. Say the results are
$p_{1}$, $q$, $p_{2}$ (in this order), then clearly $p_{1}$ and $p_{2}$ need
not be the same since $P$ and $Q$ do not commute. So from B's point of view it
seems that something disturbed the system between his two measurements of $P$.
However, in our interpretation it is actually B's information that has been
invalidated by A and C's measurement of $Q$. This is not too strange, since B
and C are merely A's measuring instruments. One could ask what would happen if
A wasn't there. Would B then get $p_{1}=p_{2}$ with probability one? In the
absence of A, does it even make sense to talk of the time order $P$, $Q$, $P$
if B and C are not aware of each other? In our interpretation time ordering
should probably be viewed as in some way defined by information received by an
observer, and in this case it seems possible that B would get $p_{1}=p_{2}$
with probability one in the absence of A and no other way to define the time
ordering. (Note that in the two-slit experiment, for example, there is a time
ordering in the sense that a measurement on a particle at the slits is
performed before a measurement on the same particle at the screen, even if the
measurements are performed by two different observers not aware of each other,
so the interference pattern at the screen can still be destroyed in this
setup.) The idea of defining time ordering in terms of a series of events (an
event in our case being the reception of information by an observer) was
introduced in \textbf{[Fi1]}.$\blacksquare$

\bigskip We have now seen that the general structure of quantum mechanics as
presented in Section 1.4 is essentially a mathematical framework for handling
noncommutative information. Based on this, we make the following two remarks:

\bigskip\noindent\textbf{1.6.9 The structure of spacetime. }If we assume that
information in our physical world is described by quantum mechanics, then we
are lead to conclude that information is actually a noncommutative phenomenon.
Perhaps this means that since information ``lives'' in spacetime (and possibly
in some way defines spacetime structure as was alluded to in 1.6.8), spacetime
itself is noncommutative, as has been suggested in attempts to construct
quantum spacetime and quantum gravity; see for example \textbf{[DFR]}. (This
opens the possibility that spacetime is discrete like many other quantum
phenomena; see for example \textbf{[Sm]} for a popular account.) On the other
extreme, the term ``noncommutative information'' may be a ``purely grammatical
trick'' of the sort \textbf{[Ma}, p. 188\textbf{]} mused might ``be the
ultimate solution of the quantum measurement problem''; this possibility seems
somewhat less interesting however.$\blacksquare$

\bigskip\noindent\textbf{1.6.10 The linear structure of quantum mechanics.
}The general structure of classical mechanics in Sections 1.3 and 1.4 is
linear since it is nothing more than probability theory, even though it can be
applied to physical systems where nonlinear aspects might be involved. It is
the statistical point of view that makes everything linear (essentially this
boils down to the use of averages, which are integrals and hence linear). The
same goes for quantum mechanics. Its linear structure should not be viewed as
an approximation to an underlying nonlinear world, but simply as a result of
the fact that it is a mathematical framework for probability theory (i.e.
statistics, averages), where the information involved happens to be
noncommutative. The appearance of a Hilbert space as the state space is simply
a mathematical way of representing the algebraic structure in Section 1.4. So
the linearity of (and hence superpositions in) the state space is just a
convenient way to express the fact that a measurement can invalidate the
information the observer had before the measurement, or more precisely, to
express noncommutative conditional probabilities. (Also see \textbf{[Fi2}, p.
175\textbf{]} and \textbf{[Ha}, p. 309\textbf{]} for similar remarks
concerning the linearity of quantum mechanics.)$\blacksquare$

\bigskip Fuchs and others have also argued convincingly that information
theoretic ideas are of great importance for the foundations of quantum
mechanics, in particular that a quantum state represents an observer's
information rather than having an objective reality (see \textbf{[FuP]},
\textbf{[Fu]} and \textbf{[CFS]}). Refer to \textbf{[St]} for a review of
quantum mechanics viewed as a generalization of classical probability theory.

We cannot claim that this ``noncommutative information interpretation'' solves
all of the conceptual problems of quantum mechanics, but for the case of a
physical system being observed by an observer not considered to be part of the
system, it does seem to clarify many issues without causing any new problems
(except if you consider the idea of noncommutative information itself to be a problem).

\section{A quantum analogue of Liouville's Theorem}

In Section 1.2 to 1.4 we saw that in purely C*-algebraic terms, quantum
mechanics and classical mechanics are identical, except of course for the fact
that the classical observable algebra is abelian while this is not in general
true for quantum mechanics. This suggests that it might be possible to find a
quantum mechanical analogue of Liouville's Theorem, a search we pursue in this
section for reasons explained in the Introduction, and simply because it is an
interesting possibility in its own right (see Proposition 1.7.5 for the final
result). Our first clue in this direction is the following simple proposition
(where for a $\sigma$-algebra in a set $X$, we denote by $B_{\infty}(\Sigma
)$\ the C*-algebra of all bounded complex-valued $\Sigma$-measurable functions
on $X$, with the sup-norm, its operations defined pointwise, and its
involution given by complex conjugation, as for the special case $B_{\infty
}(\mathbb{R}^{2n})$ in Section 1.3):

\bigskip\noindent\textbf{1.7.1 Proposition. }\textit{Let }$(X,\Sigma,\mu)
$\textit{\ be a measure space with }$\mu(X)<\infty$\textit{, and let
}$T:X\rightarrow X$\textit{\ be a mapping such that }$T^{-1}(S)\in\Sigma
$\textit{\ for all }$S\in\Sigma$\textit{. Define }$\tau$\textit{\ and
}$\varphi$\textit{\ by }$\tau(g)=g\circ T$\textit{\ and }$\varphi(g)=\int
gd\mu$\textit{\ for all }$g\in B_{\infty}(\Sigma)$\textit{. Then }$\mu
(T^{-1}(S))\leq\mu(S)$\textit{\ for all }$S\in\Sigma$\textit{\ if and only if
}$\varphi(\tau(g)^{\ast}\tau(g)))\leq\varphi(g^{\ast}g)$\textit{\ for all
}$g\in B_{\infty}(\Sigma)$. \textit{Also, }$\mu(T^{-1}(S))=\mu(S)$%
\textit{\ for all }$S\in\Sigma$\textit{\ if and only if }$\varphi
(\tau(g))=\varphi(g)$\textit{\ for all }$g\in B_{\infty}(\Sigma)$\textit{. }

\bigskip\noindent\emph{Proof. }We use standard measure theoretic arguments
(refer to \textbf{[Rud]}).

Suppose $\varphi\left(  \tau(g)^{\ast}\tau(g)\right)  \leq\varphi(g^{\ast}g)$
for all $g\in B_{\infty}(\Sigma)$, then it holds in particular for $g=\chi
_{S}$, where $S\in\Sigma$, and so
\begin{align*}
\mu\left(  T^{-1}(S)\right)   &  =\varphi\left(  \chi_{T^{-1}(S)}\right)
=\varphi\left(  \left(  \chi_{T^{-1}(S)}\right)  ^{\ast}\chi_{T^{-1}%
(S)}\right) \\
&  =\varphi\left(  \left(  \chi_{S}\circ T\right)  ^{\ast}\chi_{S}\circ
T\right)  =\varphi\left(  \tau\left(  \chi_{S}\right)  ^{\ast}\tau\left(
\chi_{S}\right)  \right) \\
&  \leq\varphi\left(  \left(  \chi_{S}\right)  ^{\ast}\chi_{S}\right)
=\varphi\left(  \chi_{S}\right) \\
&  =\mu(S)\text{.}%
\end{align*}
Similarly for the case of equality.

Conversely, suppose $\mu\left(  T^{-1}(S)\right)  \leq\mu(S)$ for all
$S\in\Sigma$. This is equivalent to having $\int\chi_{S}\circ Td\mu\leq
\int\chi_{S}d\mu$ for all $S\in\Sigma$. By Lebesgue's Monotone Convergence
Theorem this extends to all positive measurable functions, namely
\[
\int f\circ Td\mu\leq\int fd\mu
\]
for positive $f\in B_{\infty}(\Sigma)$ by considering an increasing sequence
$\left(  f_{n}\right)  $ of positive simple measurable functions converging
pointwise to $f$, since then $\left(  f_{n}\circ T\right)  $ is an increasing
sequence of positive simple measurable functions converging pointwise to
$f\circ T$. Setting $f=g^{\ast}g$ for any $g\in B_{\infty}(\Sigma)$, we
obtain
\[
\varphi\left(  \tau(g)^{\ast}\tau(g)\right)  =\int\left(  g^{\ast}g\right)
\circ Td\mu\leq\int g^{\ast}gd\mu=\varphi\left(  g^{\ast}g\right)  \text{.}%
\]
Similarly for the case of equality, and this then extends by linearity to
$\varphi\left(  \tau(g)\right)  =\varphi(g)$ for all $g\in B_{\infty}(\Sigma
)$.$\blacksquare$

\bigskip Consider a classical system whose phase point is confined to a Borel
set $F$ of finite volume in the phase space $\mathbb{R}^{2n}$. That is to say
$\lambda(F)<\infty$, where $\lambda$ is the Lebesgue measure on $\mathbb{R}%
^{2n}$.

\bigskip\noindent\textbf{1.7.2 Remark.} If the phase point is confined to a
set $F\subset\mathbb{R}^{2n}$, then we can view $F$ as the phase space of the
system (whether $F$ has finite volume or not), taking the $\sigma$-algebra
$\Sigma$ of measurable sets in $F$ as the intersections of the Borel sets of
$\mathbb{R}^{2n}$ with $F$. (In Section 1.3 we simply used the Borel sets of
$\mathbb{R}^{2n}$ as the $\sigma$-algebra of measurable sets in the phase
space.) We then replace the Lebesgue measure by its restriction to $F$
(assuming $F$ is Lebesgue measurable), and we use probability measures on $F$,
instead of on $\mathbb{R}^{2n}$. Also, the observables will be represented by
$\Sigma$-measurable functions on $F$, and the observable algebra will be
$B_{\infty}(F):=B_{\infty}(\Sigma)$. The whole of Section 1.3 can then be
repeated with $F$ in the place of $\mathbb{R}^{2n}$.$\blacksquare$

\bigskip We define a measure $\nu$ on the Borel sets of $\mathbb{R}^{2n}$ by
\[
\nu(S)=\lambda(S\cap F)\text{.}%
\]
Using Proposition 1.7.1 we see that Liouville's theorem for this system can
then be expressed in C*-algebraic terms by stating that
\begin{equation}
\varphi(\tau_{t}(g))=\varphi(g) \tag{7.1}%
\end{equation}
for all $g$ in $B_{\infty}(\mathbb{R}^{2n})$, where $\tau$ is given by
equation (3.3), and $\varphi(g)=\int gd\nu$ (so $\varphi$ is a positive linear
functional on $B_{\infty}(\mathbb{R}^{2n})$). This is because $\nu
(T_{-t}(S))=\lambda(T_{-t}(S)\cap F)=\lambda(T_{-t}(S)\cap T_{-t}%
(F))=\lambda(T_{-t}(S\cap F))=\lambda(S\cap F)=\nu(S)$, since we have
$T_{t}(F)\subset F$ for all $t\in\mathbb{R}$ (the phase point is confined to
$F$) and so $F\subset(T_{-t})^{-1}(F)=T_{t}(F)$, which means that $T_{t}%
(F)=F$. Note that the condition $\mu(X)<\infty$ in Proposition 1.7.1 can be
dropped if we only consider positive elements of $B_{\infty}(\Sigma)$. Hence
(7.1) would express Liouville's Theorem for systems not necessarily bounded in
phase space if we were to use $\lambda$ instead of $\nu$, and only consider
positive elements $g$ of $B_{\infty}(\mathbb{R}^{2n})$. (In this case
$\varphi$ could assume infinite values, and it would not be a linear mapping
on $B_{\infty}(\mathbb{R}^{2n})$ any more.) We will only work with the bounded
case though, since then the measure can be normalized to give a probability
measure, which is what we will use when studying recurrence and ergodicity..

Since quantum mechanics has the same C*-algebraic structure as classical
mechanics, we now suspect that a quantum mechanical analogue of Liouville's
Theorem should have the same form as (7.1). Let's look at this from a
different angle. In the Hilbert space setting for quantum mechanics, the state
space $\mathfrak{H}$ can be viewed as the analogue of the classical phase
space $\mathbb{R}^{2n}$. $\mathfrak{H}$ is a Hilbert space while we view
$\mathbb{R}^{2n}$ purely as a measurable space. Apart from dynamics, we saw in
Sections 1.2 to 1.4 that the central objects in both quantum and classical
mechanics are the projections. A projection defined on $\mathfrak{H}$ is
equivalent to a Hilbert subspace of $\mathfrak{H}$ (namely the range of the
projection). A projection defined on $\mathbb{R}^{2n}$ is a Borel measurable
characteristic function, and is therefore equivalent to a Borel set in
$\mathbb{R}^{2n}$. Liouville's Theorem is based on the existence of a natural
way of measuring the size of a Borel set in $\mathbb{R}^{2n}$, namely the
Lebesgue measure $\lambda$. We would therefore like to have a natural way of
measuring the size of a Hilbert subspace of $\mathfrak{H}$ in order to get a
quantum analogue of Liouville's Theorem. An obvious candidate is the (Hilbert)
dimension $\dim$. For the Hamiltonian flow $T_{t}$, Liouville's Theorem states
that $\lambda(T_{-t}(S))=\lambda(S)$ for every Borel set $S$. (We use
$T_{-t}(S)$ instead of $T_{t}(S)$, since this corresponds to the action of
$T_{t}$ on the observable algebra rather than on the states, namely $\chi
_{S}\circ T_{t}=\chi_{T_{-t}(S)}$.) In the state space time-evolution is given
by a one-parameter unitary group $U_{t}$ on $\mathfrak{H} $, and for any
Hilbert subspace $\mathfrak{K}$ of $\mathfrak{H}$ we have $\dim(U_{t}^{\ast
}\mathfrak{K})=\dim(U_{-t}\mathfrak{K})=\dim(\mathfrak{K})$. This is clearly
similar to Liouville's theorem. For a finite dimensional state space we will
in fact view this as a quantum analogue of Liouville's Theorem. (This remark
is also made on p. 83-84 of \textbf{[Ba]}.) However, since state spaces are
usually infinite dimensional, we would like to work with something similar to
$\dim$ which does not assume infinite values.

This leads us naturally to the C*-algebras known as finite von Neumann
algebras (see for example \textbf{[KR2]}), since for such an algebra there is
a dimension function, defined on the projections of the algebra, which does
not assume infinite values. This function is in fact the restriction of a
so-called trace defined on the whole algebra, so we might as well work with
this trace. We now explain this in more detail.

Let $\mathfrak{M}$ denote a finite von Neumann algebra on a Hilbert space
$\mathfrak{H}$, and let $\mathfrak{M}^{\prime}$ be its commutant. Then there
is a unique positive linear mapping tr$:\mathfrak{M\rightarrow M\cap
M}^{\prime}$ such that tr$(AB)=$ tr$(BA)$ and tr$(C)=C$ for all $A,B\in
\mathfrak{M}$ and $C\in\mathfrak{M\cap M}^{\prime}$. We call tr \textit{the
trace }of $\mathfrak{M}$. This trace is faithful, that is to say tr$(A^{\ast
}A)>0$ for $A\neq0$. (Conversely, if such a faithful trace exists on a von
Neumann algebra $\mathfrak{N}$, then $\mathfrak{N}$ is finite \textbf{[KR2},
Section 8.1\textbf{]}, and hence this could be taken as the definition of a
\emph{finite} von Neumann algebra.) We mention that in the special case where
$\mathfrak{M=L(H)}$, with $\mathfrak{H}$ finite dimensional, tr is just the
usual trace (sum of eigenvalues) normalized such that tr$(1)=1$.

For a projection $P\in\mathfrak{M}$ of $\mathfrak{H}$ onto the Hilbert
subspace $\mathfrak{K}$, we see that $U_{t}^{\ast}PU_{t}$ is the projection of
$\mathfrak{H}$ onto $U_{t}^{\ast}\mathfrak{K}$, where $U_{t}$ is a
one-parameter unitary group on $\mathfrak{H}$. So in the framework of finite
von Neumann algebras we would like to replace the equation $\dim(U_{t}^{\ast
}\mathfrak{K})=\dim(\mathfrak{K})$ mentioned above by tr$(U_{t}^{\ast}%
PU_{t})=$ tr$(P)$ as a quantum analogue of Liouville's Theorem.

If a self-adjoint (possibly unbounded) operator $A$ in $\mathfrak{H}$ is an
observable and $\mathfrak{M}$ an observable algebra of a physical system, then
we want the spectral projections $\chi_{V}(A)$ of $A$ to be contained in
$\mathfrak{M}$, where $V$ is any Borel set in $\mathbb{R}$, since these
projections are the projections of the yes/no experiments that can be
performed on the system. But then $f(A)\in\mathfrak{M}$ for any bounded
complex-valued Borel function $f\mathbb{\ }$on $\mathbb{R}$. (Our argument
here is roughly that there is a bounded sequence of bounded simple functions
$s_{n}$ converging pointwise to $f$, which implies that $s_{n}(A)x\rightarrow
f(A)x$ for all $x\in\mathfrak{H}$, i.e. $s_{n}(A)$ converges strongly to
$f(A)$. Since a von Neumann algebra is strongly closed, it follows that
$f(A)\in\mathfrak{M}$. See \textbf{[SZ}, 9.10, 9.11 and 9.32\textbf{]}.) In
particular $e^{-iAt}\in\mathfrak{M}$ for all real $t$.

For these reasons we will consider physical systems of the following nature:

\bigskip\noindent\textbf{1.7.3 Definition. }\textit{A }\textbf{bounded quantum
system}\textit{\ is a quantum mechanical system for which we can take the
observable algebra as a finite von Neumann algebra }$\mathfrak{M}$ \textit{on
a Hilbert space }$\mathfrak{H}$ \textit{such that the Hamiltonian }%
$H$\textit{\ of the system can be represented as a self-adjoint (possibly
unbounded) linear operator in }$\mathfrak{H}$\textit{\ with }$e^{-iHt}%
\in\mathfrak{M}$ \textit{for real }$t$\textit{. We denote this system by
}$(\mathfrak{M},\mathfrak{H},H)$.

\bigskip The reason for the term ``bounded'' will become clear in Section 1.9.

\bigskip\noindent\textbf{1.7.4 Remark. }If for a bounded quantum system
$(\mathfrak{M},\mathfrak{H},H)$ the unit vectors of $x\in\mathfrak{H}$ are
pure states of the system, that is to say $\left\langle x,\cdot x\right\rangle
$ is a pure state on $\mathfrak{M}$ for such $x$, then $\mathfrak{H}$ can be
viewed as the state space of the system (this happens for example when
$\mathfrak{M=L(H)}$ with $\mathfrak{H}$ finite dimensional). However, the unit
elements of $\mathfrak{H}$ need not be pure states of the system, as we will
now show, in which case $\mathfrak{H}$ is \emph{not} the state space of the
system, but merely acts as a ``carrier'' for the observable algebra
$\mathfrak{M}$.

Let $\mathfrak{G}$ be a finite dimensional Hilbert space, and consider a mixed
(i.e. not pure) faithful normal state $\omega$ on $\mathfrak{L(G)}$, for
example a Gibbs state $\omega(A)=$ Tr$(\rho A)$ where $\rho=e^{-\beta G}%
/$Tr$(e^{-\beta G})$ with $G\in\mathfrak{L(}\mathfrak{G)}$ the Hamiltonian of
some system with state space $\mathfrak{G}$, and $\beta$ the inverse
temperature of the system (see \textbf{[D1}, Proposition 2.3.9\textbf{]} for
example). Here \emph{normal} refers to the form Tr$(\rho\cdot)$ of the state,
where $\rho$ is a density operator, while \emph{faithful} means that
$\omega(A^{\ast}A)>0$ if $A\neq0$.

Let $(\mathfrak{H},\pi,\Omega)$ be a cyclic representation of $(\mathfrak{L(}%
\mathfrak{G)},\omega)$ as in Section 1.2. Let $\mathfrak{M}:=$ $\pi
(\mathfrak{L(}\mathfrak{G)})$ and $H:=\pi(G)$, then we prove that
$(\mathfrak{M},\mathfrak{H},H)$ is a bounded quantum system.

First, $\mathfrak{M}$ is a von Neumann algebra, since $\mathfrak{L(G)}$ is a
von Neumann algebra and $\omega$ is normal \textbf{[BR}, Theorem
2.4.24\textbf{]}. Furthermore, $\pi$ is a $\ast$-isomorphism since $\omega$ is
faithful \textbf{[BR}, Proposition 2.5.6\textbf{]}. (Also see \textbf{[D1},
Proposition 4.4.9\textbf{]}, for the same results.) It is known that
$\mathfrak{L(G)}^{\prime}=\mathbb{C}$ (see \textbf{[D1}, Proposition
1.4.7\textbf{]}), and since $\mathfrak{M}$ is $\ast$-isomorphic to
$\mathfrak{L(G)}$, this means that the elements of $\mathfrak{M}$ which
commute with $\mathfrak{M}$ are also just the multiples of unity, that is to
say $\mathfrak{M}\cap\mathfrak{M}^{\prime}=\mathbb{C}$. Since $\pi$ is
injective and $\pi(1)=1$, we can therefore define a trace
$\mathfrak{M\rightarrow M}\cap\mathfrak{M}^{\prime}$ (in the sense described
above) by tr$(\pi(A)):=$ tr$(A)$, where tr on the right is the (normalized)
trace of $\mathfrak{L(G)}$. This trace is faithful on $\mathfrak{M}$ since the
trace on $\mathfrak{L(G)}$ is faithful. Hence $\mathfrak{M}$ is finite (see
above). Since $\pi$ is a $\ast$-homomorphism from a Banach $\ast$-algebra to a
C*-algebra, it is continuous \textbf{[Mu}, Theorem 2.1.7\textbf{]}. Hence
\[
e^{-iHt}=e^{-i\pi(G)t}=\pi(e^{-iGt})\in\mathfrak{M}\text{.}%
\]
This proves that $(\mathfrak{M},\mathfrak{H},H)$ is a bounded quantum system.
(As an example of the situation in Proposition 1.7.5 below, note that
$e^{-iHt}$ gives the time-evolution of the system in terms of $\mathfrak{M}$
rather than $\mathfrak{L(G)}$, namely
\[
\pi(e^{iGt}Ae^{-iGt})=e^{iHt}\pi(A)e^{-iHt}%
\]
for $A\in\mathfrak{L(G)}$.)

However, the state $\left\langle \Omega,\cdot\Omega\right\rangle =\omega
\circ\pi^{-1}$ is not pure on $\mathfrak{M}$, since $\omega$ is not pure (see
\textbf{[BR}, Definition 2.3.14\textbf{]} for the formal mathematical
definition of a pure state on a C*-algebra). In other words $\Omega$ is not a
pure state of the system, and therefore $\mathfrak{H}$ is not the state space
of the system.$\blacksquare$

\bigskip We now propose a quantum analogue of Liouville's Theorem based on the
intuitive arguments in terms of dimension given above. We give it in the form
of a proposition (its proof is easy; the work went into finding a sensible
candidate for such an analogue):

\bigskip\noindent\textbf{1.7.5 Proposition.}\textit{\ Consider a bounded
quantum system }$(\mathfrak{M},\mathfrak{H},H)$\textit{. By Stone's Theorem
}$U_{t}=e^{-iHt}$\textit{\ is a one-parameter unitary group on }$\mathfrak{H}%
$\textit{. Let }$\tau$\textit{\ be the time-evolution of the system, i.e.
}$\tau_{t}(A)=U_{t}^{\ast}AU_{t}$\textit{\ for all }$A\in\mathfrak{M}%
$\textit{. Then}
\begin{equation}
\text{tr}(\tau_{t}(A))=\text{ tr}(A) \tag{7.2}%
\end{equation}
\textit{for all }$A$\textit{\ in }$\mathfrak{M}$\textit{, where }%
tr\textit{\ is the trace of }$\mathfrak{M}$\textit{. (This last statement is
our quantum analogue of Liouville's theorem.)}

\bigskip\noindent\textit{Proof. }Since $U_{t}\in\mathfrak{M}$, we have
tr$(\tau_{t}(A))=$ tr$(U_{t}^{*}AU_{t})=$ tr$(U_{t}U_{t}^{*}A)=$ tr$(A)$.
$\blacksquare$

\bigskip As we suspected, our quantum analogue of Liouville's theorem,
expressed by (7.2), is of the same form as the C*-algebraic formulation of the
classical Liouville Theorem as given by (7.1), with $\varphi$ replaced by tr.
Remember that $\varphi$ and tr are both positive linear mappings on the
respective observable algebras.

A somewhat different approach to a quantum analogue of Liouville's Theorem is
described in \textbf{[AM]}.

\bigskip\noindent\textbf{1.7.6 Remark.}\textit{\ }The classical Liouville
Theorem can also be expressed in terms of the Liouville equation
\[
\frac{\partial\rho}{\partial t}=\{\rho,H\}
\]
where $\rho:\mathbb{R}^{2n}\times\mathbb{R\rightarrow R}$ is the density
function, $H$ the classical Hamiltonian, and $\{\cdot,\cdot\}$ the Poisson
bracket. This equation can be seen as describing the flow of a fluid in phase
space such that at any point moving along with the fluid, the density of the
fluid remains constant. So besides giving the time-evolution, this equation
also states a property of the time-evolution, namely that it conserves volume
in phase space. In quantum mechanics we have the analogous von Neumann
equation
\[
\frac{d\rho}{dt}=i[\rho,H]
\]
where $\rho:\mathbb{R\rightarrow}\mathfrak{L(H)}$ is the density operator as a
function of time (note that here the derivative with respect to time is total
instead of partial). This equation merely gives the time-evolution
$\rho(t)=\tau_{-t}(\rho(0))$ of the density operator, where $\tau$ is the
time-evolution on the observable algebra here viewed as acting on the state
instead of the observables. Von Neumann's equation by itself should therefore
not be regarded as a quantum mechanical analogue of Liouville's
Theorem.$\blacksquare$

\section{The state of no information}

In (b) of Section 1.4, we said that the state of a system is constructed from
information gained during measurements previously performed on the system. If
the observer hasn't performed any measurements on the system, then he has no
information regarding the system (however, the observable algebra is assumed
to be known, i.e. the observer knows what the system is). Can we describe this
situation by a state on the observable algebra of the system? It turns out
that we can in the framework of Section 1.7 (namely for bounded quantum
systems and for classical systems with phase space $F\subset\mathbb{R}^{2n}$
of finite volume). Such a state on the observable algebra can then be called a
\textit{state of no information}.

\bigskip\noindent\textbf{1.8.1 Classical mechanics. }Let's first consider a
classical system. Assume that its phase point is confined to a (Borel) set $F$
of finite volume in the phase space $\mathbb{R}^{2n}$, i.e. $\lambda
(F)<\infty$. (So we can view $F$ as the system's phase space; see Remark
1.7.2.) We now argue that practical matters force us to assume $\lambda(F)>0
$: In practice it is impossible to measure any of the position or momentum
coordinates of the system precisely, so it is safe to assume that each of
these coordinates can at best be determined only up to some interval of
positive length, and hence $F$ must contain the product of these intervals,
which implies $\lambda(F)>0$. If $F$ did not contain this product, it would
not make sense for us to use $F$ as the phase space of the system, since we
would not even know if the system's phase point is contained in $F$.

We can therefore normalize $\lambda$ on $F$ by defining a probability measure
$\lambda^{\prime}$ on the Borel sets of $\mathbb{R}^{2n}$ by
\[
\lambda^{\prime}(S)=\lambda(S\cap F)/\lambda(F)\text{.}%
\]
If we now view $\lambda^{\prime}$ as describing a state of the system (as
explained in Section 1.3), then it essentially says that every part of $F$ is
equally likely to contain the phase point of the system. Mathematically this
boils down to the fact that the Lebesgue measure $\lambda$ is translation
invariant, which means that it is the same everywhere, so $\lambda^{\prime}$
can be viewed as a uniform probability distribution. In other words, when the
observer knows nothing about where the phase point of the system is (aside
from the fact that it is in $F$), then we can describe the observer's
information by $\lambda^{\prime}$, or in C*-algebraic terms by the state
$\varphi$ on $B_{\infty}(\mathbb{R}^{2n})$ defined by
\[
\varphi(g)=\int gd\lambda^{\prime}.
\]
Since Lebesgue measure is the unique (up to some normalization factor)
translation invariant Borel measure on $\mathbb{R}^{2n}$ assuming finite
values on compact sets (which are bounded and therefore should have finite
volumes), we can view $\varphi$ as \textit{the }state of no information.
(Refer to \textbf{[Rud]} for an exposition of the properties of the Lebesgue measure.)

For this state of no information to make sense, it has to be compatible with
the time-evolution of the system in the following sense: If the observer has
no information regarding the system at time $0$, and he performs no
measurements on the system up to some later time $t$, then at time $t$ he
still has no information regarding the system. This means that if we apply the
time-evolution $\tau$ of the system to the state $\varphi$ instead of to the
observable algebra, to obtain the state $\varphi\circ\tau_{t}$ at time $t$,
then this state still has to represent the state of no information. That is to
say, we must have $\varphi\circ\tau_{t}=\varphi$. But this is exactly what
Liouville's Theorem states (see equation (7.1)). So we see that Liouville's
Theorem is intimately related to the idea of information, in the sense that it
ensures that the state of no information is compatible with the system's
time-evolution. We can say that Liouville's Theorem makes the state of no
information dynamically sensible. We can also view this as a special case of a
group invariance defining a probability distribution, in this case invariance
under time-evolution defining the state of no information (see \textbf{[J]}
for more on this idea).

\bigskip\noindent\textbf{1.8.2 Quantum mechanics. }Now we turn to a bounded
quantum system as defined in 1.7.3, namely $(\mathfrak{M},\mathfrak{H},H)$
where we assume that $\mathfrak{M}$ is a \emph{factor} (that is to say
$\mathfrak{M\cap M}^{\prime}=\mathbb{C}1$), which means that we can take tr to
be complex-valued. (In general we will refer to a finite von Neumann algebra
which is a factor, as a \emph{finite factor}.) The reason for assuming
$\mathfrak{M}$ to be a factor is that tr is then a state on $\mathfrak{M}$,
since we know that tr is positive and normalized. This means that tr can in
principle represent a physical state as described in Section 1.2.

In Section 1.7 we saw that tr can be viewed as a quantum analogue of
integration over a bounded set in phase space with respect to Lebesgue measure
$\lambda$, in other words, as a quantum analogue of $\varphi$ in 1.8.1. The
basic intuition here is that our quantum analogue of Liouville's Theorem is
expressed in terms of tr in precisely the same form as that in which
Liouville's Theorem is expressed in terms of $\varphi$, namely tr$(\tau
_{t}(A))=$ tr$(A)$ as compared to $\varphi(\tau_{t}(g))=\varphi(g)$. By this
analogy between tr and $\varphi$ we would expect tr to be the state of the
bounded quantum system when the observer knows nothing about the system, in
other words that tr is a state of no information. This is indeed true in the
special case where $\mathfrak{H}$ is finite dimensional and $\mathfrak{M=L(H)}%
$, since for any rank one projection $Q$ in $\mathfrak{M}$ we then have
tr$(Q)=1/\dim(\mathfrak{H})$ which tells us that if the state is tr, then all
eigenvalues are equally probable when an observable is measured (assuming the
observable has no degenerate eigenvalues).

As mentioned in Section 1.7, tr is the unique state on $\mathfrak{M}$ such
that tr$(AB)=$ tr$(BA)$ for all $A,B\in\mathfrak{M}$, but this is in fact
equivalent to the condition that tr$(U^{\ast}PU)=$ tr$(P)$ for all unitary
$U\in\mathfrak{M}$ and all projections $P\in\mathfrak{M}$ (see \textbf{[KR2},
Proposition 8.1.1 and its proof\textbf{]}). We can view unitary operators as
rotations in the state space of the quantum system, so tr$(AB)=$ tr$(BA)$
tells us that rotations of the state space preserve the ``size'' of Hilbert
subspaces (which correspond to projections), where ``size'' here refers to the
dimension function on the projections of $\mathfrak{M}$, mentioned in Section
1.7. This is the quantum mechanical equivalent of the classical situation
where translations preserve Lebesgue measure, since as described in Section
1.7, the dimension of Hilbert subspaces of the state space should correspond
to Lebesgue measure as a measure of the size of Borel sets (which correspond
to projections in the classical case). In the same way as in the classical
case in 1.8.1, we can therefore view tr as \emph{the} state of no information
of a bounded quantum system.

As explained in 1.8.1, Liouville's Theorem is central in the concept of a
state of no information, since it makes such a state dynamically sensible. The
same argument applies to our quantum analogue of Liouville's Theorem
(Proposition 1.7.5) to see that it ensures that the state of no information tr
is compatible with the system's time evolution, namely tr$\circ\tau_{t}= $ tr.

Furthermore, since tr is ultraweakly continuous, it is a normal state and
hence it is given by a density operator (see \textbf{[KR2}, Theorem 8.2.8,
Proposition 7.4.5, Theorem 7.1.12\textbf{]} and \textbf{[BR}, Theorem
2.4.21\textbf{]}), as one might expect for a physically meaningful state (keep
in mind, however, that this density operator is defined on $\mathfrak{H}$,
which is not necessarily the state space of the system; see Remark 1.7.4). We
therefore suggest the following hypothesis:

\bigskip\noindent\textbf{1.8.3 Postulate. }\textit{Consider a bounded quantum
system }$(\mathfrak{M},\mathfrak{H},H)$,\textit{\ where }$\mathfrak{M}%
$\textit{\ is a factor. If the observer has no information regarding the
system, then the state of the system is given by the trace }tr \textit{of
}$\mathfrak{M}$\textit{.}

\section{Bounded quantum systems}

In this section we discuss the possible physical significance of bounded
quantum systems, using the analogy with classical systems built up in Sections
1.7 and 1.8. What we want to know is which physical systems can be
mathematically described as bounded quantum systems with the observable
algebras being factors, since this is the type of system considered in
Postulate 1.8.3.

In Sections 1.7 and 1.8 we considered the case of a classical system whose
phase point is confined to a set $F$ of finite volume, which meant that we
could view $F$ as the phase space of the system. A special case of this is
where the phase space is bounded (i.e. contained in some ball in
$\mathbb{R}^{2n}$). Bounded sets are indeed less general than sets of finite
volume, as witnessed for the set $F=\{(x,y)\in\mathbb{R}^{2}:0\leq y\leq
e^{-x},0\leq x<\infty\}$ which is an unbounded closed (and hence Borel) set
which has a part of positive measure lying outside any ball in $\mathbb{R}%
^{2}$ (we might call this set \emph{Lebesgue unbounded}, since the part that
goes to infinity does not have zero Lebesgue measure), but even so $F$ has a
finite Lebesgue measure of $1$. (We will not pursue the question of whether a
Lebesgue unbounded phase space of finite volume actually occurs in any
physical system, since our arguments here will be heuristic and based on the
idea of boundedness.)

From a physical standpoint the phase space is bounded if the system itself is
confined to a finite volume in space, and it is isolated from outside
influences (which could increase its energy content), to prevent any of its
momentum components to go to infinity. To see that this is the case, use
Cartesian coordinates. Here we assume that each potential of the form $-1/r$
or the like has some ``cut-off'' at small values of $r$, since for example
particles are of finite size and collide when they get too close. The point of
this is that there is not an infinite amount of potential energy available in
the system (potentials do not go to $-\infty$). Consider as illustration a
potential with the general shape given by $-1/r+0.0015/r^{4}$, where the
$0.0015/r^{4}$ term causes the cut-off, that is to say for ``large'' $r$ the
potential looks like $-1/r$, but as $r>0$ decreases, the potential deviates
from $-1/r$, reaches a minimum, and then goes to $+\infty$.

Based on the analogy between bounded quantum systems and classical systems
with bounded phase space presented in Sections 1.7 and 1.8, we might now guess
that quantum systems bounded in space and isolated from outside influences can
be described as bounded quantum systems in the sense of Definition 1.7.3 with
$\mathfrak{M}$ a factor.

Of course, the analogy actually extends to the more general case of classical
systems with phase space of finite volume, but since we have no hard evidence
apart from this analogy, it is probably best not to push it to its limits. (We
will find some additional indirect evidence supporting our guess when we
discuss recurrence for quantum systems in Section 3.1.) Also, it is not
exactly clear how the idea of a finite volume of phase space should be
translated to quantum mechanics; possibly one could approach this problem by
considering a quantum system which is a quantization of a classical system
whose phase space has finite volume, however, the argument by analogy that
this system too is a bounded quantum system, is becoming more and more
tenuous. This seems to be related to the nuclearity requirement in quantum
field theory (see \textbf{[Ha]}), where a finite volume in classical phase
space is intuitively thought of as corresponding to a finite dimensional
subspace of quantum state space. Since a quantum system whose state space
$\mathfrak{H}$ is finite dimensional is clearly a bounded quantum system (the
observable algebra $\mathfrak{L(H)}$ is a finite factor in this case), our
guess certainly does not seem too far-fetched from this point of view.

We state our guesswork as a conjecture:

\bigskip\noindent\textbf{1.9.1 Conjecture. }\textit{A quantum mechanical
system bounded in space, and isolated from outside influences, can be
mathematically described as a bounded quantum system in the sense of
Definition 1.7.3, with the observable algebra }$\mathfrak{M}$\textit{\ a
factor. }

\bigskip\noindent\textbf{1.9.2 Remark.} A bounded quantum system
$(\mathfrak{M},\mathfrak{H},H)$ as defined in 1.7.3, with $\mathfrak{M}$ a
factor, deviates from the usual ``type I'' quantum mechanics (see
\textbf{[Ha}, Section VII.2\textbf{]}), in that the former does not
necessarily have ``finest'' yes/no experiments. This refers to the fact that
the range of the dimension function (on the projections of $\mathfrak{M}$) can
be the whole interval $[0,1]$ which has no minimum non-zero value, in which
case $\mathfrak{M}$ is called a type II$_{1}$ factor. It should be noted
though, that a bounded quantum system always has pure states (states of
maximal information), since any non-zero C*-algebra (and in particular a
finite factor) has pure states (see \textbf{[Mu}, Theorem 5.1.11\textbf{]}),
as is physically required, since nonmaximal information is a result of the
observer's lack of precision rather than a property of the system. Loosely
this means that although an observer can always do a finer measurement than
the ones he already did, such a measurement will not necessarily improve his
information, it might simply give new information invalidating his old
information (noncommuting observables), but giving a ``smaller'' subspace in
the state space, not contained in the subspace corresponding to his old
information, since $\dim(P_{2})\leq\dim(P_{1})$ does not imply $P_{2}\leq
P_{1}$ in the C*-algebraic partial order. (Keep in mind that $\mathfrak{H}$ is
not necessarily the state space, it \ just acts as a ``carrier'' for
$\mathfrak{M}$; see Remark 1.7.4.)

In type I quantum mechanics the observable algebra is simply taken as the type
I factor $\mathfrak{L(H)}$ where a separable Hilbert space $\mathfrak{H}$ is
the state space of the system. The dimension function on the projections of
$\mathfrak{L(H)}$ is simply the dimension of the range of a projection, and
hence it has the minimum non-zero value $1$; see \textbf{[Co}, p.
455\textbf{]} for example. The projections with dimension one represent the
finest yes/no experiments that can be performed on the system.$\blacksquare$

\bigskip\noindent\textbf{1.9.3 Example.} A one-dimensional quantum harmonic
oscillator has a discrete unbounded energy spectrum consisting of equally
spaced values
\[
E_{n}=(2n+1)E_{0}%
\]
for $n=0,1,2,...$ where $E_{0}>0$ is the lowest energy value (see
\textbf{[CDL}, Section V.B\textbf{]} or \textbf{[Kre}, Example
11.3-1\textbf{]}). In the state of no information each of these energy values
should be equally likely, but that would mean that all of them have
probability zero, which doesn't make physical sense, since if the oscillator's
energy is measured, some value must be obtained, and so this value does not
have zero probability. Therefore the state of no information does not exist as
a state on the observable algebra in this case, which means that the
oscillator is not a bounded quantum system. This makes sense, since the energy
eigenstate in $L^{2}(\mathbb{R})$ corresponding to $E_{n}$ is a ``Gaussian
tapered'' Hermite polynomial of the form
\[
e^{-\gamma^{2}x^{2}/2}H_{n}(\gamma x)
\]
(where $x$ is the position, and $\gamma$ a constant deriving from the physical
properties of the oscillator, namely mass and frequency), which has a steadily
increasing non-negligible spatial extension as $n$ increases, corresponding to
the classical situation where the amplitude in space increases as the energy
increases (\textbf{[CDL}, Section V.C.2\textbf{]} or \textbf{[Kre}, Example
11.3-1\textbf{]}). So if all the energy values are allowed, then the system is
not bounded in space.

An approximate description of a quantum harmonic oscillator bounded in space
as a bounded quantum system, could be to take the state space $\mathfrak{H}$
as the finite dimensional subspace of $L^{2}(\mathbb{R})$ spanned by energy
eigenstates corresponding to $E_{0},...,E_{N}$ for some $N$, and then using
the finite factor $\mathfrak{L}(\mathfrak{H})$ as the observable algebra.
However, a careful analysis from the ground up would be necessary to see if an
isolated quantum harmonic oscillator bounded in space is indeed a bounded
quantum system.$\blacksquare$

\chapter{ Recurrence and ergodicity in $\ast$-algebras}

In this chapter (based on \textbf{[DS]}), results concerning recurrence and
ergodicity are proved in an abstract Hilbert space setting based on the proof
of Khintchine's recurrence theorem for sets, and on the Hilbert space
characterization of ergodicity. These results are carried over to a
noncommutative $\ast$-algebraic setting using the GNS-construction. This
generalizes the corresponding measure theoretic results, in particular a
variation of Khintchine's Theorem for ergodic systems, where the image of one
set overlaps with another set, instead of with itself.

\section{Introduction}

The inspiration for this chapter is the following theorem of Khintchine dating
from 1934 (see \textbf{[Pete]} for a proof):

\bigskip\noindent\textbf{2.1.1 Khintchine's Theorem.}\textsc{\ }\emph{Let
}$(X,\Sigma,\mu)$\emph{\ be a probability space (that is to say, }$\mu
$\emph{\ is a measure on a }$\sigma$\emph{-algebra }$\Sigma$\emph{\ of subsets
of a set }$X$\emph{, with }$\mu(X)=1$\emph{), and consider a mapping
}$T:X\rightarrow X$\emph{\ such that }$T^{-1}(S)\in\Sigma$\emph{\ and }%
$\mu(T^{-1}(S))\leq\mu(S)$\emph{\ for all }$S\in\Sigma$\emph{. Then for any
}$A\in\Sigma$\emph{\ and }$\varepsilon>0$\emph{, the set}
\[
E=\left\{  k\in\mathbb{N}:\mu\left(  A\cap T^{-k}(A)\right)  >\mu
(A)^{2}-\varepsilon\right\}
\]
\emph{is relatively dense in }$\mathbb{N}=\{1,2,3,...\}$\emph{.}

\bigskip We will call $(X,\Sigma,\mu,T)$, as given above, a \emph{measure
theoretic dynamical system}. Recall that the relatively denseness of $E$ in
$\mathbb{N}$ means that there exists an $n\in\mathbb{N}$ such that
$E\cap\{j,j+1,...,j+n-1\}$ is non-empty for every $j\in\mathbb{N}$.
Khintchine's Theorem is an example of a recurrence result. It tells us that
for every $k\in E$, the set $A$ contains a set $A\cap T^{-k}(A)$ of measure
larger than $\mu(A)^{2}-\varepsilon$ which is mapped back into $A$ by $T^{k}$.

A question that arises from Khintchine's Theorem is whether, given
$A,B\in\Sigma$ and $\varepsilon>0$, the set
\[
F=\left\{  k\in\mathbb{N}:\mu\left(  A\cap T^{-k}(B)\right)  >\mu
(A)\mu(B)-\varepsilon\right\}
\]
is relatively dense in $\mathbb{N}$. This is clearly not true in general, for
example if $T$ is the identity and $A$, $B$ and $\varepsilon$ are chosen such
that $\mu(A)\mu(B)>\varepsilon$ while $A\cap B$ is empty, then $F$ is empty.
$T$ has to ``mix'' the measure space sufficiently for $F$ to be non-empty. In
\textbf{[Wa]} it is shown for the case where $\mu(T^{-1}(S))=\mu(S)$ for all
$S\in\Sigma$, that if for every pair $A,B\in\Sigma$ of positive measure there
exists some $k\in\mathbb{N}$ such that $\mu\left(  A\cap T^{-k}(B)\right)
>0$, then the dynamical system is ergodic. Ergodicity therefore seems like the
natural concept to use when considering the question posed above. This is
indeed what we will do.

The notion of ergodicity originally developed as a way to characterize systems
in classical statistical mechanics for which the time mean and the phase space
mean of any observable are equal. For our purposes it will be most convenient
to define ergodicity of a measure theoretic dynamical system $(X,\Sigma
,\mu,T)$ as follows (refer to \textbf{[Pete]}, for example): $(X,\Sigma
,\mu,T)$ is called \emph{ergodic }if the fixed points of the linear Hilbert
space operator $U:L^{2}(\mu)\rightarrow L^{2}(\mu):f\mapsto f\circ T$ form a
one-dimensional subspace of $L^{2}(\mu)$. Keep in mind that $L^{2}(\mu)$
consists of equivalence classes of functions, with two functions equivalent if
they are equal almost everywhere, but it is easy to see that $U$ is
well-defined on $L^{2}(\mu)$, that is to say, if $f$ and $g$ are measurable
functions equal almost everywhere, then $f\circ T$ and $g\circ T$ are equal
almost everywhere. Also, for $f\in L^{2}(\mu)$ we have
\[
\int\left|  f\circ T\right|  ^{2}d\mu=\int\left|  f\right|  ^{2}\circ
Td\mu=\int\left|  f\right|  ^{2}d(\mu\circ T^{-1})\leq\int\left|  f\right|
^{2}d\mu<\infty
\]
and so $f\circ T\in L^{2}(\mu)$. Furthermore this inequality says that
$\left\|  U\right\|  \leq1$. Here $\mu\circ T^{-1}$ is the measure on $\Sigma$
defined by $\left(  \mu\circ T^{-1}\right)  (S):=\mu\left(  T^{-1}(S)\right)
\leq\mu(S)$.

As we shall see, the ideas we have discussed so far are not really measure
theoretic in nature. This is in large part due to the fact that the proof of
Khintchine's Theorem is essentially a Hilbert space proof using the Mean
Ergodic Theorem. This proof can for the most part be written purely in Hilbert
space terms, hence giving an abstract Hilbert space result. Along with the
Hilbert space characterization of ergodicity given above, this means that a
fair amount of ergodic theory can be done purely in an abstract Hilbert space
setting. This is the approach taken in Section 2.4, using the Mean Ergodic
Theorem as the basic tool.

Having built up some ergodic theory in abstract Hilbert spaces, nothing is to
stop us from applying the results to mathematical structures other than
measure theoretic dynamical systems. The mathematical structure we will
consider is much more general than measure theoretic dynamical systems and can
easily be motivated as follows: From a measure theoretic dynamical system
$(X,\Sigma,\mu,T)$ we obtain the unital $\ast$-algebra $B_{\infty}(\Sigma)$ of
all bounded complex-valued measurable functions defined on $X$, and two linear
mappings
\[
\varphi:B_{\infty}(\Sigma)\rightarrow\mathbb{C}:f\mapsto\int fd\mu
\]
and
\begin{equation}
\tau:B_{\infty}(\Sigma)\rightarrow B_{\infty}(\Sigma):f\mapsto f\circ T
\tag{1.1}%
\end{equation}
with the following properties: $\varphi(1)=1$, $\varphi(f^{\ast}f)\geq0$,
$\tau(1)=1$ and $\varphi(\tau(f)^{\ast}\tau(f))\leq\varphi(f^{\ast}f)$ for all
$f\in B_{\infty}(\Sigma)$ by Proposition 1.7.1, where $f^{\ast}=\overline{f}$
defines the involution on $B_{\infty}(\Sigma)$, making it a $\ast$-algebra. We
can view this abstractly by replacing $B_{\infty}(\Sigma)$ with any unital
$\ast$-algebra and considering linear mappings $\varphi$ and $\tau$ on it with
the properties mentioned above. (A \textit{unital }$\ast$\textit{-algebra}
$\mathfrak{A}$ is an algebra with an involution, and a unit element denoted by
$1$, that is to say $1A=A=A1$ for all $A\in\mathfrak{A}$. We will only work
with the case of complex scalars.) The most obvious generalization this brings
is that the unital $\ast$-algebra need not be commutative, for example the
bounded linear operators on a Hilbert space. Also note that $\tau$ in (1.1) is
a $\ast$-homomorphism of $B_{\infty}(\Sigma)$, but we will not need this
property of $\tau$ in the abstract $\ast$-algebraic setting. We describe the
$\ast$-algebraic setting in more detail in Section 2.3, and in Section 2.5 the
Hilbert space results are applied to this setting using the GNS-construction
(treated in Section 2.2). In Section 2.6 we obtain the measure theoretic
results as a special case, and also briefly discuss another special case,
namely von Neumann algebras.

In Section 2.7 an alternative approach to recurrence is described where
$\varphi$ is not required to be linear (which precludes the use of the
GNS-construction), and can even assume values in a unital C*-algebra. Section
2.7 is independent from the rest of the work in this chapter.

\bigskip\noindent\textbf{2.1.2 Remark. }In Chapter 1 the observable algebra of
a physical system was assumed to be a unital C*-algebra, rather than merely a
unital $\ast$-algebra. This assumption is not restrictive, since the
representations $\mathfrak{L(H)}$ and $B_{\infty}(\Sigma)$, and also any von
Neumann algebra, are indeed C*-algebras. In the general structure of mechanics
given by (i)-(iv) of Section 1.4 (in other words the abstract probabilistic
description of noncommutative information; see 1.6.1) we can take the
observable algebra $\mathfrak{A}$ as merely a unital $\ast$-algebra without
losing any of the ideas involved. But for more specific topics we need more
structure, for example in the quantum analogue of Liouville's Theorem
described in Section 1.7, where a finite von Neumann algebra is used as the
observable algebra. Also, in the GNS-construction, used in Section 1.2, a
C*-algebra delivers more than a mere $\ast$-algebra (see Remark 2.2.3).
However, in this chapter we will use as few assumptions as possible to build
the theory, and in Sections 2.2 to 2.5 we only need unital $\ast
$-algebras.$\blacksquare$

\section{Cyclic representations}

By a \emph{state} on a unital $\ast$-algebra $\mathfrak{A}$ we mean a linear
functional $\varphi$ on $\mathfrak{A}$ which is positive (i.e. $\varphi
(A^{\ast}A)\geq0$ for all $A\in\mathfrak{A}$) with $\varphi(1)=1$. Let $L(V)$
denote the algebra of all linear operators $V\rightarrow V$ on the vector
space $V$.

\bigskip\noindent\textbf{2.2.1 Definition.}\textsc{\ }\textit{Let }$\varphi
$\textit{\ be a state on a unital }$\ast$\textit{-algebra }$\mathfrak{A}
$\textit{. A }\textbf{cyclic representation} \textit{of }$(\mathfrak{A}%
,\varphi) $\textit{\ is a triple }$(\mathfrak{G},\pi,\Omega)$\textit{, where
}$\mathfrak{G} $\textit{\ is an inner product space, }$\pi:\mathfrak{A}%
\rightarrow L(\mathfrak{G})$\textit{\ is linear with }$\pi(1)=1$\textit{,
}$\pi(AB)=\pi(A)\pi(B)$\textit{, }$\Omega\in\mathfrak{G}$\textit{, }%
$\pi(\mathfrak{A})\Omega=\mathfrak{G}$\textit{, and }$\left\langle
\pi(A)\Omega,\pi(B)\Omega\right\rangle =\varphi(A^{\ast}B)$\textit{, for all
}$A,B\in\mathfrak{A}$\textit{.}

\bigskip A cyclic representation as in Definition 2.2.1 exists by the
GNS-construction (given below), but we will not actually need the property
$\pi(AB)=\pi(A)\pi(B)$ in this chapter. The term ``cyclic'' refers to the fact
that $\pi(\mathfrak{A})\Omega=\mathfrak{G}$. Note that
\begin{equation}
\iota:\mathfrak{A}\rightarrow\mathfrak{G}:A\mapsto\pi(A)\Omega\tag{2.1}%
\end{equation}
is a linear surjection such that $\iota(1)=\Omega$. Also, $\left\|
\Omega\right\|  ^{2}=\varphi(1^{\ast}1)=1$. We define a seminorm $\left\|
\cdot\right\|  _{\varphi}$ on $\mathfrak{A}$ by
\[
\left\|  A\right\|  _{\varphi}=\sqrt{\varphi\left(  A^{\ast}A\right)
}=\left\|  \iota(A)\right\|
\]
for all $A\in\mathfrak{A}$.

\bigskip\noindent\textbf{2.2.2 The GNS-construction.} \textit{Let }%
$\varphi:\mathfrak{A}\rightarrow\mathbb{C}$\textit{\ be a positive linear
functional on a }$\ast$\textit{-algebra }$\mathfrak{A}$.

\textbf{(i)} \textit{Then there exists a inner product space }$\mathfrak{G}%
$\textit{, a linear surjection }$\iota:\mathfrak{A}\rightarrow\mathfrak{G}%
$\textit{,} \textit{and a linear mapping} $\pi:\mathfrak{A}\rightarrow
L(\mathfrak{G})$\textit{, such that}
\[
\left\langle \iota(A),\iota(B)\right\rangle =\varphi(A^{\ast}B)
\]%
\[
\pi(A)\iota(B)=\iota(AB)
\]
\textit{and}
\[
\pi(AB)=\pi(A)\pi(B)
\]
\textit{for all }$A,B\in\mathfrak{A}$.

\textbf{(ii)} \textit{Now assume that $\mathfrak{A}$ is unital, and set
}$\Omega=\iota(1)$\textit{. From (i) it then follows that }
\[
\pi(A)\Omega=\iota(A)
\]%
\[
\pi(1)=1
\]
\textit{\ }
\[
\pi(\mathfrak{A})\Omega=\mathfrak{G}%
\]
\textit{and }
\[
\left\langle \pi(A)\Omega,\pi(B)\Omega\right\rangle =\varphi(A^{\ast
}B)=\left\langle \Omega,\pi(A^{\ast}B)\Omega\right\rangle
\]
\textit{for all }$A,B\in\mathfrak{A}$\textit{. In particular }
\[
\varphi(A)=\left\langle \Omega,\pi(A)\Omega\right\rangle
\]
\textit{for all }$A\in\mathfrak{A}$\textit{.}

\bigskip\noindent\textit{Proof.} We have to construct $\mathfrak{G}$, $\iota$
and $\pi$. This construction is called the Gelfand-Naimark-Segal (GNS) construction.

\textbf{(i)} Consider the vector subspace $\mathfrak{I}=\{A\in\mathfrak{A}%
:\left\|  A\right\|  _{\varphi}=0\}$ of $\mathfrak{A}$. Note that
$\mathfrak{I}$ is indeed a vector space, since for $A,B\in\mathfrak{I}$ we
have
\begin{align*}
\left\|  A+B\right\|  _{\varphi}^{2}  &  =\left\|  A\right\|  _{\varphi}%
^{2}+\varphi(A^{\ast}B)+\varphi(B^{\ast}A)+\left\|  B\right\|  _{\varphi}%
^{2}\\
&  \leq\left|  \varphi(A^{\ast}B)\right|  +\left|  \varphi(B^{\ast}A)\right|
\\
&  \leq\left\|  A\right\|  _{\varphi}\left\|  B\right\|  _{\varphi}+\left\|
B\right\|  _{\varphi}\left\|  A\right\|  _{\varphi}\\
&  =0
\end{align*}
by the Cauchy-Schwarz inequality (\textbf{[BR}, Lemma 2.3.10\textbf{]}). Then
$\mathfrak{G}:=\mathfrak{A}/\mathfrak{I}$ is also a vector space, on which we
can define an inner product by
\[
\left\langle \iota(A),\iota(B)\right\rangle :=\varphi(A^{\ast}B)
\]
where $\iota:\mathfrak{A}\rightarrow\mathfrak{G}$ is defined by
\[
\iota(A):=A+\mathfrak{I}%
\]
for all $A\in\mathfrak{A}$. Note that $\iota$ is a surjection by definition,
and that it is linear. We show that this inner product is well-defined:

Say $\iota(C)=\iota(A)$ and $\iota(D)=\iota(B)$, and set $I:=C-A$ and
$J:=D-B$. Then
\[
\varphi(C^{\ast}D)=\varphi(A^{\ast}B)+\varphi(A^{\ast}J)+\varphi(I^{\ast
}B)+\varphi(I^{\ast}J)
\]
but $\left|  \varphi(A^{\ast}J)\right|  \leq\left\|  A\right\|  _{\varphi
}\left\|  J\right\|  _{\varphi}=0$ by the Cauchy-Schwarz inequality, since
$J\in\mathfrak{I}$. Similarly $\varphi(I^{\ast}B)=\varphi(I^{\ast}J)=0$, hence
$\varphi(C^{\ast}D)=\varphi(A^{\ast}B)$, proving that the inner product is well-defined.

That $\left\langle \cdot,\cdot\right\rangle $ is indeed an inner product on
$\mathfrak{G}$ follows from the definitions given, and the fact that
\[
\overline{\left\langle \iota(A),\iota(B)\right\rangle }=\overline
{\varphi(A^{\ast}B)}=\varphi(B^{\ast}A)=\left\langle \iota(B),\iota
(A)\right\rangle
\]
(see \textbf{[BR}, Lemma 2.3.10\textbf{]}).

Define $\pi:\mathfrak{A}\rightarrow L(\mathfrak{G})$ by
\[
\pi(A)\iota(B)=\iota(AB)\text{.}%
\]
$\pi(A)$ is a well-defined element of $L(\mathfrak{G})$, since $\iota$ is a
linear surjection, and if $\iota(C)=\iota(B)$, then $I:=C-B\in\mathfrak{I}$,
and therefore by the Cauchy-Schwarz inequality
\[
\left\|  AI\right\|  _{\varphi}^{2}=\left|  \varphi\left(  (A^{\ast}AI)^{\ast
}I\right)  \right|  \leq\left\|  A^{\ast}AI\right\|  _{\varphi}\left\|
I\right\|  _{\varphi}=0
\]
which means that $AI\in\mathfrak{I}$, i.e. $\mathfrak{I}$ is a \emph{left
ideal} of $\mathfrak{A}$,and this in turn implies that $\iota(AC)=\iota
(AB)+\iota(AI)=\iota(AB)+\mathfrak{I}=\iota(AB)$, since $\mathfrak{I}$ is the
zero element of $\mathfrak{G}$. Since $\iota$ is linear, so is $\pi$. Also
note that for any $A,B,C\in\mathfrak{A}$,
\[
\pi(AB)\iota(C)=\iota(ABC)=\pi(A)\iota(BC)=\pi(A)\pi(B)\iota(C)
\]
so $\pi(AB)=\pi(A)\pi(B)$.

\textbf{(ii)} By (i) we have $\pi(A)\Omega=\pi(A)\iota(1)=\iota(A1)=\iota(A)$
and $\pi(1)\iota(A)=\iota(1A)=\iota(A)$ for all $A$. Since $\iota$ is
surjective, it follows that $\pi(\mathfrak{A})\Omega=\iota(\mathfrak{A}%
)=\mathfrak{G}$ and $\pi(1)=1$. Furthermore,
\begin{align*}
\left\langle \pi(A)\Omega,\pi(B)\Omega\right\rangle  &  =\left\langle
\iota(A),\iota(B)\right\rangle =\varphi(A^{\ast}B)\\
&  =\varphi\left(  1^{\ast}(A^{\ast}B)\right) \\
&  =\left\langle \iota(1),\iota(A^{\ast}B)\right\rangle \\
&  =\left\langle \Omega,\pi(A^{\ast}B)\Omega\right\rangle .
\end{align*}

In particular, setting $A=1$, we have $\varphi(B)=\left\langle \Omega
,\pi(B)\Omega\right\rangle $.$\blacksquare$

\bigskip\noindent\textbf{2.2.3 Remark. }If $\mathfrak{A}$ in 2.2.2 is a
C*-algebra, then we can replace $L(\mathfrak{G})$ by $\mathfrak{L(G)}$, and
using this boundedness, each $\pi(A)$ can be uniquely extended to an element
of $\mathfrak{L(H)}$, where $\mathfrak{H}$ is the completion of $\mathfrak{G}%
$. This is what was used in Section 1.2. See \textbf{[BR}, Section
2.3.3\textbf{]} for details.$\blacksquare$

\section{$*$-dynamical systems and ergodicity}

Motivated by our remarks in Section 2.1, we give the following definition:

\bigskip\noindent\textbf{2.3.1 Definition.}\textsc{\ }\textit{Let }$\varphi
$\textit{\ be a state on a unital }$\ast$\textit{-algebra }$\mathfrak{A}
$\textit{. Consider any linear function }$\tau:\mathfrak{A}\rightarrow
\mathfrak{A}$\textit{\ such that }
\[
\tau(1)=1
\]
\textit{and }
\[
\varphi\left(  \tau(A)^{\ast}\tau(A)\right)  \leq\varphi(A^{\ast}A)
\]
\textit{for all }$A\in\mathfrak{A}$\textit{. Then we call }$(\mathfrak{A}%
,\varphi,\tau)$\textit{\ a }$\ast$\textbf{-dynamical system}.

\bigskip\noindent Note that for $\tau$ as in Definition 2.3.1 and $\iota$
given by equation (2.1),
\begin{equation}
U_{0}:\mathfrak{G}\rightarrow\mathfrak{G}:\iota(A)\mapsto\iota(\tau(A))
\tag{3.1}%
\end{equation}
is a well-defined linear operator with $\left\|  U_{0}\right\|  \leq1$, since
$\left\|  \iota(\tau(A))\right\|  ^{2}=\varphi(\tau(A)^{\ast}\tau
(A))\leq\varphi(A^{\ast}A)=\left\|  \iota(A)\right\|  ^{2}$.

We now want to define the concept of ergodicity for a $\ast$-dynamical system.

\bigskip\noindent\textbf{2.3.2 Definition.}\textsc{\ }\textit{A }$\ast
$\textit{-dynamical system }$(\mathfrak{A},\varphi,\tau)$\textit{\ is called}
\textbf{ergodic}\textit{\ if it has the following property: For any sequence
}$(A_{n})$\textit{\ in }$\mathfrak{A}$\textit{\ such that }$\left\|
\tau(A_{n})-A_{n}\right\|  _{\varphi}\rightarrow0$\textit{\ and such that for
any }$\varepsilon>0$\textit{\ there exists an }$N\in\mathbb{N}$\textit{\ for
which }$\left\|  A_{m}-A_{n}\right\|  _{\varphi}\leq\varepsilon$\textit{\ if
}$m>N$\textit{\ and }$n>N$\textit{, it follows that }$\left\|  A_{n}%
-\alpha\right\|  _{\varphi}\rightarrow0$\textit{\ for some }$\alpha
\in\mathbb{C}$\textit{.}

\bigskip In Section 2.5 we will give a simple example of an ergodic $\ast
$-dynamical system whose $\ast$-algebra is noncommutative. Recall that for any
vectors $x$ and $y$ in a Hilbert space $\mathfrak{H}$, we denote by $x\otimes
y$ the bounded linear operator $\mathfrak{H\rightarrow H}$ defined by
$(x\otimes y)z=x\left\langle y,z\right\rangle $. The motivation for Definition
2.3.2 is the following proposition:

\bigskip\noindent\textbf{2.3.3 Proposition.}\textsc{\ }\emph{Consider a }%
$\ast$\emph{-dynamical system }$(\mathfrak{A},\varphi,\tau)$\emph{\ and let
}$U_{0}$\emph{\ be given by (3.1) in terms of any cyclic representation of
}$(\mathfrak{A},\varphi)$\emph{. Let }$U:\mathfrak{H}\rightarrow\mathfrak{H}%
$\emph{\ be the bounded linear extension of }$U_{0}$\emph{\ to the completion
}$\mathfrak{H}$\emph{\ of }$\mathfrak{G}$\emph{, and let }$P$\emph{\ be the
projection of }$\mathfrak{H}$\emph{\ onto the subspace of fixed points of }%
$U$\emph{. Then }$(\mathfrak{A},\varphi,\tau)$ \emph{is ergodic if and only if
}$P=\Omega\otimes\Omega$\emph{, or equivalently, if and only if the fixed
points of }$U$\emph{\ form a one-dimensional subspace of }$\mathfrak{H}%
$\emph{.\bigskip}

\noindent\emph{Proof. }Since $\left\|  \Omega\right\|  ^{2}=\varphi(1^{\ast
}1)=1$, we know that $\Omega\otimes\Omega$ is the projection of $\mathfrak{H}$
onto the one-dimensional subspace $\mathbb{C}\Omega$. Also note that
$U\Omega=\Omega$, since $\Omega=\iota(1)$, hence $\mathbb{C}\Omega\subset
P\mathfrak{H}$.

Suppose $(\mathfrak{A},\varphi,\tau)$ is ergodic and let $x$ be a fixed point
of $U$. Consider any sequence $(x_{n})$ in $\mathfrak{G}$ such that
$x_{n}\rightarrow x$, say $x_{n}=\iota(A_{n})$. Then $\left\|  \tau
(A_{n})-A_{n}\right\|  _{\varphi}=\left\|  Ux_{n}-x_{n}\right\|  \rightarrow
0$, since $U$ is continuous, while for any $\varepsilon>0$ there exists some
$N$ for which $\left\|  A_{m}-A_{n}\right\|  _{\varphi}=\left\|  x_{m}%
-x_{n}\right\|  <\varepsilon$ if $m>N$ and $n>N$. Since $(\mathfrak{A}%
,\varphi,\tau)$ is ergodic, it follows that $\left\|  x_{n}-\iota
(\alpha)\right\|  =\left\|  A_{n}-\alpha\right\|  _{\varphi}\rightarrow0$ for
some $\alpha\in\mathbb{C}$, but then $x=\iota(\alpha)=\alpha\Omega$. Therefore
$P\mathfrak{H}=\mathbb{C}\Omega$ which means that $P=\Omega\otimes\Omega$.

Conversely, suppose $P=\Omega\otimes\Omega$ and consider any sequence
$(A_{n})$ in $\mathfrak{A}$ such that $\left\|  \tau(A_{n})-A_{n}\right\|
_{\varphi}\rightarrow0$ and such that for any $\varepsilon>0$ there exists
some $N$ for which $\left\|  A_{m}-A_{n}\right\|  _{\varphi}<\varepsilon$ if
$m>N$ and $n>N$. Then $x_{n}=\iota(A_{n})$ is a Cauchy sequence and hence
convergent in $\mathfrak{H}$, since $\left\|  x_{m}-x_{n}\right\|  =\left\|
A_{m}-A_{n}\right\|  _{\varphi}$. Say $x_{n}\rightarrow x$, then
$Ux_{n}\rightarrow Ux$ since $U$ is continuous. Since $\left\|  Ux_{n}%
-x_{n}\right\|  =\left\|  \tau(A_{n})-A_{n}\right\|  _{\varphi}\rightarrow0$,
it follows that $Ux_{n}\rightarrow x$, hence $Ux=x $. This means that $x\in
P\mathfrak{H}$ which implies that $x=\alpha\Omega$ for some $\alpha
\in\mathbb{C}$. Therefore $\left\|  A_{n}-\alpha\right\|  _{\varphi}=\left\|
x_{n}-\alpha\Omega\right\|  \rightarrow0$, and so we conclude that
$(\mathfrak{A},\varphi,\tau)$ is ergodic.$\blacksquare$

\bigskip Proposition 2.3.3 tells us that Definition 2.3.2 includes the measure
theoretic definition as a special case. This can be seen as follows: From a
measure theoretic dynamical system $(X,\Sigma,\mu,T)$ we obtain the $\ast
$-dynamical system $(B_{\infty}(\Sigma),\varphi,\tau)$, where $\varphi(f)=\int
fd\mu$ and $\tau(f)=f\circ T$ for all $f\in B_{\infty}(\Sigma)$. A cyclic
representation of $(B_{\infty}(\Sigma),\varphi,\tau)$ is $(\mathfrak{G}%
,\pi,\Omega)$ with $\mathfrak{G}=\left\{  [g]:g\in B_{\infty}(\Sigma)\right\}
$, $\pi(f)[g]=[fg]$ for all $f,g\in B_{\infty}(\Sigma)$, and $\Omega=[1]$,
where $[g]$ denotes the equivalence class of all measurable complex-valued
functions on the measure space that are almost everywhere equal to $g$. Note
that $\iota$ defined by equation (2.1), now becomes $\iota(f)=[f]$. The
completion of $\mathfrak{G}$ is $L^{2}(\mu)$ by the following :

\bigskip\noindent\textbf{2.3.4 Proposition.} \textit{Let }$\mu$\textit{\ be a
measure on a }$\sigma$\textit{-algebra }$\Sigma$\textit{\ of subsets of a set
}$X$\textit{. Then }$\mathfrak{G}:=\left\{  [g]:g\in B_{\infty}(\Sigma
)\right\}  $\textit{\ is dense in }$L^{2}(\mu)$\textit{.}

\bigskip\noindent\textit{Proof.} For any $\Sigma$-measurable $g:X\rightarrow
\mathbb{C}$ with $g\geq0$, we know that a sequence of simple $\Sigma
$-measurable functions $s_{n}$ exist such that $0\leq s_{1}\leq s_{2}%
\leq...\leq g$ and $s_{n}(x)\rightarrow g(x)$ for all $x\in X$ (see
\textbf{[Rud}, Theorem 1.17\textbf{]}). So $\left|  s_{n}(x)-g(x)\right|
^{2}\rightarrow0$ for all $x\in X$, while of course $s_{n}\in B_{\infty
}(\Sigma)$, and so $[s_{n}]\in\mathfrak{G}$, for all $n$. Clearly $\left|
s_{n}-g\right|  ^{2}\leq\left|  g\right|  ^{2}$, so if we assume that $[g]\in
L^{2}(\mu)$, then $\left|  g\right|  ^{2}\in L^{1}(\mu)$, and we conclude by
Lebesgue's Dominated Convergence Theorem \textbf{[Rud}, 1.34\textbf{]} that
\[
\left\|  \lbrack s_{n}]-[g]\right\|  _{2}=\int\left|  s_{n}-g\right|  ^{2}%
d\mu\rightarrow\int0d\mu=0
\]
which means that $[g]$ is contained in the closure of $\mathfrak{G}$ in
$L^{2}(\mu)$. For an arbitrary $[g]\in L^{2}(\mu)$, we have the standard
representation $g=u^{+}-u^{-}+iv^{+}-iv^{-}$ where $u^{+},u^{-},v^{+}%
,v^{-}\geq0$ are $\Sigma$-measurable (\textbf{[Rud}, 1.9(b) and
1.14(b)\textbf{]}). Note that $[u^{+}],[u^{-}],[v^{+}],[v^{-}]\in L^{2}(\mu)$,
for example $\left|  u^{+}\right|  =u^{+}\leq u^{+}+u^{-}=\left|  u\right|
\leq\left|  g\right|  $ where $u=u^{+}-u^{-}$. Since $[u^{+}],[u^{-}%
],[v^{+}],[v^{-}]$ are then contained in $\mathfrak{G}$'s closure, so is
$[g]=[u^{+}]-[u^{-}]+i[v^{+}]-i[v^{-}]$.$\blacksquare$

\bigskip The operator $U$ in Proposition 2.3.3 is now given by $U[f]=[f\circ
T]$ or, dropping the $[\cdot]$ notation as is standard for $L^{2}$-spaces,
\[
Uf=f\circ T
\]
for all $f\in L^{2}(\mu)$, where $f$ and $f\circ T$ now denote equivalence
classes of functions. Proposition 2.3.3 tells us that $(B_{\infty}%
(\Sigma),\varphi,\tau)$ is ergodic if and only if the fixed points of $U$ form
a one dimensional subspace of $L^{2}(\mu)$, in other words if and only if
$(X,\Sigma,\mu,T)$ is ergodic, as was mentioned in Section 2.1.

Finally we remark that we use Definition 2.3.2 as the definition of
ergodicity, since it is formulated purely in terms of the objects
$\mathfrak{A}$, $\varphi$ and $\tau$ appearing in the $\ast$-dynamical system
$(\mathfrak{A},\varphi,\tau)$, unlike Proposition 2.3.3 which involves a
cyclic representation of these objects. However, as a characterization of
ergodicity, Proposition 2.3.3 is generally easier to use. Of course, one might
wonder if Definition 2.3.2 could not be simplified by using a single element
rather than a sequence. With $U$ as in Proposition 2.3.3, and $x=\iota(A)$ for
some $A\in\mathfrak{A}$, we have $Ux=x$ if and only if $\left\|  Ux-x\right\|
=0$, which is equivalent to $\left\|  \tau(A)-A\right\|  _{\varphi}=0$. For
ergodicity we need this to imply that $x=\alpha\Omega$ for some $\alpha
\in\mathbb{C}$, which is equivalent to $\left\|  A-\alpha\right\|  _{\varphi
}=\left\|  x-\alpha\Omega\right\|  =0$. However, we cannot define ergodicity
as ``$\left\|  \tau(A)-A\right\|  _{\varphi}=0$ implies that $\left\|
A-\alpha\right\|  _{\varphi}=0$ for some $\alpha\in\mathbb{C}$'', since
Proposition 2.3.3 would no longer hold: There would be examples of ergodic
$\ast$-dynamical systems for which the fixed points of $U$ do not form a
one-dimensional subspace of $\mathfrak{H}$. (In Appendix A.1 we give such an
example.) Our theory would then fall apart, since much of our later work is
based on the fact that for ergodic systems the fixed point space of $U$ is
one-dimensional. For example, the characterization of ergodicity in terms of
the equality of means of the sort mentioned in Section 2.1 (but extended to
$\ast$-dynamical systems), implies this one-dimensionality. Also, this
one-dimensionality is used in our proof of the variation of Khintchine's
Theorem mentioned in Section 2.1. (See Sections 2.4 and 2.5 for details.) The
use of a sequence rather than a single element is therefore necessary in
Definition 2.3.2.

\section{Some ergodic theory in Hilbert spaces}

Our main tool in this section is the following:

\bigskip\noindent\textbf{2.4.1 The Mean Ergodic Theorem.}\textsc{\ }%
\emph{Consider a linear operator }$U:\mathfrak{H}\rightarrow\mathfrak{H}%
$\emph{\ with }$\left\|  U\right\|  \leq1$\emph{\ on a Hilbert space
}$\mathfrak{H}$\emph{\ . Let }$P$\emph{\ be the projection of }$\mathfrak{H}%
$\emph{\ onto the subspace of fixed points of }$U$\emph{. For any }%
$x\in\mathfrak{H}$\emph{\ we then have }
\[
\frac{1}{n}\sum_{k=0}^{n-1}U^{k}x\rightarrow Px
\]
\emph{as }$n\rightarrow\infty$\emph{.}

\bigskip Refer to \textbf{[Pete]} for a proof. We now state and prove a
generalized Hilbert space version of Khintchine's Theorem:

\bigskip\noindent\noindent\textbf{2.4.2 Theorem.}\textsc{\ }\emph{Let
}$\mathfrak{H}$\emph{, }$U$\emph{\ and }$P$\emph{\ be as in the Mean Ergodic
Theorem above. Consider any }$x,y\in\mathfrak{H}$\emph{\ and }$\varepsilon
>0$\emph{. Then the set }
\[
E=\left\{  k\in\mathbb{N}:\left|  \left\langle x,U^{k}y\right\rangle \right|
>\left|  \left\langle x,Py\right\rangle \right|  -\varepsilon\right\}
\]
\emph{is relatively dense in }$\mathbb{N}$\emph{.}

\bigskip\noindent\emph{Proof. }The proof is essentially the same as that of
Khintchine's Theorem. By the Mean Ergodic Theorem there exists an
$n\in\mathbb{N}$ such that
\[
\left\|  \frac{1}{n}\sum_{k=0}^{n-1}U^{k}y-Py\right\|  <\frac{\varepsilon
}{\left\|  x\right\|  +1}\text{.}%
\]
Since $UPy=Py$ and $\left\|  U\right\|  \leq1$, it follows for any
$j\in\mathbb{N}$ that
\[
\left\|  \frac{1}{n}\sum_{k=j}^{j+n-1}U^{k}y-Py\right\|  \leq\left\|
\frac{1}{n}\sum_{k=0}^{n-1}U^{k}y-Py\right\|  <\frac{\varepsilon}{\left\|
x\right\|  +1}%
\]
and therefore
\[
\left|  \left\langle x,\frac{1}{n}\sum_{k=j}^{j+n-1}U^{k}y-Py\right\rangle
\right|  \leq\left\|  x\right\|  \left\|  \frac{1}{n}\sum_{k=j}^{j+n-1}%
U^{k}y-Py\right\|  <\varepsilon\text{.}%
\]
Hence
\[
\left|  \left\langle x,Py\right\rangle \right|  -\varepsilon<\left|
\frac{1}{n}\sum_{k=j}^{j+n-1}\left\langle x,U^{k}y\right\rangle \right|
\leq\frac{1}{n}\sum_{k=j}^{j+n-1}\left|  \left\langle x,U^{k}y\right\rangle
\right|
\]
and so $\left|  \left\langle x,U^{k}y\right\rangle \right|  >\left|
\left\langle x,Py\right\rangle \right|  -\varepsilon$ for some $k\in
\{j,j+1,...,j+n-1\}$, in other words $E$ is relatively dense in $\mathbb{N}%
$.$\blacksquare$

\bigskip Khintchine's Theorem corresponds to the case where $y=x$ (see Theorem
2.5.1). The following two propositions are the Hilbert space building blocks
for two characterizations of ergodicity to be considered in the next section.

\bigskip\noindent\textbf{2.4.3 Proposition.}\textsc{\ }\emph{Let
}$\mathfrak{H} $\emph{, }$U$\emph{\ and }$P$\emph{\ be as in the Mean Ergodic
Theorem above. Consider an }$\Omega\in\mathfrak{H}$\emph{\ and let
}$\mathfrak{T}$\emph{\ be any total set in }$\mathfrak{H}$\emph{. Then the
following hold:}

\textbf{(i)}\emph{\ If }$P=\Omega\otimes\Omega$\emph{, then}
\begin{equation}
\left\|  \frac{1}{n}\sum_{k=0}^{n-1}U^{k}y-\Omega\left\langle \Omega
,y\right\rangle \right\|  \rightarrow0 \tag{4.1}%
\end{equation}
\emph{as }$n\rightarrow\infty$\emph{, for every }$y\in\mathfrak{H}$\emph{.}

\textbf{(ii)}\emph{\ If (4.1) holds for every }$y\in\mathfrak{T}$\emph{, then
}$P=\Omega\otimes\Omega$\emph{.}

\bigskip\noindent\emph{Proof. }By the Mean Ergodic Theorem we know that
\begin{equation}
\left\|  \frac{1}{n}\sum_{k=0}^{n-1}U^{k}y-Py\right\|  \rightarrow0 \tag{4.2}%
\end{equation}
for every $y\in\mathfrak{H}$ as $n\rightarrow\infty$, but for $P=\Omega
\otimes\Omega$ we have $Py=\Omega\left\langle \Omega,y\right\rangle $ and this
proves (i).

To prove (ii), consider any $y\in\mathfrak{T}$. From (4.1) and (4.2) it then
follows that $Py=\Omega\left\langle \Omega,y\right\rangle =(\Omega
\otimes\Omega)y$. Since by definition the linear span of $\mathfrak{T}$ is
dense in $\mathfrak{H}$, and since $P$ and $\Omega\otimes\Omega$ are bounded
(and hence continuous) linear operators on $\mathfrak{H}$, we conclude that
$P=\Omega\otimes\Omega$.$\blacksquare$

\bigskip\noindent\textbf{2.4.4 Proposition.}\textsc{\ }\emph{Let
}$\mathfrak{H} $\emph{, }$U$\emph{\ and }$P$\emph{\ be as in the Mean Ergodic
Theorem above. Consider an }$\Omega\in\mathfrak{H}$\emph{\ and let
}$\mathfrak{S}$\emph{\ and }$\mathfrak{T}$\emph{\ be total sets in
}$\mathfrak{H}$\emph{. Then the following hold:}

\textbf{(i)}\emph{\ If }$P=\Omega\otimes\Omega$\emph{, then}
\begin{equation}
\frac{1}{n}\sum_{k=0}^{n-1}\left\langle x,U^{k}y\right\rangle \rightarrow
\left\langle x,\Omega\right\rangle \left\langle \Omega,y\right\rangle
\tag{4.3}%
\end{equation}
\emph{as }$n\rightarrow\infty$\emph{, for all }$x,y\in\mathfrak{H}$\emph{.}

\textbf{(ii)}\emph{\ If (4.3) holds for all }$x\in\mathfrak{S}$\emph{\ and
}$y\in\mathfrak{T}$\emph{, then }$P=\Omega\otimes\Omega$\emph{.}

\bigskip\noindent\emph{Proof. }Statement (i) follows immediately from
Proposition 2.4.3(i) by simply taking the inner product of $x$ with the
expression inside the norm in (4.1).

To prove (ii), consider any $x\in\mathfrak{S}$ and $y\in\mathfrak{T}$. From
the Mean Ergodic Theorem it follows that
\[
\frac{1}{n}\sum_{k=0}^{n-1}\left\langle x,U^{k}y\right\rangle \rightarrow
\left\langle x,Py\right\rangle
\]
as $n\rightarrow\infty$. Combining this with (4.3) we see that $\left\langle
x,Py\right\rangle =\left\langle x,\Omega\right\rangle \left\langle
\Omega,y\right\rangle =\left\langle x,(\Omega\otimes\Omega)y\right\rangle $.
Since the linear span of $\mathfrak{S}$ is dense in $\mathfrak{H}$, this
implies that $Py=(\Omega\otimes\Omega)y$. Hence $P=\Omega\otimes\Omega$ as in
the proof of Proposition 2.4.3(ii).$\blacksquare$

\bigskip The reason for using total sets will become clear in Sections 2.5 and 2.6.

\section{Ergodic results for $*$-dynamical systems}

In this section we carry the results of Section 2.4 over to $\ast$-dynamical
systems using cyclic representations. Firstly we give a $\ast$-dynamical
generalization of Khintchine's Theorem which follows from Theorem 2.4.2:

\bigskip\noindent\textbf{2.5.1 Theorem.}\textsc{\ }\emph{Let }$(\mathfrak{A}%
,\varphi,\tau)$\emph{\ be a }$\ast$\emph{-dynamical system, and consider any
}$A\in\mathfrak{A}$\emph{\ and }$\varepsilon>0$\emph{. Then the set }
\[
E=\left\{  k\in\mathbb{N}:\left|  \varphi\left(  A^{\ast}\tau^{k}(A)\right)
\right|  >\left|  \varphi(A)\right|  ^{2}-\varepsilon\right\}
\]
\emph{is relatively dense in }$\mathbb{N}$\emph{.}

\bigskip\noindent\emph{Proof. }Let $U$ and $P$ be defined as in Proposition
2.3.3 in terms of any cyclic representation of $(\mathfrak{A},\varphi)$. Set
$x=\iota(A)$. From equation (3.1) it is clear that $\Omega=\iota(1)$ is a
fixed point of $U$, so $\left\langle \Omega,x\right\rangle =\left\langle
P\Omega,x\right\rangle =\left\langle \Omega,Px\right\rangle $. It follows that
$\left|  \varphi(A)\right|  =\left|  \varphi(1^{\ast}A)\right|  =\left|
\left\langle \Omega,x\right\rangle \right|  \leq\left\|  \Omega\right\|
\left\|  Px\right\|  =\left\|  Px\right\|  $. We also have $\varphi(A^{\ast
}\tau^{k}(A))=\left\langle x,U^{k}x\right\rangle $. Hence by Theorem 2.4.2,
with $y=x$, the set $E$ is relatively dense in $\mathbb{N}$.$\blacksquare$

\bigskip A C*-algebraic version of Theorem 2.5.1 was previously obtained in
\textbf{[NSZ]}. Next we use Theorem 2.4.2 to prove a variant of Theorem 2.5.1:

\bigskip\noindent\textbf{2.5.2 Theorem.}\textsc{\ }\emph{Let }$(\mathfrak{A}%
,\varphi,\tau)$\emph{\ be an ergodic }$\ast$\emph{-dynamical system, and
consider any }$A,B\in\mathfrak{A}$\emph{\ and }$\varepsilon>0$\emph{. Then the
set }
\[
E=\left\{  k\in\mathbb{N}:\left|  \varphi\left(  A\tau^{k}(B)\right)  \right|
>\left|  \varphi(A)\varphi(B)\right|  -\varepsilon\right\}
\]
\emph{is relatively dense in }$\mathbb{N}$\emph{.}

\bigskip\noindent\emph{Proof. }Let $U$ and $P$ be defined as in Proposition
2.3.3 in terms of any cyclic representation of $(\mathfrak{A},\varphi)$. Set
$x=\iota(A^{\ast})$ and $y=\iota(B)$. By Proposition 2.3.3 we have
$Px=\alpha\Omega$ and $Py=\beta\Omega$ where $\overline{\alpha}=\overline
{\left\langle \Omega,x\right\rangle }=\left\langle x,\Omega\right\rangle
=\varphi(A^{\ast\ast}1)=\varphi(A)$ and similarly $\beta=\varphi(B)$.
Therefore $\left|  \left\langle x,Py\right\rangle \right|  =\left|
\left\langle Px,Py\right\rangle \right|  =\left|  \overline{\alpha}%
\beta\right|  \left\|  \Omega\right\|  ^{2}=\left|  \varphi(A)\varphi
(B)\right|  $. Furthermore, $\varphi(A\tau^{k}(B))=\left\langle x,U^{k}%
y\right\rangle $. Hence $E$ is relatively dense in $\mathbb{N}$ by Theorem
2.4.2.$\blacksquare$

\bigskip We are now going to prove two characterizations of ergodicity using
Propositions 2.4.3 and 2.4.4 respectively. But first we need to consider a
notion of totality of a set in a unital $\ast$-algebra. (Remember that an
abstract unital $\ast$-algebra has no norm.)

\bigskip\noindent\textbf{2.5.3 Definition.}\textsc{\ }\textit{Let }$\varphi
$\textit{\ be a state on a unital }$\ast$\textit{-algebra }$\mathfrak{A} $.
\textit{A subset }$\mathfrak{T}$ \textit{of }$\mathfrak{A}$\textit{\ is called
}$\varphi$\textbf{-dense}\textit{\ in }$\mathfrak{A}$\textit{\ if it is dense
in the seminormed space }$(\mathfrak{A},\left\|  \cdot\right\|  _{\varphi}%
)$\textit{. A subset }$\mathfrak{T}$\textit{\ of }$\mathfrak{A}$\textit{\ is
called }$\varphi$\textbf{-total}\textit{\ in }$\mathfrak{A}$\textit{\ if the
linear span of }$\mathfrak{T}$\textit{\ is }$\varphi$\textit{-dense in
}$\mathfrak{A}$.

\bigskip Trivially, a unital $*$-algebra is $\varphi$-total in itself for any
state $\varphi$.

\bigskip\noindent\textbf{2.5.4 Lemma.}\textsc{\ }\emph{Let }$\varphi
$\emph{\ be a state on a unital }$\ast$\emph{-algebra }$\mathfrak{A}$\emph{,
and consider any subset }$\mathfrak{T}$\emph{\ of }$\mathfrak{A}$\emph{. Let
}$\iota$ \emph{be given by (2.1) in terms of any cyclic representation of
}$(\mathfrak{A},\varphi)$\emph{, and let }$\mathfrak{H}$\emph{\ be the
completion of }$\mathfrak{G} $\emph{. Then }$\mathfrak{T}$ \emph{is }$\varphi
$\emph{-total in }$\mathfrak{A}$\emph{\ if and only if} $\iota(\mathfrak{T}%
)$\emph{\ is total in }$\mathfrak{H}$\emph{.}

\bigskip\noindent\emph{Proof. }Suppose $\mathfrak{T}$ is $\varphi$-total in
$\mathfrak{A}$, that is to say the linear span $\mathfrak{B}$ of
$\mathfrak{T}$ is $\varphi$-dense in $\mathfrak{A}$. Then $\iota
(\mathfrak{B})$ is dense in $\mathfrak{G}=\iota(\mathfrak{A})$, since for any
$A\in\mathfrak{A}$ there exists a sequence $(A_{n})$ in $\mathfrak{B}$ such
that $\left\|  \iota(A_{n})-\iota(A)\right\|  =\left\|  A_{n}-A\right\|
_{\varphi}\rightarrow0$. But by definition $\mathfrak{G} $ is dense in
$\mathfrak{H}$, hence $\iota(\mathfrak{B})$ is dense in $\mathfrak{H}$. Since
$\iota$ is linear, this means that $\iota(\mathfrak{T})$ is total in
$\mathfrak{H}$.

Conversely, suppose $\iota(\mathfrak{T})$ is total in $\mathfrak{H}$, then
$\iota(\mathfrak{B})$ is dense in $\mathfrak{H}$. It follows that
$\mathfrak{B}$ is $\varphi$-dense in $\mathfrak{A}$, since for any
$A\in\mathfrak{A}$ there exists a sequence $(A_{n})$ in $\mathfrak{B}$ such
that $\left\|  A_{n}-A\right\|  _{\varphi}=\left\|  \iota(A_{n})-\iota
(A)\right\|  \rightarrow0$. In other words, $\mathfrak{T}$ is $\varphi$-total
in $\mathfrak{A}$.$\blacksquare$

\bigskip\noindent\textbf{2.5.5 Proposition.}\textsc{\ }\emph{Let
}$(\mathfrak{A},\varphi,\tau)$\emph{\ be a }$\ast$\emph{-dynamical system, and
consider any }$\varphi$\emph{-total set }$\mathfrak{T}$\emph{\ in
}$\mathfrak{A}$\emph{. Then the following hold:}

\textbf{(i)}\emph{\ If }$(\mathfrak{A},\varphi,\tau)$\emph{\ is ergodic,
then}
\begin{equation}
\left\|  \frac{1}{n}\sum_{k=0}^{n-1}\tau^{k}(A)-\varphi(A)\right\|  _{\varphi
}\rightarrow0 \tag{5.1}%
\end{equation}
\emph{as }$n\rightarrow\infty$\emph{, for every }$A\in\mathfrak{A}$\emph{.}

\textbf{(ii)}\emph{\ If (5.1) holds for every }$A\in\mathfrak{T}$\emph{, then
}$(\mathfrak{A},\varphi,\tau)$\emph{\ is ergodic.}

\bigskip\noindent\emph{Proof. }Let $U$ and $P$ be defined as in Proposition
2.3.3 in terms of any cyclic representation of $(\mathfrak{A},\varphi)$.
Suppose $(\mathfrak{A},\varphi,\tau)$ is ergodic. For any $A\in\mathfrak{A}$
we then have
\begin{equation}
\left\|  \frac{1}{n}\sum_{k=0}^{n-1}\tau^{k}(A)-\varphi(A)\right\|  _{\varphi
}=\left\|  \frac{1}{n}\sum_{k=0}^{n-1}U^{k}\iota(A)-\iota\left(
\varphi(A)\right)  \right\|  \rightarrow0 \tag{5.2}%
\end{equation}
as $n\rightarrow\infty$, by Proposition 2.4.3(i) and Proposition 2.3.3, since
$\iota\left(  \varphi(A)\right)  =\iota(1)\varphi(A)=\Omega\varphi(1^{\ast
}A)=\Omega\left\langle \Omega,\iota(A)\right\rangle $. This proves (i).

Now suppose (5.1), and therefore (5.2), hold for every $A\in\mathfrak{T}$.
Since $\iota(\mathfrak{T})$ is total in $\mathfrak{H}$ according to Lemma
2.5.4, it follows from Proposition 2.4.3(ii) and the identity $\iota\left(
\varphi(A)\right)  =\Omega\left\langle \Omega,\iota(A)\right\rangle $, that
$P=\Omega\otimes\Omega$. So $(\mathfrak{A},\varphi,\tau)$ is ergodic by
Proposition 2.3.3, confirming (ii).$\blacksquare$

\bigskip In the spirit of the original motivation behind the concept of
ergodicity, this proposition characterizes ergodic $\ast$-dynamical systems as
those for which the \emph{time mean} of each element $A$ of the $\ast$-algebra
converges in the seminorm $\left\|  \cdot\right\|  _{\varphi}$ to the ``phase
space'' mean $\varphi(A)$. A better name for the latter would be the
\emph{system mean} in this case, since there is no phase space involved. For a
measure theoretic dynamical system $(X,\Sigma,\tau,\mu)$, the state $\varphi$
is given by $\varphi(f)=\int fd\mu$ which is indeed the phase space mean of
$f\in B_{\infty}(\Sigma)$, where $X$ is the phase space. We will come back to
this in Section 2.6.

For any subset $\mathfrak{S}$ of a $*$-algebra, we write $\mathfrak{S}%
^{*}=\{A^{*}:A\in\mathfrak{S}\}$.

\bigskip\noindent\textbf{2.5.6 Proposition.}\textsc{\ }\emph{Let
}$(\mathfrak{A},\varphi,\tau)$\emph{\ be a }$\ast$\emph{-dynamical system, and
consider any }$\varphi$\emph{-total sets $\mathfrak{S}$ and }$\mathfrak{T}%
$\emph{\ in }$\mathfrak{A}$\emph{. Then the following hold:}

\textbf{(i)}\emph{\ If }$(\mathfrak{A},\varphi,\tau)$\emph{\ is ergodic,
then}
\begin{equation}
\frac{1}{n}\sum_{k=0}^{n-1}\varphi\left(  A\tau^{k}(B)\right)  \rightarrow
\varphi(A)\varphi(B) \tag{5.3}%
\end{equation}
\emph{as }$n\rightarrow\infty$\emph{, for all }$A,B\in\mathfrak{A}$\emph{.}

\textbf{(ii)}\emph{\ If (5.3) holds for all }$A\in\mathfrak{S}^{\ast}%
$\emph{\ and }$B\in\mathfrak{T}$\emph{, then }$(\mathfrak{A},\varphi,\tau
)$\emph{\ is ergodic.}

\bigskip\noindent\emph{Proof. }Let $U$ and $P$ be defined as in Proposition
2.3.3 in terms of any cyclic representation of $(\mathfrak{A},\varphi)$.
Suppose $(\mathfrak{A},\varphi,\tau)$ is ergodic. Then $P=\Omega\otimes\Omega$
by Proposition 2.3.3, and so by Proposition 2.4.4(i) it follows that
\begin{equation}
\frac{1}{n}\sum_{k=0}^{n-1}\varphi\left(  A\tau^{k}(B)\right)  =\frac{1}%
{n}\sum_{k=0}^{n-1}\left\langle \iota(A^{\ast}),U^{k}\iota(B)\right\rangle
\rightarrow\varphi(A)\varphi(B) \tag{5.4}%
\end{equation}
as $n\rightarrow\infty$, since $\left\langle \iota(A^{\ast}),\Omega
\right\rangle =\varphi(A)$ and $\left\langle \Omega,\iota(B)\right\rangle
=\varphi(B)$, as in the proof of Theorem 2.5.2. This proves (i).
(Alternatively, (i) can be derived from Proposition 2.5.5(i) using the
Cauchy-Schwarz inequality $\left|  \varphi(AC)\right|  \leq\left\|  A^{\ast
}\right\|  _{\varphi}\left\|  C\right\|  _{\varphi}$ with $C=\frac{1}{n}%
\sum_{k=0}^{n-1}\tau^{k}(B)-\varphi(B)$. This is essentially how Proposition
2.4.4(i) was derived from Proposition 2.4.3(i).)

Now suppose (5.3), and therefore (5.4), hold for all $A\in\mathfrak{S}^{\ast}$
and $B\in\mathfrak{T}$. Since $\iota(\mathfrak{S})$ and $\iota(\mathfrak{T})$
are total in $\mathfrak{H}$ according to Lemma 2.5.4, it follows from
Proposition 2.4.4(ii) and the identities $\left\langle \iota(A^{\ast}%
),\Omega\right\rangle =\varphi(A)$ and $\left\langle \Omega,\iota
(B)\right\rangle =\varphi(B)$, that $P=\Omega\otimes\Omega$. So $(\mathfrak{A}%
,\varphi,\tau) $ is ergodic by Proposition 2.3.3, confirming
(ii).$\blacksquare$

\bigskip This characterizes ergodicity in terms of \emph{mixing}. We now give
a simple example of an ergodic $\ast$-dynamical system whose $\ast$-algebra is noncommutative:

\bigskip\noindent\textbf{2.5.7 Example.}\textsc{\ }Let $\mathfrak{A}$ be the
unital $\ast$-algebra of $2\times2$-matrices with entries in $\mathbb{C}$, the
involution being the conjugate transpose. Let $\varphi$ be the normalized
trace on $\mathfrak{A}$, that is to say $\varphi=\frac{1}{2}$Tr. Define
$\tau:\mathfrak{A}\rightarrow\mathfrak{A}$ by
\[
\tau\left(
\begin{array}
[c]{ll}%
a_{11} & a_{12}\\
a_{21} & a_{22}%
\end{array}
\right)  =\left(
\begin{array}
[c]{cc}%
a_{22} & c_{1}a_{12}\\
c_{2}a_{21} & a_{11}%
\end{array}
\right)
\]
for some fixed $c_{1},c_{2}\in\mathbb{C}$ with $\left|  c_{1}\right|  \leq1$,
$\left|  c_{2}\right|  \leq1$, $c_{1}\neq1$ and $c_{2}\neq1$. The conditions
$\left|  c_{1}\right|  \leq1$ and $\left|  c_{2}\right|  \leq1$ are necessary
and sufficient for $(\mathfrak{A},\varphi,\tau)$ to be a $\ast$-dynamical
system. Note that for any $c\in\mathbb{C}$ with $\left|  c\right|  \leq1$, it
follows from the Mean Ergodic Theorem 2.4.1 that
\[
\frac{1}{n}\sum_{k=0}^{n-1}c^{k}%
\]
converges to $0$ if $c\neq1$, and to $1$ otherwise. Using this fact and
Proposition 2.5.6(ii) with $\mathfrak{S}=\mathfrak{T}=\mathfrak{A}$ (and some
calculations), it can be verified that the conditions $c_{1}\neq1$ and
$c_{2}\neq1$ are necessary and sufficient for $(\mathfrak{A},\varphi,\tau)$ to
be ergodic, assuming that $\left|  c_{1}\right|  \leq1$ and $\left|
c_{2}\right|  \leq1$. See Appendix A.2 for more details, and Appendix B for a
physically motivated example of an ergodic $\ast$-dynamical
system.$\blacksquare$

\bigskip\noindent\textbf{2.5.8 Open Problem.} As mentioned in Section 2.1, the
converse of Theorem 2.5.2 holds in the measure theoretic case. In general the
question is as follows (also see Proposition 2.5.6(ii)): Consider a $\ast
$-dynamical system $(\mathfrak{A},\varphi,\tau)$, and $\varphi$-total sets
$\mathfrak{S}$ and $\mathfrak{T}$ in $\mathfrak{A}$, such that for every
$A\in\mathfrak{S}^{\ast}$ and $B\in\mathfrak{T}$ with $\varphi(A)\neq0$ and
$\varphi(B)\neq0$, there exists a $k\in\mathbb{N}$ for which $\varphi
(A\tau^{k}(B))\neq0$. Is $(\mathfrak{A},\varphi,\tau)$ necessarily
ergodic?$\blacksquare$

\section{Measure theory and von Neumann algebras}

As was mentioned in Section 2.3, from a measure theoretic dynamical system
$(X,\Sigma,\mu,T)$ we obtain the $\ast$-dynamical system $(B_{\infty}%
(\Sigma),\varphi,\tau)$, where $\varphi(f)=\int fd\mu$ and $\tau(f)=f\circ T$.
This allows us to apply the results of Section 2.5 to measure theoretic
dynamical systems. For example, if $(X,\Sigma,\mu,T)$ is ergodic, then we know
from Section 2.3 that $(B_{\infty}(\Sigma),\varphi,\tau)$ is ergodic. Hence
for this $\ast$-dynamical system Theorem 2.5.2 tells us that for any
$A,B\in\Sigma$ and $\varepsilon>0$, the set
\[
\left\{  k\in\mathbb{N}:\left|  \varphi\left(  \chi_{A}\tau^{k}(\chi
_{B})\right)  \right|  >\left|  \varphi(\chi_{A})\varphi(\chi_{B})\right|
-\varepsilon\right\}
\]
is relatively dense in $\mathbb{N}$, but this set is exactly the set $F$ from
Section 2.1. (Here $\chi$ denotes characteristic functions, as before.) So we
have answered our original question:

\bigskip\noindent\textbf{2.6.1 Corollary.}\textsc{\ }\emph{Let }$(X,\Sigma
,\mu,T)$\emph{\ be an ergodic measure theoretic dynamical system. Then for any
}$A,B\in\Sigma$\emph{\ and }$\varepsilon>0$\emph{, the set}
\[
F=\left\{  k\in\mathbb{N}:\mu\left(  A\cap T^{-k}(B)\right)  >\mu
(A)\mu(B)-\varepsilon\right\}
\]
\emph{is relatively dense in }$\mathbb{N}$\emph{.}

\bigskip This result says that for every $k\in F$, the set $A$ contains a set
$A\cap T^{-k}(B)$ of measure larger than $\mu(A)\mu(B)-\varepsilon$, which is
mapped into $B$ by $T^{k}$. Using a similar argument, Khintchine's Theorem
follows from Theorem 2.5.1.

Likewise, Propositions 2.5.5 and 2.5.6 can be applied to the measure theoretic
case. For example, Proposition 2.5.5(i) tells us that if $(X,\Sigma,\mu,T)$ is
ergodic, then
\begin{equation}
\int\left|  \frac{1}{n}\sum_{k=0}^{n-1}f\circ T^{k}-\varphi(f)\right|
^{2}d\mu\rightarrow0 \tag{6.1}%
\end{equation}
as $n\rightarrow\infty$, for every $f\in B_{\infty}(\Sigma)$. Note that this
result is not pointwise and is therefore not quite as strong as the usual
measure theoretic statement of equality of the time mean and the phase space
mean. This is of course where Birkhoff's Pointwise Ergodic Theorem comes into
play (see for example \textbf{[Pete]}).

What about the converse? Well, in order to effectively apply Propositions
2.5.5(ii) and 2.5.6(ii) to the measure theoretic case, we need to know what
the measure theoretic significance of a $\varphi$-total set in $B_{\infty
}(\Sigma)$ is. The basic fact we will use is the following simple proposition:

\bigskip\noindent\textbf{2.6.2 Proposition.}\textsc{\ }\emph{Let }%
$(X,\Sigma,\mu)$\emph{\ be a probability space and set }$\varphi(f)=\int
fd\mu$\emph{\ for all }$f\in B_{\infty}(\Sigma)$\emph{. Then the set
}$\mathfrak{T}=\{\chi_{S}:S\in\Sigma\}$\emph{\ is }$\varphi$\emph{-total in
}$B_{\infty}(\Sigma)$\emph{.}

\bigskip\noindent\textit{Proof.} The same argument as in the proof of
Proposition 2.3.4, keeping in mind that $\left\|  f\right\|  _{\varphi
}=\left(  \int\left|  f\right|  ^{2}d\mu\right)  =\left\|  [f]\right\|  _{2}$
for all $f\in B_{\infty}(\Sigma)$, shows that for any $g\in B_{\infty}%
(\Sigma)$ there is a sequence simple functions $s_{n}$ such that $\left\|
s_{n}-g\right\|  _{\varphi}\rightarrow0$. However, by definition a simple
function is a linear combination of elements of $\mathfrak{T}$, so we conclude
that the linear span of $\mathfrak{T}$ is $\varphi$-dense in $B_{\infty
}(\Sigma)$, which completes the proof.$\blacksquare$

\bigskip From this we see that if (6.1) holds for all measurable
characteristic functions $f$, then $(B_{\infty}(\Sigma),\varphi,\tau)$ is
ergodic by Proposition 2.5.5(ii), hence $(X,\Sigma,\mu,T)$ is ergodic as
mentioned in Section 2.3.

Finally, with reference to Proposition 2.5.6(ii), we note that $\mathfrak{T}%
^{\ast}=\mathfrak{T}$ for $\mathfrak{T}$ as in Proposition 2.6.2.

Next we briefly look at von Neumann algebras, as they are well-known examples
of unital $\ast$-algebras. Consider a von Neumann algebra $\mathfrak{M} $ and
suppose $(\mathfrak{M},\varphi,\tau)$ is a $\ast$-dynamical system. For
example, $\tau$ might be a $\ast$-homomorphism leaving $\varphi$ invariant,
that is to say, $\varphi(\tau(A))=\varphi(A)$ for all $A\in\mathfrak{M}$. Then
the results of Section 2.5 can be applied directly to $(\mathfrak{M}%
,\varphi,\tau)$. As a more explicit (and ergodic) example, we note that
$\mathfrak{A}$ in Example 2.5.7 is a von Neumann algebra on the Hilbert space
$\mathbb{C}^{2}$. We can also mention that $\tau$ in Example 4.7 is not a
homomorphism (see Appendix A.2).

We now describe one suitable choice for the $\varphi$-total sets appearing in
Propositions 2.5.5 and 2.5.6. Let $\mathfrak{P}$ be the projections of
$\mathfrak{M}$. It is known that $\mathfrak{M}$ is the norm closure of the
linear span of $\mathfrak{P}$, as is mentioned for example on p. 326 of
\textbf{[KR1]}. Since any state $\varphi$ on $\mathfrak{M}$ is continuous by
virtue of being positive (see \textbf{[BR}, Proposition 2.3.11\textbf{]}), it
follows that $\mathfrak{P}$ is $\varphi$-total in $\mathfrak{M}$. Note also,
regarding Proposition 2.5.6(ii), that $\mathfrak{P}^{\ast}=\mathfrak{P}$. This
is all very similar to the measure theoretic case in Proposition 2.6.2, since
the measurable characteristic functions on $X$ are exactly the projections of
$B_{\infty}(\Sigma)$. This similarity should not be too surprising, since the
theory of von Neumann algebras is often described as ``noncommutative measure
theory'' because of the close analogy with measure theory.

\section{An alternative approach to recurrence}

In this section (which is based on work contained in \textbf{[D2]}) we discuss
an alternative approach to recurrence which does not require $\varphi$ to be
linear or complex-valued as in Definition 2.3.1. The lack of linearity in this
approach however precludes the use of the GNS construction and Hilbert spaces,
and because of this it does not give any quantitative result as in
Khintchine's Theorem and its noncommutative generalization Theorem 2.5.1.

As we shall see, the theory is surprisingly close to the usual measure
theoretic setting. It therefore seems appropriate to briefly review a
Poincar\'{e}-like probabilistic recurrence result. Consider a measure space
$(X,\Sigma,\mu)$ with $\mu(X)<\infty$, and let $T:X\rightarrow X$ be a mapping
such that $\mu(T^{-1}$\noindent$(S))=\mu(S)$ for all $S$ in $\Sigma$. This is
merely an abstraction of Liouville's theorem. For some $S\in\Sigma$, suppose
that $\mu(S\cap T^{-n}(S))=0$ for all $n\in\mathbb{N}$. For all $n,k\in
\mathbb{N}$ we then have $\mu(T^{-k}(S)\cap T^{-(n+k)}(S))=\mu(T^{-k}(S\cap
T^{-n}(S)))=\mu(S\cap T^{-n}(S))=0$. So $\mu(T^{-m}(S)\cap T^{-n}(S))=0$ for
all $m,n\in\mathbb{N}$ with $m\neq n$. It follows that
\begin{equation}
\mu(X)\geq\mu\left(  \bigcup_{k=1}^{n}T^{-k}(S)\right)  =\sum_{k=1}^{n}%
\mu(T^{-k}(S))=\sum_{k=1}^{n}\mu(S)=n\mu(S). \tag{7.1}%
\end{equation}
Note that the weaker condition $\mu(T^{-1}$\noindent$(S))\leq\mu(S)$ appearing
in Khintchine's Theorem 2.1.1 would not be good enough to ensure this
inequality. Letting $n\rightarrow\infty$ it follows that $\mu(S)=0$. This is a
recurrence result, namely if $\mu(S)>0$, then there exists a positive integer
$n$ such that $\mu(S\cap T^{-n}(S))>0$. It tells us that $S $ contains a set
$S\cap T^{-n}(S)$ of positive measure which is mapped back into $S$ by $T^{n}%
$. From (7.1) it is clear that the intuitive idea is simply that we cannot fit
an infinite number of sets the size of $S$ into $X$ without the sets
overlapping, since $X$ is of finite size (where the size of a set is its
measure). This is similar to the pigeon hole principle.

Note that the mapping $g\mapsto\tau(g)=g\circ T$ is a $\ast$-homomorphism of
the $\ast$-algebra $B_{\infty}(\Sigma)$ into itself such that $\varphi
(\tau(g))=\varphi(g)$ by Proposition 1.7.1, and $\mu(S\cap T^{-n}%
(S))=\varphi\left(  \chi_{S}\tau^{n}(\chi_{S})\right)  $ for $S\in\Sigma$,
where $\varphi(g)=\int gd\mu$ for all $g\in B_{\infty}(\Sigma)$. Using this
notation the recurrence result above can be stated as follows: If
$\varphi(\chi_{S})>0$, then there exists a positive integer $n$ such that
$\varphi\left(  \chi_{S}\tau^{n}(\chi_{S})\right)  >0$. The general $\ast
$-algebraic approach will now be modelled after this situation. We also get
some inspiration from Postulate 1.2.1, for reasons which will become clear in
Section 3.1.

For an element $A$ of a $\ast$-algebra $\mathfrak{A}$, we write $A\geq0$ if
$A=R^{\ast}R$ for some $R\in\mathfrak{A}$. If also $A\neq0$, we write $A>0$.
By $A\leq B$ we mean that $B-A\geq0$.

\bigskip\noindent\textbf{2.7.1 Definition. }\textit{Let }$\mathfrak{A}%
$\textit{\ be a }$\ast$\textit{-algebra, and }$\mathfrak{B}$\textit{\ a unital
}$\ast$\textit{-algebra. Let }$\varphi:\mathfrak{A\rightarrow B}$ \textit{be a
positive mapping \ (i.e. }$\varphi(A^{\ast}A)\geq0$ \textit{for all}
$A\in\mathfrak{A}$\textit{). We call }$\varphi$ \textbf{additive}%
\textit{\ if}
\[
\sum_{k=1}^{n}\varphi\left(  P_{k}\right)  \leq1
\]
\textit{for any projections }$P_{1},...,P_{n}\in\mathfrak{A}$ \textit{for
which} $\varphi(P_{k}P_{l}P_{k})=0$ \textit{if }$k<l$.\textit{\ We call
}$\varphi$ \textbf{faithful}\textit{\ if it is linear, $\mathfrak{A}$ is
unital, }$\varphi(1)=1$\textit{, and }$\varphi(A^{\ast}A)>0$\textit{\ for all
non-zero }$A$\textit{\ in }$\mathfrak{A}$\textit{\ (note that this requires
that }$A^{\ast}A\neq0$\textit{\ for }$A\neq0$\textit{, which is true for
example in any C*-algebra).}

\bigskip\noindent\textbf{2.7.2 Proposition.} \textit{If the positive mapping
}$\varphi$\textit{\ given in Definition 2.7.1 is faithful, then it is also additive.}

\bigskip\noindent\textit{Proof. }Let $P_{1},...,P_{n}\in\mathfrak{A}$ be any
projections for which $\varphi(P_{k}P_{l}P_{k})=0$ if $k<l$. For $k<l$ we then
have $\varphi\left(  (P_{l}P_{k})^{\ast}P_{l}P_{k}\right)  =0$, so $P_{l}%
P_{k}=0$, and therefore $P_{k}P_{l}=(P_{l}P_{k})^{\ast}=0$. This implies that
\[
\sum_{k=1}^{n}P_{k}\leq1
\]
since the left-hand side is a projection in $\mathfrak{A}$. Thus
\[
\sum_{k=1}^{n}\varphi\left(  P_{k}\right)  =\varphi\left(  \sum_{k=1}^{n}%
P_{k}\right)  \leq\varphi(1)=1
\]
as promised.$\blacksquare$

\bigskip\noindent\textbf{2.7.3 Remark.} In the measure theoretic setting
described above, we can assume without loss of generality that $\mu(X)=1$.
Then $\varphi:B_{\infty}(\Sigma)\rightarrow\mathbb{C}$ is a linear additive
mapping, since
\[
\sum_{k=1}^{n}\varphi\left(  \chi_{S_{k}}\right)  =\sum_{k=1}^{n}\mu
(S_{k})=\mu\left(  \bigcup_{k=1}^{n}S_{k}\right)  \leq\mu(X)=1
\]
for any $S_{1},...,S_{n}\in\Sigma$ such that $\varphi\left(  \chi_{S_{k}}%
\chi_{S_{l}}\right)  =\mu\left(  S_{k}\cap S_{l}\right)  =0$ if $k\neq l$.
However, $\varphi$ need not be faithful, since there can be a non-empty set
$S$ of measure zero (giving $\varphi\left(  \chi_{S}^{\ast}\chi_{S}\right)
=0$ even though $\chi_{S}\neq0$), which is why we introduced the notion of
additivity.$\blacksquare$

\bigskip We now state and prove a $\ast$-algebraic version of the recurrence
result described above:

\bigskip\noindent\textbf{2.7.4 Theorem. }\textit{Consider a }$\ast
$\textit{-algebra }$\mathfrak{A}$ \textit{and a unital C*-algebra
}$\mathfrak{B}$,\textit{\ and let} $\varphi:\mathfrak{A\rightarrow B}$
\textit{be an additive mapping. Let }$\tau:\mathfrak{A\rightarrow A}$
\textit{be a }$\ast$\textit{-homomorphism such that} $\varphi(\tau
(PQP))=\varphi(PQP)$ \textit{for all projections }$P,Q\in\mathfrak{A}$.
\textit{Then, for any projection }$P\in\mathfrak{A}$\textit{\ such that
}$\varphi(P)>0$\textit{, there exists a positive integer }$n$\textit{\ such
that} $\varphi(P\tau^{n}(P)P)>0$\textit{.}

\bigskip\noindent\textit{Proof.} Note that $\varphi(P\tau^{n}(P)P)=\varphi
\left(  (\tau^{n}(P)P)^{\ast}\tau^{n}(P)P\right)  \geq0$ for all
$n\in\mathbb{N}$, since $\tau$ is a $\ast$-homomorphism. We now imitate the
measure theoretic proof.

Suppose $\varphi(P\tau^{n}(P)P)=0$ for all $n\in\mathbb{N}$. For all
$k,n\in\mathbb{N}$ we then have
\[
\varphi\left(  \tau^{k}(P)\tau^{n+k}(P)\tau^{k}(P)\right)  =\varphi\left(
\tau^{k}\left(  P\tau^{n}(P)P\right)  \right)  =\varphi\left(  P\tau
^{n}(P)P\right)  =0
\]
since $\tau$ is a homomorphism and $P$ and therefore $\tau^{n}(P)$ are
projections. Since $\varphi$ is additive, it follows for any $n\in\mathbb{N}$
that
\[
\sum_{k=1}^{n}\varphi\left(  \tau^{k}(P)\right)  \leq1.
\]
Furthermore,
\[
\sum_{k=1}^{n}\varphi\left(  \tau^{k}(P)\right)  =\sum_{k=1}^{n}\varphi\left(
P\right)  =n\varphi(P)\geq0
\]
since $P=PPP$, $\varphi$ is positive and $P=P^{\ast}P$. Hence $0\leq
n\varphi(P)\leq1$, and therefore $n\left\|  \varphi(P)\right\|  \leq1$ since
$\mathfrak{B}$ is a C*-algebra (see \textbf{[Mu}, Theorem 2.2.5(3)\textbf{]}).
Letting $n\rightarrow\infty$, it follows that $\varphi(P)=0$.$\blacksquare$

\bigskip It is clear that because of Remark 2.7.3, the measure theoretic
recurrence result described above is just a special case of Theorem 2.7.4,
since the projections of the $\ast$-algebra $B_{\infty}(\Sigma)$ are exactly
the characteristic functions $\chi_{S}$, where $S\in\Sigma$.

Note that the trace tr$:\mathfrak{M}\rightarrow\mathfrak{M}\cap\mathfrak{M}%
^{\prime}$of a finite von Neumann algebra is faithful in the sense of
Definition 2.7.1, hence we have the following corollary of Theorem 2.7.4 and
Proposition 2.7.2, which will be used in Section 3.1:

\bigskip\noindent\textbf{2.7.5 Corollary. }\textit{Consider a finite von
Neumann algebra }$\mathfrak{M}$, \textit{and let} tr \textit{be its trace. Let
}$\tau:\mathfrak{M\rightarrow M}$ \textit{be a }$\ast$\textit{-homomorphism
such that} tr$(\tau(A))=$ tr$(A)$ \textit{for all }$A$\textit{\ in
}$\mathfrak{M}$. \textit{Then, for any projection }$P\in\mathfrak{M}%
$\textit{\ such that }tr$(P)>0$\textit{, there exists a positive integer }%
$n$\textit{\ such that} tr$(P\tau^{n}(P))>0$\textit{.}\noindent

\bigskip We conclude this chapter with an open problem inspired by Theorem 2.7.4:

\bigskip\noindent\textbf{2.7.6 Open Problem.} Does Theorem 2.5.1 still hold if
we only assume that $\varphi$ is $\mathfrak{B}$-valued, instead of
complex-valued, where $\mathfrak{B}$ is any unital C*-algebra? In fact, we can
ask if we can obtain the whole theory in Sections 2.3 and 2.5 if in Definition
2.3.1 we generalized the framework to $\varphi$ being $\mathfrak{B}$-valued
instead of complex-valued. A possible line of attack is to use Hilbert
C*-modules (see \textbf{[La]}).$\blacksquare$

\chapter{Recurrence and ergodicity in mechanics}

In this chapter we discuss recurrence and ergodicity in certain physical
systems (quantum and classical). In Section 3.1 (which is based on
\textbf{[D2]}) it is shown that recurrence takes place in a probabilistic
sense in exactly the same way in bounded quantum systems as in classical
systems with finite volume phase space. In Section 3.2 we show under
physically reasonable assumptions that quantum and classical systems are not
ergodic in the sense of Definition 2.3.2 (or, equivalently, in terms of the
characterization in Proposition 2.5.5), if the state of the system allows more
than one energy level to be obtained in a measurement (i.e. if more than one
energy level has a nonzero probability).

\section{Recurrence}

Consider a bounded quantum system $(\mathfrak{M},\mathfrak{H},H)$ and assume
that $\mathfrak{M}$ is a factor. Let $\tau$ be the system's time-evolution, as
in Proposition 1.7.5. Fix any $t>0$. Since the trace tr of $\mathfrak{M}$ is
faithful, Corollary 2.7.5 and Proposition 1.7.5 tell us that for any nonzero
projection $P\in\mathfrak{M}$ there exists an $n(t)\in\mathbb{N}$ such that
\begin{equation}
\text{tr}\left(  P\tau_{n(t)t}(P)\right)  >0. \tag{1.1}%
\end{equation}
Note that tr$(P\tau_{n(t)t}(P))=$ tr$(P\tau_{n(t)t}(P)P)$, which has the form
of $\omega^{\prime}$ in Postulate 1.2.1, i.e. the state after a ``yes'' was
obtained in a yes/no experiment with projection $P$ when the initial state was
tr. Also remember that according to Postulate 1.8.3, tr is the state of no information.

So, to interpret (1.1), consider the case where we have no information about
the state of our bounded quantum system. By Postulate 1.8.3 the state is then
given by tr. At time $0$ we perform a yes/no experiment with projection
$P\in\mathfrak{M}$ on the system. Assuming the result is ``yes'', the state of
the system after the experiment is given by the state $\omega$ on
$\mathfrak{M}$ defined by
\[
\omega(A)=\text{ tr}(PA)/\text{tr}(P)\text{,}%
\]
according to Postulate 1.2.1. (Also recall from Section 1.2 that the
probability of getting ``yes'' is tr$(P)$, therefore tr$(P)>0$ in this case.)
By (1.1) we then have
\begin{equation}
p(t):=\omega(\tau_{n(t)t}(P))>0. \tag{1.2}%
\end{equation}
This simply tells us that if we were to repeat the above mentioned yes/no
experiment exactly at the moment $n(t)t$, when its projection is given by
$\tau_{n(t)t}(P)$ according to Section 1.4 (iv), then there is a nonzero
probability $p(t)$ that we will again get ``yes''. By replacing $t$ by
$t^{\prime}=n(t)t+1$, we see that there is in fact an unbounded set of moments
$n(t)t<n(t^{\prime})t^{\prime}<...$ for which (1.2) holds.

So we have obtained a quantum mechanical version of recurrence. Note that the
measure theoretic recurrence result described in Section 2.7 will give exactly
the same result as (1.2), with the same physical interpretation, when applied
to a classical mechanical system whose phase space (see Remark 1.7.2) has
finite Lebesgue measure; just replace $\omega$, tr, $\tau$ and $P$ by their
classical analogues described in Sections 1.3 and 1.8. In particular, tr is
replaced by integration with respect to normalized Lebesgue measure, which
then represents the state of no information. So we see that (probabilistic)
recurrence in quantum mechanics and in classical mechanics follow from the
same general result, namely Theorem 2.7.4, since Corollary 2.7.5 and the
measure theoretic recurrence result are both special cases of this theorem.

A drawback of (1.2) is that it gives no indication as to how large
$\omega(\tau_{n(t)t}(P))$ is, or how often it is positive. Theorem 2.5.1 on
the other hand, tells us that for any $\varepsilon>0$ there is in fact a
relatively dense set $M$ in $\mathbb{N}$ such that
\begin{equation}
\omega(\tau_{mt}(P))>\text{ tr}(P)-\varepsilon\tag{1.3}%
\end{equation}
for all $m\in M$, which is a quantitative improvement over (1.2), since it
says that $\omega(\tau_{mt}(P))$ is regularly (i.e. \emph{almost
periodically}) larger than tr$(P)-\varepsilon$. Since tr$(P)$ was the
probability of getting a ``yes'' during the first execution of the yes/no
experiment, we see from (1.3) that at the moments $mt$ the probability of
getting ``yes'' when doing the experiment a second time is larger or at least
arbitrarily close to the original probability of getting ``yes''. Similar
results concerning wave functions and density operators are presented in
\textbf{[HH]} and \textbf{[Perc]}. If as before we replace $\omega$, tr,
$\tau$ and $P$ by their classical counterparts, and then apply Theorem 2.5.1
again, we find the same result as (1.3) for classical mechanics, with exactly
the same interpretation as in quantum mechanics.

There is, however, a small technical problem: The probability of repeating the
yes/no experiment exactly at the moment $n(t)t$ is zero. The same goes for any
of the moments $mt$ above. The next simple proposition remedies the situation
in the quantum case:

\bigskip\noindent\textbf{3.1.1 Proposition. }\textit{Let }$\tau$\textit{\ be
as in Proposition 1.7.5, where we take }$\mathfrak{M}$\textit{\ to be a finite
factor. Then for any projection }$P$\textit{\ in }$\mathfrak{M}$\textit{, the
mapping}
\[
\mathbb{R\rightarrow R}:t\mapsto\text{ tr}(P\tau_{t}(P))
\]
\textit{is continuous, where }tr \textit{is the trace of $\mathfrak{M}$.}

\bigskip\noindent\textit{Proof. }By Stone's Theorem $U_{t}$ in Proposition
1.7.5 is strongly continuous (i.e., $t\mapsto U_{t}x$ is continuous for every
$x\in\mathfrak{H}$), so clearly the mapping $t\mapsto\tau_{t}(A)$ is weakly
continuous for every $A\in\mathfrak{M}$ (i.e., $t\mapsto\left\langle
x,\tau_{t}(A)y\right\rangle =\left\langle U_{t}x,AU_{t}y\right\rangle $ is
continuous for any $x,y\in\mathfrak{H}$). Hence $t\mapsto P\tau_{t}(P)$ is
weakly continuous. We know that tr is ultraweakly continuous (see
\textbf{[KR2}, Theorem 8.2.8\textbf{]}, for example), and therefore it is
weakly continuous on the unit ball of $\mathfrak{M}$ by \textbf{[KR2},
Proposition 7.4.5\textbf{]}. Since $\left\|  P\tau_{t}(P)\right\|  \leq1$, we
conclude that $t\mapsto$ tr$(P\tau_{t}(P))$ is continuous.$\blacksquare$

\bigskip So from (1.3) we see that for every $m\in M$ there exists a
$\delta_{m}>0$ such that
\[
\omega(\tau_{s}(P))>\text{tr}(P)-\varepsilon\qquad\text{for \qquad}%
mt-\delta_{m}<s<mt+\delta_{m}.
\]
This tells us that quantum mechanical recurrence is possible in practice,
assuming we are working with a bounded quantum system as above, since there is
a non-zero probability of repeating the yes/no experiment during one of the
time-intervals $\left(  mt-\delta_{m},mt+\delta_{m}\right)  $. It should be
mentioned though, that the elements of $M$ might be very far apart, so we
might have to wait very long after the initial yes/no experiment before the
probability tr$(P)-\varepsilon$ is reached as in (1.3).

According to Conjecture 1.9.1, a quantum mechanical system bounded in space,
and isolated from outside influences, can be mathematically described as a
bounded quantum system. So this is the physical situation for which we could
expect recurrence as above. This guess is confirmed by \textbf{[BL]} and
\textbf{[Perc]}. In classical mechanics we indeed have recurrence for systems
with finite volume phase space, in particular for a system with bounded phase
space in $\mathbb{R}^{2n}$, which corresponds to a system bounded in space and
isolated from outside influences (see Section 1.9). This fact constitutes some
additional circumstantial evidence for Conjecture 1.9.1.

\section{Ergodicity}

In Section 3.1 we saw how recurrence comes about in mechanics in terms of the
state of no information (tr in the quantum case; integration with respect to
normalized Lebesgue measure in the classical case). What is important here, is
that when we applied Theorem 2.5.1 (and Theorem 2.7.3) to mechanics, we took
$\varphi$ to be the state of no information.

Say we also want to apply Theorem 2.5.2 to mechanics to find the following
result: We consider two yes/no experiments with projections $P$ and $Q$ at
time zero, for a given system. The $P$ experiment is performed when we have no
information regarding the systems state (i.e. we start with the state of no
information $\varphi$), and a ``yes'' is obtained, changing the state to
$\omega$ defined by $\omega(A)=\varphi(PA)/\varphi(P)$. We want to know if a
subsequent execution of the $Q$ experiment (at one of the points in time from
the set $E$ in Theorem 2.5.2) will give ``yes'' with probability
$\varphi(Q)-\varepsilon$ or larger, where $E$ depends on $\varepsilon>0$. This
is a simple extension of the recurrence result we found in Section 3.1 (see in
particular equation (1.3)). However, for Theorem 2.5.2 to be applicable, we
need the system to be an ergodic $\ast$-dynamical system. In this section we
show that under physically reasonable assumptions, we do not have ergodicity.
(However, to prove that this implies that for any fixed $t>0 $ there is a pair
$P$ and $Q$ as above with $\varphi(P)>0$ and $\varphi(Q)>0$, such that the
probability for a ``yes'' in the $Q$ experiment is zero at all discrete times
$kt$, $k\in\mathbb{N}$, we would first have to solve Open Problem 2.5.8.)

\bigskip\noindent\textbf{3.2.1 Definition.} \textit{Consider a quantum or
classical mechanical system }$(\mathfrak{A},\varphi,\tau_{t})$\textit{\ where
}$\mathfrak{A}$\textit{\ is the observable algebra of the system, }$\varphi$
\textit{is the state of no information (we assume that it exists) and }%
$\tau_{t}$\textit{\ is the time-evolution. We call the system }%
\textbf{bounded}\textit{\ if it is either a bounded quantum system
}$(\mathfrak{M},$tr$,\tau_{t})$\textit{\ where }$\mathfrak{M}$\textit{\ is a
finite factor with }tr\textit{\ its trace and }$\tau_{t}$\textit{\ defined as
in Proposition 1.7.5, or a classical system }$(B_{\infty}(F),\varphi,\tau
_{t})$\textit{\ whose phase space }$F\subset\mathbb{R}^{2n}$\textit{\ (see
Remark 1.7.2) has finite Lebesgue measure, where} $\varphi(g)=\left(  \int
gd\lambda\right)  /\lambda(F)$\textit{\ with }$\lambda$\textit{\ the Lebesgue
measure on }$\mathbb{R}^{2n}$\textit{,} \textit{and }$\tau_{t}$\textit{\ is
given by equation (3.3) in Section 1.3. }

\bigskip Note that because of Liouville's Theorem (equation (7.1) in Section
1.7) and its quantum analogue, Proposition 1.7.5, a bounded mechanical system
$(\mathfrak{A},\varphi,\tau_{t})$ is a $\ast$-dynamical system as defined in
Definition 2.3.1, for any fixed $t$. Our goal in this section is therefore to
show that under physically reasonable assumptions, such a system is not
ergodic. Actually we will prove the more general result that if the state of a
system allows more than one energy level (in the sense of Definition 3.2.3),
then we do not have ergodicity.

We will work in the following general setting:

\bigskip\noindent\textbf{3.2.2 General Setting.} Let $\mathfrak{A}$ be the
observable algebra of a physical system (quantum or classical), and $H$ the
system's Hamiltonian (remember that the Hamiltonian of a system gives the
system's energy). $\mathfrak{A}$ is a unital $\ast$-algebra. In the classical
case we assume $\mathfrak{A}$ to be an algebra of bounded complex-valued
measurable functions on some measurable space $F$ with $g^{\ast}=\overline{g}$
the involution, and we assume $H$ to be a (possibly unbounded) measurable
function $F\rightarrow\mathbb{R}$. In the quantum case we assume
$\mathfrak{A}$ to be an algebra of bounded linear operators $\mathfrak{H}%
\rightarrow\mathfrak{H}$ on some Hilbert space $\mathfrak{H}$ with the
involution being the Hilbert adjoint, and we assume $H$ to be a (possibly
unbounded) self-adjoint linear operator in $\mathfrak{H}$. Keep in mind that
in the quantum case we allow the Hamiltonian to be represented in a Hilbert
space which might not be the state space, as is the case in Definition 1.7.3
and Remark 1.7.4. That is to say, $\mathfrak{H}$ is not necessarily the state
space of the quantum system. For reasons of generality, we likewise do not
assume that $F$ is the phase space of the classical system.

Furthermore, we assume that $\chi_{V}(H)\in\mathfrak{A}$ for all Borel
$V\subset\mathbb{R}$, where $\chi_{V}(H)$ is given by the Borel functional
calculus (in the classical case $\chi_{V}(H):=\chi_{V}\circ H$ as in Remark
1.4.2), and that $\chi_{V}(H)$ is the projection of the yes/no experiment ``Is
the energy in $V$?'' (Note that if we were to take $\mathfrak{A}=B_{\infty
}(F)$ for a classical system, or $\mathfrak{A=L(H)}$ for a quantum system,
then $\mathfrak{A}$ would contain all these projections in any case.)

As always, we assume the time-evolution to be a one-parameter $\ast
$-automorphism group $\tau$ of $\mathfrak{A}$ as in Section 1.4 (iv). In the
quantum case it is given by
\[
\tau_{t}(A)=e^{iHt}Ae^{-iHt}%
\]
and in the classical case by
\[
\tau_{t}(A)=A\circ T_{t}%
\]
where $T_{t}$ is an energy conserving (i.e. $H\circ T_{t}=H$) flow depending
on $H$. (If the time-evolution does not conserve energy, then it means that
the system is interacting with other systems. We could consider these systems
as part of our system to ensure conservation of energy. The time-evolution for
a quantum system as given above automatically conserves energy, since we take
$H$ to be fixed, so it does not allow interactions with other systems; see the
proof of Theorem 3.2.7.)

We then call $(\mathfrak{A},H)$ a \textbf{mechanical system}.

Where reference is made to an observable of the system, it will be assumed to
have the same mathematical form as $H$ above.$\blacksquare$

\bigskip We will assume that a bounded mechanical system is nontrivial in the
sense that it has more than one distinguishable energy level. We have to state
more clearly what we mean by this however. A simple way to do this in our
framework is as follows:

\bigskip\noindent\textbf{3.2.3 Definition.}\textit{\ Consider a state }%
$\omega$\textit{\ of a mechanical system }$(\mathfrak{A},H)$\textit{\ in the
general setting above. (So }$\omega$\textit{\ is a state on $\mathfrak{A}$.)
We say that }$\omega$\textit{\ }\textbf{allows more than one energy
level}\textit{\ if there are two open intervals }$J_{1}$\textit{\ and }$J_{2}%
$\textit{\ in }$\mathbb{R}$\textit{\ such that }$\overline{J_{1}}\cap
\overline{J_{2}}=\varnothing$\textit{, }$\omega\left(  \chi_{J_{1}}(H)\right)
>0$\textit{\ and }$\omega\left(  \chi_{J_{2}}(H)\right)  >0$\textit{, and a
bounded interval }$J$\textit{\ in }$\mathbb{R}$\textit{\ such that }%
$\omega(\chi_{J}(H))>0$\textit{. A bounded mechanical system }$(\mathfrak{A}%
,\varphi,\tau_{t})$\textit{\ with Hamiltonian }$H$ \textit{is called
}\textbf{nontrivial}\textit{\ if }$\varphi$\textit{\ allows more than one
energy level.}$\blacksquare$

\bigskip\noindent\textbf{3.2.4 Remark.} Definition 3.2.3 says that if we have
the state $\omega$ for the system, and we measure the energy, then there is a
nonzero probability of getting a value in $J_{1}$, and a nonzero probability
of getting a value in $J_{2}$. In this sense then, more than one energy level
of the system can be distinguished, since $J_{1}$ and $J_{2}$ are separated
(i.e. $\overline{J_{1}}\cap\overline{J_{2}}=\varnothing$). The existence of
the bounded interval $J$ implies that the system has at least one finite
energy level (this is a sensible assumption and not at all restrictive, since
in practice one can generally assume that a physical system does not possess
an infinite amount of energy; note that when modelling a physical system, some
useful models might have an infinite amount of energy, for example in the
thermodynamic limit \textbf{[Rue]}, but in this thesis we consider the system,
rather than a model which deviates from the system in such a nonphysical way).

If the state of no information of a bounded mechanical system does not allow
more than one energy level (in the technical sense given in Definition 3.2.3),
then it effectively means that the system only has one energy level (i.e. it
is physically trivial), since in the state of no information all energy levels
should be equally likely.$\blacksquare$

\bigskip\noindent\textbf{3.2.5 Lemma.}\textit{\ For Borel sets }%
$U,V\subset\mathbb{R}$\textit{\ with }$U\subset V$\textit{\ we have }
\[
\chi_{U}(A)\leq\chi_{V}(A)
\]
\textit{where }$A$\textit{\ is an observable of a mechanical system as in
General Setting 3.2.2.}

\bigskip\noindent\textit{Proof.} In the classical case this is easy, namely
\[
\chi_{U}(A)=\chi_{A^{-1}(U)}\leq\chi_{A^{-1}(V)}=\chi_{V}(A)
\]
since $A^{-1}(U)\subset A^{-1}(V)$. Alternatively (as harbinger to the quantum
case below), one can note that
\[
\chi_{U}(A)\chi_{V}(A)=\left(  \chi_{U}\chi_{V}\right)  (A)=\chi_{U}(A)
\]
since $U\subset V$, hence $\chi_{U}(A)\leq\chi_{V}(A)$.

In the quantum case it follows from the properties of the Borel functional
calculus \textbf{[SZ}, 9.11(v), 9.13(iii) and 9.32\textbf{]} and the fact that
a bounded linear operator on a Hilbert space is closed, that
\[
\chi_{U}(A)\chi_{V}(A)=\left(  \chi_{U}\chi_{V}\right)  (A)=\chi_{U}(A)
\]
and hence $\chi_{U}(A)\leq\chi_{V}(A)$, since $\chi_{U}(A)$ and $\chi_{V}(A)$
are projections (see \textbf{[Mu}, Theorem 2.3.2\textbf{]} for properties of
projections).$\blacksquare$

\bigskip\noindent\textbf{3.2.6 Proposition.}\textit{\ Consider a state
}$\omega$\textit{\ of a mechanical system }$(\mathfrak{A},H)$\textit{\ which
allows more than one energy level in the sense of Definition 3.2.3}%
.\textit{\ Then there exists a bounded interval }$I$\textit{\ in }$\mathbb{R}%
$\textit{\ such that }$0<\omega(\chi_{I}(H))<1$\textit{.}

\bigskip\noindent\textit{Proof.} Write $p(V):=\omega(\chi_{V}(H))$ for
all\thinspace Borel sets $V\subset\mathbb{R}$. ($p(V)$ is the probability for
a ``yes'' in the yes/no experiment ``Is the system's energy in $V$?'') Suppose
that
\begin{equation}
p(I)\in\{0,1\} \tag{2.1}%
\end{equation}
for all bounded intervals $I$ in $\mathbb{R}$. By assumption there exists a
bounded interval $I_{0}$ in $\mathbb{R}$ such that $p(I_{0})>0$, and hence
$p(I_{0})=1$. Because of Lemma 3.2.5, we can assume without loss that this
interval is of the form $I_{0}=[a_{0},b_{0})$. We now inductively construct a
sequence $I_{0},I_{1},I_{2},...$ of intervals such that $p(I_{n})=1$ for all
$n$:

Divide $I_{n}$ in its left and right halves (each of the form $[c,d)$), and
let $I_{n+1}=[a_{n+1},b_{n+1})$ be the half such that $p(I_{n+1})=1$.

Note that $I_{n+1}$ exists by induction, since if it did not, we would have
$p(L)=p(R)=0$ by (2.1), where $L$ and $R$ are the left and right halves of
$I_{n}$, and then by the properties of the Borel functional calculus (and
arguments as in the proof of Lemma 3.2.5)
\begin{equation}
0=p(L)+p(R)=\omega\left(  \chi_{L}(H)+\chi_{R}(H)\right)  =\omega\left(
(\chi_{L}+\chi_{R})(H)\right)  =p(I_{n}) \tag{2.2}%
\end{equation}
which contradicts $p(I_{0})>0$. The sequences $(a_{n})$ and $(b_{n})$ are
bounded, and increasing and decreasing respectively, while $b_{n}-a_{n}%
=(b_{0}-a_{0})/2^{n}$. This implies that they converge to the same value, say
$E$.

We can view $E$ as the only energy level of the system that can be obtained in
a measurement, since any open set $V$ containing $E$ contains an $I_{n}$, and
hence $1=p(I_{n})\leq p(V)\leq1$ by Lemma 3.2.5, so the probability for a
``yes'' in the yes/no experiment ``Is the energy in $V$?'' is one. The idea is
therefore to get a contradiction with Definition 3.2.3, which says that there
are at least two energy levels. So consider any open intervals $J_{1}$ and
$J_{2}$ in $\mathbb{R}$ with $\overline{J_{1}}\cap\overline{J_{2}}%
=\varnothing$.

\textit{Case 1.} Say $E\in J_{1}$. Then $p(J_{1})=1$ as for $p(V)$ above. It
follows that $p(J_{2})=0$, otherwise we would have
\[
p(J_{1}\cup J_{2})=p(J_{1})+p(J_{2})>1
\]
similar to (2.2), which contradicts the definition of $p$. (Similarly if we
had $E\in J_{2}$.)

\textit{Case 2.} Now suppose $E\notin J_{1}\cup J_{2}$. Since $\overline
{J_{1}}\cap\overline{J_{2}}=\varnothing$, we can assume without loss that
$E\notin\overline{J_{1}}$, which implies that an $I_{n}$ exists such that
$I_{n}\subset\mathbb{R}\backslash\overline{J_{1}}$, as for $V$ above. So by
Lemma 3.2.5 we then have $p(\mathbb{R}\backslash\overline{J_{1}})=1$ and also
$p(\mathbb{R})=1$, and therefore (in the same way as (2.2)),
\[
p\left(  \overline{J_{1}}\right)  =p(\mathbb{R})-p\left(  \mathbb{R}%
\backslash\overline{J_{1}}\right)  =0\text{.}%
\]
So, again by Lemma 3.2.5, $0\leq p(J_{1})\leq p(\overline{J_{1}})=0$.

From these two cases we see that we either have $p(J_{1})=0$ or $p(J_{2})=0$,
contradicting the assumptions. Therefore (2.1) must be wrong, which means that
$0<p(I)<1$ for some bounded interval $I$.$\blacksquare$

\bigskip\noindent\textbf{3.2.7 Theorem.}\textit{\ Consider a state }$\omega
$\textit{\ of a mechanical system }$(\mathfrak{A},H)$\textit{\ which allows
more than one energy level in the sense of Definition 3.2.3, and let }$\tau
$\textit{\ be the time-evolution of the system as in General Setting 3.2.2.
Fix any }$t\in\mathbb{R}$\textit{, and assume that }$(\mathfrak{A},\omega
,\tau_{t})$\textit{\ is a }$\ast$\textit{-dynamical system (i.e. }$\omega
(\tau_{t}(A^{\ast}A))\leq\omega(A^{\ast}A)$\textit{\ for all }$A\in
\mathfrak{A}$\textit{). Then }$(\mathfrak{A},\omega,\tau_{t})$\textit{\ is not
ergodic (in the sense of Definition 2.3.2). In particular, a nontrivial
bounded mechanical system (as in Definitions 3.2.1 and 3.2.3) is not ergodic.}

\bigskip\noindent\textit{Proof.} By Proposition 3.2.6 there is a Borel set
$V\subset\mathbb{R}$ such that $0<\omega(P)<1$ for $P:=\chi_{V}(H)$.

By conservation of energy in the classical case, we have $H\circ T_{t}=H$,
hence $\tau_{t}(P)=\chi_{V}\circ H\circ T_{t}=\chi_{V}\circ H=P$. In the
quantum mechanical case we have $\tau_{t}(P)=e^{iHt}\chi_{V}(H)e^{-iHt}%
=\left(  e^{i(\cdot)t}\chi_{V}e^{-i(\cdot)t}\right)  (H)=\chi_{V}(H)=P$ by the
properties of the Borel functional calculus \textbf{[SZ}, 9.11(v)\textbf{]},
which says that energy is conserved. So, in either case
\begin{equation}
\tau_{t}(P)=P\text{.} \tag{2.3}%
\end{equation}

Consider any $a_{1},a_{2}\in\mathbb{C}$ and set $A:=a_{1}P+a_{2}(1-P)$. Now
set
\[
B_{n}:=\frac{1}{n}\sum_{k=0}^{n-1}\tau_{t}(A)\qquad\text{and\qquad}%
C_{n}:=B_{n}-\omega(A)
\]
then $B_{n}=A$ by (2.3) since $\tau_{t}(1)=1$. Write $p:=\omega(P)$, then it
follows that
\begin{align*}
C_{n}  &  =a_{1}P+a_{2}(1-P)-a_{1}p-a_{2}(1-p)\\
&  =(a_{1}-a_{2})(P-p)
\end{align*}
and therefore
\[
\left\|  C_{n}\right\|  _{\omega}=\sqrt{\omega(C_{n}^{\ast}C_{n})}=\left|
a_{1}-a_{2}\right|  \sqrt{p(1-p)}%
\]
so
\[
\lim_{n\rightarrow\infty}\left\|  C_{n}\right\|  _{\omega}=\left|  a_{1}%
-a_{2}\right|  \sqrt{p(1-p)}\neq0
\]
if we choose $a_{1}\neq a_{2}$, since $0<p<1$. Therefore the system is not
ergodic, by Proposition 2.5.5(i).$\blacksquare$

\bigskip The system in Example 2.5.7 is ergodic despite the fact that tr is
the state of no information, simply because the ``time-evolution'' $\tau$
behaves differently from that of a physical system as in Theorem 3.2.7. In the
ergodic case, $\tau$ in Example 2.5.7 only has fixed points of the form
\[
\left(
\begin{array}
[c]{cc}%
a & 0\\
0 & a
\end{array}
\right)
\]
which is only a projection if $a\in\{0,1\}$, hence a projection $P$ as in
(2.3) with $0<$ tr$(P)<1$ does not exist. One can say that $\tau$ does not
preserve the various ``energy levels'' of the system, but only preserves the
system as a whole.

\bigskip\noindent\textbf{3.2.8 Remarks. }Essentially Theorem 3.2.7 says that
if the state is a mixture of more than one energy state (so more than one
value of energy has nonzero probability when the observer measures the
energy), then the state is not ergodic (in this context it makes more sense to
speak of an ergodic state, rather than an ergodic system, since the state
describes the observer's information about the physical system as in Section
1.6, rather than being a property of the system itself). From the statistical
point of view that we have been using since Chapter 1, this should be the
typical situation in practice, since normally an observer would not be able to
measure the energy precisely enough to give a state allowing only one energy
level. So if the observer does not have complete (or precise) information
about the system's energy, then the state describing his information isn't ergodic.

Intuitively Theorem 3.2.7 makes perfect sense. If more than one energy level
is present in the state, then we can imagine decomposing it into its various
energy ``components'' (for example, decompose the phase space into its
constant energy surfaces in the case of a classical system; see below). By the
conservation of energy, the time-evolution does not mix the various energy
components with each other. But this clearly violates the basic intuition
behind ergodicity, namely that in an ergodic system, any ``part'' is
eventually mixed with every other part (see Corollary 2.6.1 and the discussion
following it, as well as Theorem 2.5.2 and Proposition 2.5.6, which all say
that any part of an ergodic system eventually overlaps with every other part).
So it is also clear why conservation of energy plays a central role in the
proof of Theorem 3.2.7.

This result does not mean that the idea of ergodicity is in principle
irrelevant in physics. Theoretically one can still consider states allowing
only one energy level, and study whether they are ergodic or not. For example,
a state given by any probability measure on a constant energy surface (given
by $H=E$, where $E$ is the energy of the surface) of a classical system, by
definition allows only one energy level $E$, while each energy eigenstate of a
quantum system (assuming the Hamiltonian has eigenvectors) by definition
corresponds to a single energy level (also see Appendix B). Ergodicity would
then be a property of the system, rather than of the observer's information,
which in the light of Theorem 3.2.7 seems like the sensible approach to
ergodicity in physics.

In classical mechanics ergodicity arises in the sense that one would consider
systems where for almost every pure state (point) $x$ in a constant energy
surface, the time average
\[
\frac{1}{n}\sum_{k=1}^{n-1}f\circ T_{kt}(x)
\]
of any observable $f$ converges to the average $\omega(f)$ of the observable
over the constant energy surface, for any fixed $t>0$, where the state
$\omega$ of the system is given by a time-invariant probability measure on the
constant energy surface (the existence of such a measure follows from
Liouville's Theorem; see for example \textbf{[Kh}, Section 7\textbf{]} or
\textbf{[Pete}, Chapter 1, Proposition 2.2\textbf{]}, and also \textbf{[Rue},
Section 1.1\textbf{]}). Since only one energy level is involved, this is not
in conflict with Theorem 3.2.7. We can mention that in 1962-63 Sinai succeeded
in proving that a classical gas, consisting of hard spheres enclosed in a box
and interacting through pair potentials, is ergodic in this sense (refer to
\textbf{[AA}, Section 18\textbf{]} or \textbf{[Rue}, Section 1.1\textbf{]} and
references therein). Ergodicity as given by Definition 2.3.2, or equivalently
by equation (6.1) in Section 2.6, with $\varphi=\omega$ and $\mu$ the
probability measure on the constant energy surface, is a slightly weaker form
of ergodicity. Refer to \textbf{[Rud}, Theorem 3.12\textbf{]} for the
connection of this with the almost everywhere convergence mentioned above,
namely that it implies the existence of a subsequence of the time-averages
\[
\frac{1}{n}\sum_{k=1}^{n-1}f\circ T_{kt}%
\]
converging pointwise almost everywhere to $\omega(f)$, whereas for the case
above the whole sequence converges pointwise almost everywhere to $\omega(f) $.

In quantum mechanics the idea is to study states that are ergodic in some
sense, the simplest approach being to take eigenstates of the Hamiltonian (if
they exist) as ergodic, since for such an eigenstate $x$ we have
$e^{-tHt}x=e^{-iEt}x$ where $E$ is the corresponding eigenvalue (the energy),
and hence for any fixed $t$
\[
\frac{1}{n}\sum_{k=0}^{n-1}\omega\left(  \tau_{t}^{k}(A)\right)  =\frac{1}%
{n}\sum_{k=0}^{n-1}\left\langle x,\tau_{t}^{k}(A)x\right\rangle =\frac{1}%
{n}\sum_{k=0}^{n-1}\left\langle e^{-iHkt}x,Ae^{-iHkt}x\right\rangle
=\left\langle x,Ax\right\rangle =:\omega(A)
\]
which is an equality of a time average and a ``state average''. (Also see
\textbf{[T}, Remark (3.1.23;1)\textbf{]}.) This is a very primitive form of
ergodicity of a state. For a deeper approach, refer to \textbf{[T}, Sections
3.1 and 3.2\textbf{]}, and in particular \textbf{[T}, Remarks (3.2.10;6) and
(3.2.16;1)\textbf{]} for the relation between ergodicity and KMS states
(equilibrium states). Also see Appendix B for a more precise description of
the ergodicity of the Hamiltonian's eigenstates.

The unfortunate situation in quantum mechanics (as far as ergodicity goes), is
that even if the system is in a state containing only one energy level,
measuring an observable not commuting with the Hamiltonian will typically
leave the system in a state which does contain more than one energy level, in
which case it can no longer be ergodic. But as mentioned earlier, this still
doesn't stop us from studying those states which are ergodic.

For more on quantum ergodicity, see [NTW], [ENTS] and [Z].$\blacksquare$

\appendix

\chapter{Examples concerning ergodicity}

\section{On the definition of ergodicity}

This section is devoted to the construction of a $\ast$-dynamical system
$(\mathfrak{A},\varphi,\tau)$ with the property that if $\left\|
\tau(A)-A\right\|  _{\varphi}=0$, then $\left\|  A-\alpha\right\|  _{\varphi
}=0$ for some $\alpha\in\mathbb{C}$, but for which the fixed points of the
operator $U$ defined in Proposition 2.3.3 in terms of some cyclic
representation, form a vector subspace of $\mathfrak{H}$ with dimension
greater than one. This will prove the necessity of a sequence, rather than a
single element, in Definition 2.3.2, in order for Proposition 2.3.3 to hold.

First some general considerations. Consider a dense vector subspace
$\mathfrak{G} $ of a Hilbert space $\mathfrak{H}$, and let $\mathfrak{L(H)}$
be the bounded linear operators $\mathfrak{H\rightarrow H}$. Set
\[
\mathfrak{A:=}\left\{  A|_{\mathfrak{G}}:A\in\mathfrak{L(H)}\text{,
}A\mathfrak{G\subset G}\text{ and }A^{\ast}\mathfrak{G\subset G}\right\}
\]
where $A|_{\mathfrak{G}}$ denotes the restriction of $A$ to $\mathfrak{G}$,
then $\mathfrak{A}$ is clearly a vector subspace of $\mathfrak{L(G)}$. For any
$A\in\mathfrak{A}$, denote by $\overline{A}$ the (unique) bounded linear
extension of $A$ to $\mathfrak{H}$. Now define an involution on $\mathfrak{A}$
by
\[
A^{\ast}:=\overline{A}^{\ast}|_{\mathfrak{G}}%
\]
for all $A\in\mathfrak{A}$, then it is easily verified that $\mathfrak{A}$
becomes a unital $\ast$-algebra. (For $A,B\in\mathfrak{A}$ it is clear that
$AB$ is a bounded linear operator $\mathfrak{G\rightarrow G}$ which therefore
has the extension $\overline{A}.\overline{B}\in\mathfrak{L(H)}$ for which
$\overline{A}.\overline{B}\mathfrak{G\subset G}$ and $\left(  \overline
{A}.\overline{B}\right)  ^{\ast}\mathfrak{G}=\overline{B}^{\ast}\overline
{A}^{\ast}\mathfrak{G\subset G}$ by the definition of $\mathfrak{A}$. Hence
$AB\in\mathfrak{A}$, which means that $\mathfrak{A}$ is a subalgebra of
$\mathfrak{L(G)}$. Also, $\left(  AB\right)  ^{\ast}=\left(  \overline
{A}.\overline{B}\right)  ^{\ast}|_{\mathfrak{G}}=\left(  \overline{B}^{\ast
}\overline{A}^{\ast}\right)  |_{\mathfrak{G}}=\overline{B}^{\ast}\left(
\overline{A}^{\ast}|_{\mathfrak{G}}\right)  =\overline{B}^{\ast}A^{\ast
}=B^{\ast}A^{\ast}$. Similarly for the other defining properties of an
involution.) Note that for $A\in\mathfrak{A}$ and $x,y\in\mathfrak{G}$ we
have
\[
\left\langle x,Ay\right\rangle =\left\langle x,\overline{A}y\right\rangle
=\left\langle \overline{A}^{\ast}x,y\right\rangle =\left\langle A^{\ast
}x,y\right\rangle \text{.}%
\]
For a given norm one vector $\Omega\in\mathfrak{G}$ we define a state
$\varphi$ on $\mathfrak{A}$ by
\[
\varphi(A)=\left\langle \Omega,A\Omega\right\rangle \text{.}%
\]
Next we construct a cyclic representation of $(\mathfrak{A},\varphi)$. Let
\[
\pi:\mathfrak{A\rightarrow}L(\mathfrak{G}):A\mapsto A
\]
then clearly $\pi$ is linear with $\pi(1)=1$ and $\pi(AB)=\pi(A)\pi(B)$. Note
that for any $x,y\in\mathfrak{G}$ we have $(x\otimes y)^{\ast}=y\otimes x$,
hence $(x\otimes y)\mathfrak{G\subset G}$ and $(x\otimes y)^{\ast
}\mathfrak{G\subset G}$, so $(x\otimes y)|_{\mathfrak{G}}\in\mathfrak{A}$.
Now, $\pi\left(  (x\otimes\Omega)|_{\mathfrak{G}}\right)  \Omega=x\left\langle
\Omega,\Omega\right\rangle =x$, hence $\pi(\mathfrak{A})\Omega=\mathfrak{G}$.
Furthermore, $\left\langle \pi(A)\Omega,\pi(B)\Omega\right\rangle
=\left\langle A\Omega,B\Omega\right\rangle =\left\langle \Omega,A^{\ast
}B\Omega\right\rangle =\varphi(A^{\ast}B)$. Thus $(\mathfrak{G},\pi,\Omega)$
is a cyclic representation of $(\mathfrak{A},\varphi)$.

Suppose we have a unitary operator $U:\mathfrak{H}\rightarrow\mathfrak{H}$
such that $U\mathfrak{G=G}$ and $U\Omega=\Omega$. Then $U^{\ast}%
\mathfrak{G}=U^{-1}\mathfrak{G=G}$, so $V:=U|_{\mathfrak{G}}\in\mathfrak{A}$,
and $V^{\ast}=U^{\ast}|_{\mathfrak{G}}$. It follows that $VAV^{\ast}%
\in\mathfrak{A}$ for all $A\in\mathfrak{A}$, hence we can define a linear
function $\tau:\mathfrak{A}\rightarrow\mathfrak{A}$ by
\[
\tau(A)=VAV^{\ast}\text{.}%
\]
Clearly $V^{\ast}V=1=VV^{\ast}$, so $\tau(1)=1$ and $\varphi\left(
\tau(A)^{\ast}\tau(A)\right)  =\varphi\left(  VA^{\ast}AV^{\ast}\right)
=\left\langle U^{\ast}\Omega,A^{\ast}AU^{\ast}\Omega\right\rangle
=\varphi(A^{\ast}A)$, since $U^{\ast}\Omega=U^{-1}\Omega=\Omega$. Therefore
$(\mathfrak{A},\varphi,\tau)$ is a $\ast$-dynamical system. Note that
$U|_{\mathfrak{G}}$ satisfies equation (3.1) of Section 2.3, namely
$U\pi(A)\Omega=UA\Omega=UAU^{\ast}\Omega=\tau(A)\Omega=\pi\left(
\tau(A)\right)  \Omega$, hence $U$ is the operator which appears in
Proposition 2.3.3.

Assume $\left\{  x\in\mathfrak{G}:Ux=x\right\}  =\mathbb{C}\Omega$. If
$\left\|  \tau(A)-A\right\|  _{\varphi}=0$, it then follows for $x=\iota(A)$,
with $\iota$ given by equation (2.1) of Section 2.2, that $\left\|
Ux-x\right\|  =\left\|  \iota\left(  \tau(A)-A\right)  \right\|  =\left\|
\tau(A)-A\right\|  _{\varphi}=0$, so $x=\alpha\Omega$ for some $\alpha
\in\mathbb{C}$. Therefore $\left\|  A-\alpha\right\|  _{\varphi}=\left\|
\iota(A-\alpha)\right\|  =\left\|  x-\alpha\Omega\right\|  =0$.

In other words, assuming that the fixed points of $U$ in $\mathfrak{G}$ form
the one-dimensional subspace $\mathbb{C}\Omega$, it follows that $\left\|
\tau(A)-A\right\|  _{\varphi}=0$ implies that $\left\|  A-\alpha\right\|
_{\varphi}=0$ for some $\alpha\in\mathbb{C}$.

It remains to construct an example of a $U$ with all the properties mentioned
above, whose fixed point space in $\mathfrak{H}$ has dimension greater than
one. The following example was constructed by L. Zsid\'{o}:

Let $\mathfrak{H}$ be a separable Hilbert space with an orthonormal basis of
the form
\[
\{\Omega\,,\,y\}\cup\left\{  u_{k}\,:\,k\in\mathbb{Z}\right\}
\]
(that is to say, this is a total orthonormal set in $\mathfrak{H}$) and define
the linear operator $U:\mathfrak{H}\longrightarrow\mathfrak{H}$ via bounded
linear extension by
\begin{align*}
U\Omega &  =\Omega\,,\\
Uy  &  =y\,,\\
Uu_{k}  &  =u_{k+1}\,,\quad k\in\mathbb{Z}\text{.}%
\end{align*}
Clearly $U$ is isometric, while $U\mathfrak{H}$ is dense in $\mathfrak{H}$,
hence $U$ is surjective, since $\mathfrak{H}$ is complete. Since $U$ is a
surjective isometry, it is unitary. Let $\mathfrak{G}$ be the linear span of
\[
\{\Omega\}\cup\left\{  y+u_{k}:\,k\in\mathbb{Z}\right\}  \text{.}%
\]
Then $U\mathfrak{G}=\mathfrak{G}$. Furthermore, $\mathfrak{G}$ is dense in
$\mathfrak{H}\,$. Indeed,
\[
\Vert y-\frac{1}{n}\sum_{k=1}^{n}(y+u_{k})\Vert=\frac{1}{n}\Vert\sum_{k=1}%
^{n}u_{k}\Vert=\frac{1}{\sqrt{n}}\longrightarrow0
\]
implies that $y\in\overline{\mathfrak{G}}\,$, the closure of $\mathfrak{G}$,
hence also
\[
u_{k}=(y+u_{k})-y\in\overline{\mathfrak{G}}\,
\]
for $k\in\mathbb{Z}$.

Next we show that
\begin{equation}
\{x\in\mathfrak{G}\,:\,Ux=x\}=\mathbb{C}\Omega. \tag{1.1}%
\end{equation}
If $\alpha\Omega+\sum\limits_{k=-n}^{n}\beta_{k}(y+u_{k})\in\mathfrak{G}$ is
left fixed by $U$, then
\[
\alpha\Omega+\sum\limits_{k=-n}^{n}\beta_{k}y+\sum\limits_{k=-n}^{n}\beta
_{k}u_{k+1}=\alpha\Omega+\sum\limits_{k=-n}^{n}\beta_{k}y+\sum\limits_{k=-n}%
^{n}\beta_{k}u_{k}%
\]
and it follows that $\beta_{-n}=0$, and that $\beta_{k+1}=\beta_{k}$ for
$k=-n,...,n-1$. Thus
\[
\alpha\Omega+\sum\limits_{k=-n}^{n}\beta_{k}(y+u_{k})=\alpha\Omega\,
\]
proving (1.1).

On the other hand,
\[
\{x\in\mathfrak{H}\,:\,Ux=x\}
\]
clearly contains the two-dimensional vector space spanned by $\Omega$ and
$y\,$.\pagebreak

\section{An example of an ergodic system}

Here we give the proof that Example 2.5.7 is indeed ergodic. It is clear that
$\tau$ is linear and that $\tau(1)=1$. Let
\[
A=\left(
\begin{array}
[c]{cc}%
a_{11} & a_{12}\\
a_{21} & a_{22}%
\end{array}
\right)
\]
and
\[
B=\left(
\begin{array}
[c]{cc}%
b_{11} & b_{12}\\
b_{21} & b_{22}%
\end{array}
\right)
\]
be complex matrices. Then
\[
\tau(A)^{\ast}=\left(
\begin{array}
[c]{cc}%
\overline{a_{22}} & \overline{c_{2}a_{21}}\\
\overline{c_{1}a_{12}} & \overline{a_{11}}%
\end{array}
\right)
\]
and
\[
\tau(A)^{\ast}\tau(A)=\left(
\begin{array}
[c]{cc}%
\left|  a_{22}\right|  ^{2}+\left|  c_{2}a_{21}\right|  ^{2} & \overline
{a_{22}}c_{1}a_{12}+\overline{c_{2}a_{21}}a_{11}\\
\overline{c_{1}a_{12}}a_{22}+\overline{a_{11}}c_{2}a_{21} & \left|
c_{1}a_{12}\right|  ^{2}+\left|  a_{11}\right|  ^{2}%
\end{array}
\right)
\]
while
\[
A^{\ast}=\left(
\begin{array}
[c]{cc}%
\overline{a_{11}} & \overline{a_{21}}\\
\overline{a_{12}} & \overline{a_{22}}%
\end{array}
\right)
\]
and
\[
A^{\ast}A=\left(
\begin{array}
[c]{cc}%
\left|  a_{11}\right|  ^{2}+\left|  a_{21}\right|  ^{2} & \overline{a_{11}%
}a_{12}+\overline{a_{21}}a_{22}\\
\overline{a_{12}}a_{11}+\overline{a_{22}}a_{21} & \left|  a_{12}\right|
^{2}+\left|  a_{22}\right|  ^{2}%
\end{array}
\right)
\]
so
\begin{align*}
\varphi\left(  \tau(A)^{\ast}\tau(A)\right)   &  =\frac{1}{2}\left(  \left|
a_{22}\right|  ^{2}+\left|  c_{2}a_{21}\right|  ^{2}+\left|  c_{1}%
a_{12}\right|  ^{2}+\left|  a_{11}\right|  ^{2}\right) \\
&  \leq\frac{1}{2}\left(  \left|  a_{22}\right|  ^{2}+\left|  a_{21}\right|
^{2}+\left|  a_{12}\right|  ^{2}+\left|  a_{11}\right|  ^{2}\right) \\
&  =\varphi\left(  A^{\ast}A\right)
\end{align*}
for all $A$, meaning that $(\mathfrak{A},\varphi,\tau)$ is a $\ast$-dynamical
system, if and only if $\left|  c_{1}\right|  \leq1$ and $\left|
c_{2}\right|  \leq1$, which is what we will assume.

Next we prove that it is ergodic. For even $k\geq0$ we have
\[
\tau^{k}(B)=\left(
\begin{array}
[c]{cc}%
b_{11} & c_{1}^{k}b_{12}\\
c_{2}^{k}b_{21} & b_{22}%
\end{array}
\right)
\]
and therefore
\[
A\tau^{k}(B)=\left(
\begin{array}
[c]{cc}%
a_{11}b_{11}+a_{12}c_{2}^{k}b_{21} & a_{11}c_{1}^{k}b_{12}+a_{12}b_{22}\\
a_{21}b_{11}+a_{22}c_{2}^{k}b_{21} & a_{21}c_{1}^{k}b_{12}+a_{22}b_{22}%
\end{array}
\right)
\]
which means
\[
\varphi\left(  A\tau^{k}(B)\right)  =\frac{1}{2}\left(  a_{11}b_{11}%
+a_{12}c_{2}^{k}b_{21}+a_{21}c_{1}^{k}b_{12}+a_{22}b_{22}\right)  \text{.}%
\]
For odd $k>0$ we then get
\[
\varphi\left(  A\tau^{k}(B)\right)  =\frac{1}{2}\left(  a_{11}b_{22}%
+a_{12}c_{2}^{k}b_{21}+a_{21}c_{1}^{k}b_{12}+a_{22}b_{11}\right)
\]
by switching $b_{11}$ and $b_{22}$. For $c\in\mathbb{C}$ it is clear that
$U:\mathbb{C}\rightarrow\mathbb{C}:x\mapsto cx$ is a linear operator with
$\left\|  U\right\|  \leq1$ if and only if $\left|  c\right|  \leq1$, and for
$c\neq1$ the only fixed point of $U$ is $0$, in which case
\[
\frac{1}{n}\sum_{k=0}^{n-1}c^{k}x=\frac{1}{n}\sum_{k=0}^{n-1}U^{k}%
x\longrightarrow0
\]
for all $x\in\mathbb{C}$ as $n\rightarrow\infty$, by the Mean Ergodic Theorem
2.4.1. Hence, for $c_{1}\neq1$ and $c_{2}\neq1$ it follows that
\begin{align*}
&  \lim_{n\rightarrow\infty}\frac{1}{n}\sum_{k=0}^{n-1}\varphi\left(
A\tau^{k}(B)\right) \\
&  =\lim_{n\rightarrow\infty}\frac{1}{n}%
\genfrac{\{}{\}}{0pt}{0}{\frac{n}{2}\left[  \frac{1}{2}(a_{11}b_{11}%
+a_{22}b_{22})+\frac{1}{2}(a_{11}b_{22}+a_{22}b_{11})\right]  \text{ for
}n\text{ even}}{\frac{n-1}{2}\left[  \frac{1}{2}(a_{11}b_{11}+a_{22}%
b_{22})+\frac{1}{2}(a_{11}b_{22}+a_{22}b_{11})\right]  +\frac{1}{2}%
(a_{11}b_{11}+a_{22}b_{22})\text{ for }n\text{ odd}}%
\\
&  =\left(  \frac{a_{11}+a_{22}}{2}\right)  \left(  \frac{b_{11}+b_{22}}%
{2}\right) \\
&  =\varphi(A)\varphi(B)
\end{align*}
which means that $(\mathfrak{A},\varphi,\tau)$ is ergodic, by Proposition 2.5.6(ii).

On the other hand, if $c_{1}=1$ and $c_{2}\neq1$, then we have by a similar
calculation that
\[
\lim_{n\rightarrow\infty}\frac{1}{n}\sum_{k=0}^{n-1}\varphi\left(  A\tau
^{k}(B)\right)  =\varphi(A)\varphi(B)+\frac{a_{21}b_{12}}{2}\text{.}%
\]
Likewise for the other cases where either $c_{1}$ or $c_{2}$ or both are equal
to $1$. So $(\mathfrak{A},\varphi,\tau)$ is ergodic if and only if $c_{1}%
\neq1$ and $c_{2}\neq1$.

\bigskip\noindent\textbf{A.2.1 Remark.} It is easily seen that $\tau$ is not a
homomorphism, namely
\[
\tau(AB)=\left(
\begin{array}
[c]{cc}%
a_{21}b_{12}+a_{22}b_{22} & c_{1}(a_{11}b_{12}+a_{12}b_{22})\\
c_{2}(a_{21}b_{11}+a_{22}b_{21}) & a_{11}b_{11}+a_{12}b_{21}%
\end{array}
\right)
\]
while
\[
\tau(A)\tau(B)=\left(
\begin{array}
[c]{cc}%
a_{22}b_{22}+c_{1}c_{2}a_{12}b_{21} & c_{1}(a_{22}b_{12}+a_{12}b_{11})\\
c_{2}(a_{21}b_{22}+a_{11}b_{21}) & c_{1}c_{2}a_{21}b_{12}+a_{11}b_{11}%
\end{array}
\right)  \text{.}%
\]
In fact, unless $c_{1}c_{2}=1$, it follows that we don't even have $\tau
(A^{2})=\tau(A)^{2}$ for all $A$. Nor, for that matter, do we have
$\tau(A^{\ast})=\tau(A)^{\ast}$ for all $A$, unless $c_{2}=\overline{c_{1}}$.
This is opposed to the situation for a measure theoretic dynamical system as
defined in Section 2.1, where $\tau$ in equation (1.1) of that section is
always a $\ast$-homomorphism. It therefore makes sense not to assume that
$\tau$ is a $\ast$-homomorphism in Definition 2.3.1, since we now have an
example where it isn't.$\blacksquare$

\bigskip\noindent\textbf{A.2.2 Remark.} We note that $\varphi(\tau
(A))=\varphi(A)$, i.e. $\varphi$ is $\tau$-invariant, but this fact in itself
does not imply that $\varphi(\tau(A)^{\ast}\tau(A))\leq\varphi(A^{\ast}A)$,
since $\tau$ is not a $\ast$-homomorphism, by Remark A.2.1.

Furthermore, $\varphi(AB)=\varphi(BA)$ for all $A,B\in\mathfrak{A}$, so
$\varphi$ is commutative (so to speak) even though $\mathfrak{A}$ is not.
Also, while $\tau(AB)\neq\tau(BA)$ for some $A,B\in\mathfrak{A}$, we still
have $\varphi(\tau(AB))=\varphi(AB)=\varphi(BA)=\varphi(\tau(BA))$, so $\tau$
is noncommutative (so to speak), but with respect to $\varphi$ it is again
commutative. We conclude that while $\mathfrak{A}$ is noncommutative,
$(\mathfrak{A},\varphi,\tau)$ is still in many respects commutative simply
because $\varphi(AB)=\varphi(BA)$ for all $A$ and $B$.$\blacksquare$

\bigskip\noindent\textbf{A.2.3 Question.} Is there an example of an ergodic
$\ast$-dynamical system $(\mathfrak{A},\varphi,\tau)$ in which $\varphi
(AB)\neq\varphi(BA)$ for some $A,B\in\mathfrak{A}$? Yes, see the example in
Appendix B.$\blacksquare$

\chapter{Ergodicity of energy eigenstates}

In this appendix, added to the thesis in December 2003\footnote{but prepared
as part of a lecture at the SA Mathematical Society's 45th annual conference
in Stellenbosch in November 2002}, we briefly study the ergodicity of energy
eigenstates, and in the process exhibit another example of an ergodic system.

Consider any Hilbert space $\mathfrak{H}$, and let $\tau
:\mathfrak{L(H)\rightarrow L(H)}$ be given by
\[
\tau(A):=e^{iH}Ae^{-iH}%
\]
where $H$ is a (possibly unbounded) self-adjoint linear operator in
$\mathfrak{H}$. Consider the state $\omega$ on $\mathfrak{L(H)}$ given by
\begin{equation}
\omega(A)=\left\langle \Omega,A\Omega\right\rangle \tag{1}%
\end{equation}
for some unit vector $\Omega\in\mathfrak{H}$. Then $(\mathfrak{H}%
,$id$_{\mathfrak{L(H)}},\Omega)$ is a cyclic representation of
$(\mathfrak{L(H)},\omega)$, since $(x\otimes\Omega)\Omega=x$, so
$\mathfrak{L(H)}\Omega=\mathfrak{H}$. Now set
\[
\iota:\mathfrak{L(H)\rightarrow H}:A\mapsto A\Omega.
\]
as in Section 2.2. If we assume that
\begin{equation}
e^{-iH}\Omega=e^{-iE}\Omega\tag{2}%
\end{equation}
for some $E\in\mathbb{R}$, then $\omega\circ\tau=\omega$, making
$(\mathfrak{L(H)},\omega,\tau)$ a $\ast$-dynamical system, and ensuring that
\[
U:\mathfrak{H\rightarrow H}:\iota(A)\mapsto\iota(\tau(A))
\]
is well-defined just as in Section 2.3 and simplifies to $UA\Omega
=e^{-iE}e^{iH}A\Omega$. For any $x\in\mathfrak{H}$ we now have
\[
Ux=U(x\otimes\Omega)\Omega=e^{-iE}e^{iH}(x\otimes\Omega)\Omega=e^{-iE}e^{iH}x
\]
so
\[
U=e^{-iE}e^{iH}.
\]

We have to look at the dimension of the fixed point space of $U$, since this
is how we decide whether the system is ergodic or not, by 2.3.3. In terms of
the projection $P$ onto this space, we have
\[
P\mathfrak{H}=\{x\in\mathfrak{H}:e^{iH}x=e^{iE}x\}.
\]
Since $e^{iH}e^{-iH}=1$, we see from (2) that $\Omega\in P\mathfrak{H}$, as we
know it must by the proof of 2.3.3. Suppose that $E$ is degenerate in the
sense that there exists a $x\in\mathfrak{H}\backslash(\mathbb{C}\Omega)$ such
that $e^{-iH}x=e^{-iE}x$, then exactly as above, $x\in P\mathfrak{H}$, so
$\dim(P\mathfrak{H})>1$, which means that $(\mathfrak{L(H)},\omega,\tau)$ is
not ergodic by 2.3.3. On the other hand, if such an $x$ does not exist, then
$\dim(P\mathfrak{H})=1$, and hence $(\mathfrak{L(H)},\omega,\tau)$ is ergodic
by 2.3.3.

\bigskip\noindent\textbf{Remarks.} So for example, suppose that $E_{1}$ and
$E_{2}$ are different eigenvalues of a Hamiltonian $H$ such that $E_{2}%
=E_{1}+2\pi n$ for some $n\in\mathbb{Z}$. Let $\Omega$ be a unit eigenvector
of $H$ corresponding to $E_{1}$, and let $x$ be an eigenvector of $H$
corresponding to $E_{2}$, so $x\notin\mathbb{C}\Omega$. Then $e^{-iH}%
\Omega=e^{-iE_{1}}\Omega$ and $e^{-iH}x=e^{-iE_{2}}x=e^{-iE_{1}}x$, which
means that $(\mathfrak{L(H)},\omega,\tau)$ is not ergodic as explained above.

Suppose however that there exists an orthonormal basis $b_{1},b_{2},b_{3},...$
for $\mathfrak{H}$, consisting of eigenvectors of $H$, with corresponding
eigenvalues $E_{1},E_{2},E_{3},...$ where $E_{k}\notin\{E_{1}+2\pi
n:n\in\mathbb{Z}\}$ for $k>1$, which means in particular that $E_{1}$ is a
nondegenerate eigenvalue (though some of the other $E_{k}$'s might be equal to
each other and hence degenerate). Let $\Omega=b_{1}$. Note that $\mathfrak{H}$
could be finite or infinite dimensional. Then the system is ergodic: Consider
any $x\in\mathfrak{H}$. By assumption $x=\sum_{k}\alpha_{k}b_{k}$ for some
$\alpha_{k}\in\mathbb{C}$. The condition $e^{-iH}x=e^{-iE_{1}}x$, or
equivalently $e^{iH}x=e^{iE_{1}}x$, then implies that
\[
\sum_{k}\alpha_{k}e^{iE_{k}}b_{k}=\sum_{k}\alpha_{k}e^{iE_{1}}b_{k}%
\]
which means that $\alpha_{k}=0$ or $e^{iE_{k}}=e^{iE_{1}}$. For $k>0$ the
latter contradicts the fact that $E_{k}\notin\{E_{1}+2\pi n:n\in\mathbb{Z}\}$,
hence $a_{k}=0$. But this means that $x=\alpha_{1}\Omega\in\mathbb{C}\Omega$,
so $P\mathfrak{H}=\mathbb{C}\Omega$ and hence $\dim(P\mathfrak{H})=1$. Note
that instead of saying the system is ergodic, we could also say that the
energy eigenvector $\Omega$ is ergodic.

This is complimentary to 3.2.7 and 3.2.8 where we saw that for a system to be
ergodic, no more than one energy level is allowed to be present in the state
(the intuition being that energy conservation would prohibit \ ``mixing'' if
more than one energy level was present). Now we see that, assuming the
existence of energy eigenvectors, there are even more restrictions for such an
eigenvector to be an ergodic state, namely the energy $E$ of the state should
have a one dimensional eigenspace, and shouldn't differ from any other energy
level by an integer multiple of $2\pi$. In practice it is however quite
possible that in a system where ``theoretically'' these conditions aren't met,
the presence of interactions and slight variations from place to place in the
system might separate a higher dimensional energy eigenspace into lower
dimensional eigenspaces by splitting the single energy level of the eigenspace
into slightly differing energy levels, some of which then could have one
dimensional eigenspaces, which would ensure ergodicity of these energy
eigenvectors if any energy differences of exactly $2\pi n$ that they might
have had with other eigenvectors are also destroyed by the splitting of the
energy levels.$\blacksquare$

\bigskip\noindent\textbf{Example of an ergodic system.} Consider a spin-1/2
particle at a fixed position in a magnetic field. Its state space is
$\mathfrak{H}=\mathbb{C}^{2}$ and its Hamiltonian
\[
H=\left(
\begin{array}
[c]{cc}%
E & 0\\
0 & -E
\end{array}
\right)
\]
where $E\in\mathbb{R}$. If we assume
\begin{equation}
E-(-E)\notin2\pi\mathbb{Z} \tag{3}%
\end{equation}
then this system $(\mathfrak{L(H)},\omega,\tau)$ with the state $\omega$ given
by (1) in terms of the energy eigenvector
\[
\Omega=\left(
\begin{array}
[c]{c}%
1\\
0
\end{array}
\right)
\]
is ergodic as explained above. This can also be checked directly by
considering
\[
A=\left(
\begin{array}
[c]{cc}%
a_{11} & a_{12}\\
a_{21} & a_{22}%
\end{array}
\right)
\]
in which case $\omega(A)=a_{11}$ and
\begin{align*}
\tau(A)  &  =e^{iH}Ae^{-iH}=\left(
\begin{array}
[c]{cc}%
e^{iE} & 0\\
0 & e^{-iE}%
\end{array}
\right)  \left(
\begin{array}
[c]{cc}%
a_{11} & a_{12}\\
a_{21} & a_{22}%
\end{array}
\right)  \left(
\begin{array}
[c]{cc}%
e^{-iE} & 0\\
0 & e^{iE}%
\end{array}
\right) \\
&  =\left(
\begin{array}
[c]{cc}%
a_{11} & a_{12}e^{2iE}\\
a_{21}e^{-2iE} & a_{22}%
\end{array}
\right)
\end{align*}%
\[
\therefore\frac{1}{n}\sum_{k=0}^{n-1}\tau^{k}(A)=\left(
\begin{array}
[c]{cc}%
a_{11} & a_{12}\frac{1}{n}\sum_{k=0}^{n}e^{2iEk}\\
a_{21}\frac{1}{n}\sum_{k=0}^{n}e^{-2iEk} & a_{22}%
\end{array}
\right)
\]
so%
\begin{align*}
&  \left\|  \frac{1}{n}\sum_{k=0}^{n-1}\tau^{k}(A)-\omega(A)\right\|
_{\omega}^{2}\\
&  =\left\|  \left(
\begin{array}
[c]{cc}%
0 & a_{12}\frac{1}{n}\sum_{k=0}^{n}e^{2iEk}\\
a_{21}\frac{1}{n}\sum_{k=0}^{n}e^{-2iEk} & a_{22}-a_{11}%
\end{array}
\right)  \right\|  _{\omega}^{2}\\
&  =\omega\left(  \left(
\begin{array}
[c]{cc}%
0 & \left(  a_{21}\frac{1}{n}\sum_{k=0}^{n}e^{-2iEk}\right)  ^{\ast}\\
\left(  a_{12}\frac{1}{n}\sum_{k=0}^{n}e^{2iEk}\right)  ^{\ast} & \left(
a_{22}-a_{11}\right)  ^{\ast}%
\end{array}
\right)  \left(
\begin{array}
[c]{cc}%
0 & a_{12}\frac{1}{n}\sum_{k=0}^{n}e^{2iEk}\\
a_{21}\frac{1}{n}\sum_{k=0}^{n}e^{-2iEk} & a_{22}-a_{11}%
\end{array}
\right)  \right) \\
&  =\omega\left(
\begin{array}
[c]{cc}%
\left|  a_{21}\right|  ^{2}\left|  \frac{1}{n}\sum_{k=0}^{n}e^{-2iEk}\right|
^{2} & ...\\
... & ...
\end{array}
\right) \\
&  =\left|  a_{21}\right|  ^{2}\left|  \frac{1}{n}\sum_{k=0}^{n}\left(
e^{-2iE}\right)  ^{k}\right|  ^{2}\\
&  \rightarrow0
\end{align*}
as $n\rightarrow\infty$, by the mean ergodic theorem 2.4.1, since
$e^{-2iE}\neq1$ by (3), and $\left|  e^{-2iE}\right|  =1$. So in this direct
way using 2.5.5, we again see that $(\mathfrak{L(H)},\omega,\tau)$ is ergodic.
Note that in this example there are $A$ and $B$ such that $\omega
(AB)\neq\omega(BA)$, as asked in A.2.3.$\blacksquare$

The remark at the end of each reference indicates where in this thesis (apart
from the Introduction) the reference appears.
\end{document}